\documentclass[prb,aps,twocolumn,showpacs]{revtex4-1}
\usepackage{amsmath,amssymb,amsfonts,float}
\usepackage[dvips]{graphicx,color}  
\usepackage{times}       
\usepackage{pifont}
 
\newcommand{\bolS}{\text{\bf S}}

\newcommand{\bolk}{\mathbf{k}}

\newcommand{\VEV}[1]{\langle #1 \rangle}  

\newcommand{\be}{\text{e}}

\newcommand{\rrangle}{\rangle\!\rangle}

\newsavebox{\dotdot}
\savebox{\dotdot}[3mm]{\shortstack{\circle*{0.8}\\ \\ \circle*{0.8}}}

\begin{document}
\title{
Hidden Order and Dynamics in Supersymmetric Valence Bond Solid States \\
-- Super-Matrix Product State Formalism --
}
\author{Kazuki~Hasebe$^{1}$ and Keisuke~Totsuka$^2$}
\affiliation{
$^1$ Kagawa National College of
Technology, Takuma-cho, Mitoyo, Kagawa 769-1192, Japan
\\
$^2$ Yukawa Institute for Theoretical Physics, 
Kyoto University, Kitashirakawa Oiwake-Cho, Kyoto 606-8502, Japan}
\begin{abstract}
Supersymmetric valence bond solid models are 
extensions of the VBS model, a paradigmatic model of  `solvable' gapped 
quantum antiferromagnets, to the case with doped fermionic holes. 
In this paper, we present a detailed analysis of physical properties of the models.  
For systematic studies, a supersymmetric version of the matrix product formalism is developed.  
On 1D chains,  we exactly evaluate the hole-doping behavior of various physical quantities, 
such as the spin/charge excitation spectrum,  superconducting order parameter.    
A generalized hidden order is proposed, and the corresponding string non-local order parameter is also calculated. 
The behavior of the string order parameter is discussed in the light of  
the entanglement spectrum. 
\end{abstract}            
\maketitle
\section{Introduction}
Valence bond solid (VBS) models introduced by Affleck, Kennedy, Lieb and Tasaki \cite{affleck1987rrv,affleck1988vbg} 
are exactly solvable models that 
exemplify the gapped ground states in integer-$S$ spin chains conjectured 
by Haldane\cite{Haldane-83a,Haldane-83b}. 
Though the VBS states, which are the exact ground states of the VBS models, 
are disordered spin liquids in the sense that 
their spin-spin  correlations are exponentially dumped with a very short correlation length,   
there still exists a certain kind of ``hidden order'' captured by the non-local 
string parameter\cite{Kennedy-T-92a,Kennedy-T-92b}.    
The existence of  the hidden order highlights the exotic features of 
the Haldane-gapped antiferromagnets 
which are considered as manifestation of the topological order of quantum spin chain%
\cite{Arovas-A-H-88,Girvin-A-89,hatsugaiJPSJ1992,Oshikawa-92,%
Totsuka-S-94,Totsuka-S-mpg-95}.
With recent increasing interests in the topological states of matter spurred by the discovery of topological insulators [See Ref.\onlinecite{qi-2010-63} for instance as a review], 
the VBS model and its variants are attracting renewed attention.  
Since the VBS states enable us to calculate many interesting quantities 
exactly, they offer a rare theoretical playground 
for the study of topological states of matter.  
Due to their peculiar features,  the VBS-type states have been investigated in 
a wide variety of  contexts like 
quantum information\cite{wolf-2006-97,perezgarcia-2007-7}, 
topological order\cite{pollmann-2009,Pollmann-T-B-O-10}, 
entanglement entropy\cite{katsura-2007-76,katsura-2008-41,xu-2008-133}, 
higher symmetric generalizations\cite{affleck1991qae,arovas2008sss,%
greiter2007ers,greiter2007vbs,schuricht2008vbs,Tu08041685,%
Tu08061839,rachel-2009-86}, and topological phase transitions 
\cite{zang-2010-81,zheng-2010}.

In this paper, we present a detailed analysis of the recently 
proposed\cite{Arovas-H-Q-Z-09} supersymmetric generalization of 
valence bond solid (sVBS) states. \footnote{ sVBS states are the spin-chain 
counterpart of the supersymmetric quantum Hall effect\cite{hasebe2005PRL}.}  
The sVBS states are a precise mathematical realization of Anderson's scenario
of high-$T_\text{c}$ conductivity\cite{anderson1987rvb} and  the idea
of symmetry unification of superconductivity and 
antiferromagnetism\cite{Zhang1997Science}.  
The sVBS states are hole-pair doped VBS states containing  
both the charge sector and the spin sector;   
depending on the magnitude of the hole-doping parameter, they  
exhibit both insulating and superconducting behaviors in the charge sector, 
while in the spin sector it always displays short-range spin 
correlations\cite{Arovas-H-Q-Z-09}.  

The effects of mobile holes in the spin-gapped background are interesting 
in their own right not only in purely theoretical context\cite{Zhang-A-89} but  
also in the experimental point of view\cite{Penc-S-95,xu2000hqs}.    
However, the impact of mobile holes on the (hidden) topological properties 
has been little studied.  
In what follows, we will show that 
the sVBS states possess a kind of non-local topological order in the spin sector 
as well as local superconducting order in the charge sector, 
the latter of which is already known.  
While various (ordinary) correlation functions have been investigated 
already in Ref.\onlinecite{Arovas-H-Q-Z-09}, dynamical properties,  
as exemplified by magnetic- (triplon) and  charge (spinon-hole pair, specifically) 
excitations, are yet to be understood and will also be addressed in this work.   

In the sVBS models, supersymmetry (SUSY), 
i.e. rotational symmetry of boson and fermion, 
is realized as the symmetry of bosonic spins and fermionic holes. 
Such SUSY of the sVBS states is  
 exact  $\it{regardless}$ $\it{of}$ the magnitude of hole-doping parameter, and their parent Hamiltonians can be readily constructed based on such (super)symmetry. 
Thus, the sVBS models enable us to systematically study hole-doped antiferromagnets on such a firm mathematical background. 
To this end, we develop a supersymmetric version of  
 the matrix product state (MPS) representation of the VBS-type states\cite{Totsuka-S-mpg-95}.  
 Since the sVBS states generally contain fermionic degrees of freedom, 
 we generalize the MPS formalism to include both fermionic and bosonic operators. 
This supersymmetric MPS (sMPS) representation is useful not only in the sense of computational efficiency, 
but also from the topological-order point of view as the emergent edge degrees of freedom, 
which characterize the topological features, are automatically incorporated in the MPS 
formalism\cite{Fannes-N-W-92,Klumper-S-Z-92,Totsuka-S-mpg-95}. 
It should also be mentioned that the MPS formalism, which has been 
introduced\cite{Fannes-N-W-92} originally 
as a special class of quantum ground states with short-range correlations, 
is now believed to be a natural  
framework to represent entangled quantum many-body states 
in 1D\cite{Verstraete-C-06,Hastings-area-law-07}.   
In a similar sense, the sMPS formalism would be applicable not only to the sVBS states 
to be investigated in this paper but to a wider class of entangled many-body states 
that contain fermionic degrees of freedom. 
 
This paper is structured as follows. 
In section \ref{sect:generalizehiddenorders}, we introduce type I and type II sVBS states 
and summarize some basic features. 
In section \ref{sec:SUSY-AKLT-1}, by including fermionic degrees of freedom,  we develop the sMPS formalism, and apply it to the calculations of physical quantities of the type I sVBS states. 
The generalized hidden order is proposed and the string order parameter is evaluated in section \ref{sec:hidden-order}. 
In section \ref{sec:singlemodeappro}, we calculate the gapped excitation spectra 
of the magnetic- and the charge (i.e. hole-pair) excitations on sVBS chains within the single-mode approximation.   
In section \ref{sec:SUSY-VBS-2}, we proceed to the analysis of type II sVBS states and derive 
the hole-doping behavior of various physical quantities (e.g. superconducting order parameter 
and string correlation).  
The stability of the hidden `topological' order found in these states is discussed from the point of view of 
the entanglement structure in section VII. 
Section \ref{sec:summary} is devoted to summary and discussions.  

\section{Basic properties}
\label{sect:generalizehiddenorders}
Before proceeding to the detail analysis, 
we quickly review  the basic features of the sVBS states in this section. 
\subsection{Type I SUSY VBS states }
In what follows, we analyze two types of sVBS states.  The first is 
the sVBS states with UOSp(1$|$2) supersymmetry \footnote{%
In Ref.\onlinecite{Arovas-H-Q-Z-09}, the symmetry is referred to OSp(1$|$2), but OSp(1$|$2) 
can also denote a non-compact supergroup whose bosonic subgroup is 
Sp(2,$\mathbb{R}$)$\simeq$SU(1,1) or Sp(2,$\mathbb{C}$)$\simeq$SO(3,1). 
To avoid possible confusions, we utilize the more precise terminology, 
UOSp(1$|$2), in the present paper. } proposed recently in Ref.\onlinecite{Arovas-H-Q-Z-09} 
(see Appendix \ref{crashsusy}, for a very brief summary of supersymmetry), 
which we shall call type I: 
\begin{equation}
|\text{sVBS-I}\rangle=\prod_{\langle ij\rangle} 
(a_i^{\dagger}b_j^{\dagger}-b_i^{\dagger}a_j^{\dagger}
-r f_i^{\dagger}f_j^{\dagger})^M|\text{vac} \rangle,
\label{sVBSstateI}
\end{equation}
where $\langle ij\rangle$ signifies a pair of adjacent sites $(i,j)$ and 
$r$ stands for the hole doping parameter.  The operators $a_i$, $b_i$ and $f_i$ 
respectively are a pair of the standard Schwinger bosons satisfying 
$[a_i,a^{\dagger}_j]=[b_i,b^{\dagger}_j]=\delta_{ij}$ 
and a (spinless) fermion satisfying
$\{f_i,f^{\dagger}_j\}=\delta_{ij}$.   
The vacuum $|\text{vac}\rangle$ is annihilated by both the boson and the fermion:  
$a|\text{vac}\rangle=b|\text{vac}\rangle=f|\text{vac}\rangle=0$. 
Since the fermions always appear in pairs of the form 
$f_i^{\dagger}f_j^{\dagger}$ ($i$, $j$ are adjacent), 
the sVBS states can be regarded as the {\em hole-pair doped} VBS states.   
One can easily see that the state $|\text{sVBS-I}\rangle$ is UOSp(1$|$2)-invariant 
from the invariance of the matrix 
\begin{equation}
\mathcal{R}_{\text{I}}=
\begin{pmatrix}
0 & 1 & 0 \\
-1 & 0 & 0 \\
0 & 0 & -1 
\end{pmatrix}
\end{equation}
used to construct $|\text{sVBS-I}\rangle$ (The parameter $r$ is absorbed 
in the renormalization of $f$. To see how the matrix $\mathcal{R}_{\text{I}}$ is 
related to $|\text{sVBS-I}\rangle$, see section ~\ref{sec:MPS-construction}), 
and hence $|\text{sVBS-I}\rangle$ has the  UOSp(1$|$2) symmetry [See Appendix \ref{subsec:uosp12} for more details]. 

The type-I sVBS states\cite{Arovas-H-Q-Z-09} (\ref{sVBSstateI}),  
that contain (fermionic) hole degrees of freedom as well as the (bosonic) spin ones,  
are a generalization 
of the standard spin-$S$ VBS states\cite{affleck1987rrv,affleck1988vbg,Arovas-A-H-88} 
. 
In the type-I SVB states (\ref{sVBSstateI}),  
the total particle number at each site is conserved:  
\begin{equation}
z M=a^{\dagger}_ia_i+b^{\dagger}_ib_i+f^{\dagger}_if_i \; ,
\end{equation}
where the lattice coordination number $z$ is $2d$ for 
the $d$-dimensional hypercubic lattice (in what follows, $z=2$ unless 
otherwise stated). The integer $zM$ plays a role of the spin quantum 
number $2S$ in the usual VBS states.  
Since $f^{\dagger}f$ takes either 0 or 1, the following 
two eigenvalues are possible for the local spin quantum number 
$S_i=\frac{1}{2}(a^{\dagger}_ia_i+b^{\dagger}_ib_i)$:  
\begin{equation}
S_i=M,~ M-\frac{1}{2}.
\end{equation}
In particular, for $M=1$, each site can take two spin values
\begin{equation}
S_i=1,~ \frac{1}{2}, 
\label{M1typeIsVBS}
\end{equation}
and the local Hilbert space is spanned by the five ($4M{+}1$, in general) basis states 
\begin{equation}
\begin{split}
&|1\rangle=\frac{1}{\sqrt{2}}{a_i^{\dagger}}^2|\text{vac}\rangle,
\;\; |0\rangle={a_i^{\dagger}b_i^{\dagger}}|\text{vac}\rangle, 
\;\; |{-}1\rangle=\frac{1}{\sqrt{2}}{b_i^{\dagger}}^2|\text{vac}\rangle, \\
&\quad\quad\quad |\!\uparrow\rangle=
 a_i^{\dagger}f_i^{\dagger}|\text{vac}\rangle, 
\;\; |\!\downarrow \rangle =b_i^{\dagger}f_i^{\dagger}|\text{vac}\rangle \; .
\end{split}
\end{equation}
Mathematically, these constitute an $\mathcal{N}{=}1$ SUSY multiplet, 
and hence we use the name `Type I'.
In addition to the local physical degrees of freedom on each site, the following emergent 
degrees of freedom localized around the edges ({\em edge states})  
will play an important role: 
\begin{equation}
|\!\uparrow\rangle\!\rangle=a^{\dagger}|\text{vac}\rangle, \;\;
|\!\downarrow\rangle\!\rangle=b^{\dagger}|\text{vac}\rangle, \;\;
|0\rangle\!\rangle=f^{\dagger}|\text{vac}\rangle.
\end{equation}
As we will see in section \ref{sec:MPS-construction}, 
the ground state of a finite {\em open} chain is 9-fold degenerate (corresponding 
to the $3\times 3$ matrix for the $M=1$ type-I sVBS states). 

The $M=1$ type-I sVBS chain interpolates between the two VBS states 
in the two extremal limits of the hole doping: 
at $r\rightarrow 0$,  $|\text{sVBS-I}\rangle$  reproduces the original spin-1 VBS state $|\text{VBS}\rangle$%
\cite{affleck1987rrv,affleck1988vbg}
\begin{equation}
|\text{sVBS-I}\rangle \rightarrow |\text{VBS}\rangle=\prod_{i}
(a^{\dagger}_i b_{i+1}^{\dagger}-b_i^{\dagger}a_{i+1})|\text{vac} \rangle, 
\end{equation}
while, in the limit $r\rightarrow \infty$,   $|\text{sVBS-I}\rangle$  reduces 
to the Majumdar-Ghosh (MG) dimer state\cite{Majumdar-G-69,Majumdar-70}  
$|\text{MG}\rangle$ 
\begin{equation}
|\text{sVBS-I}\rangle  \rightarrow 
\left\{
\prod_i  f^{\dagger}_i 
\right\}
|\text{MG}\rangle, 
\end{equation}
where $|\text{MG}\rangle$ is either of the two dimerized states of 
the MG model\footnote{%
The open boundary condition has been implicitly assumed here; 
if the periodic boundary condition had been used, the two states would have been summed 
up with a minus sign due to the anti-commutating property of the holes.
}:
\begin{equation}
|\text{MG}\rangle=
\begin{cases}
\prod_{i:\text{even}}(a^{\dagger}_i b^{\dagger}_{i+1} - b^{\dagger}_i a^{\dagger}_{i+1})|\text{vac}\rangle 
\\
\prod_{i:\text{odd}}(a^{\dagger}_i b^{\dagger}_{i+1} - b^{\dagger}_i a^{\dagger}_{i+1})|\text{vac}\rangle
\; .
\end{cases}
\end{equation}
For larger $M$, $|\text{MG}\rangle$ should be replaced with the inhomogeneous VBS 
states\cite{Arovas-A-H-88} where the number of valence bonds alternates from bond 
to bond. 

\begin{figure}[!t]
\centering
\includegraphics[width=7cm]{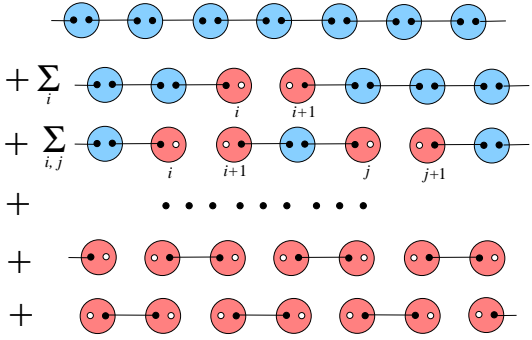}
\caption{The type I sVBS is a superposed state of hole-pair doped VBS states. 
With finite hole-doping parameter $r$, all of the hole-pair doped VBS states are superposed to form the sVBS state, and the sVBS state exhibits the superconducting property. At $r=0$, the sVBS state is reduced to the original VBS state (depicted as the first chain), while $r\rightarrow \infty$, the sVBS state is reduced to the MG dimer state (depicted as the last two chains). \label{expSUSY.fig} }
\end{figure}

According to the spin-hole coherent state formalism \cite{auerbach1994iea}, the sVBS state is expressed as 
\begin{equation}
\Psi_{\text{sVBS-I}}= \prod_{\langle ij \rangle}(u_iv_j-v_iu_j-r\theta_i\theta_j)^M, 
\end{equation}
which is simply obtained by replacing the operators $a,b,f$ with their corresponding classical counterparts $u,v,\theta$.  ($u,v$ are Grassmann even quantities, while $\theta$ is Grassmann odd.) 
From the Grassmann odd properties of $\theta$,   $\Psi_{\text{sVBS-I}}$ can be rewritten as  
\begin{equation}
\Psi_{\text{sVBS-I}}=\exp\biggl(-{Mr\sum_{\langle ij \rangle} 
\frac{\theta_i\theta_j}{u_iv_j-v_iu_j}} \biggr) \cdot \Phi_{\text{VBS}}
\end{equation}
where $\Phi_{\text{VBS}}=\prod_{\langle ij \rangle}(u_iv_j-v_iu_j)^M$ is the spin coherent state 
representation of the original VBS state.  
This expression reminds the BCS wavefunction of the superconductivity; $|\text{BCS}\rangle =\prod_k (1+g_k c_k^{\dagger}c_{-k}^{\dagger})|0\rangle = \exp(\sum_k g_k c_k^{\dagger}c_{-k}^{\dagger} )|0\rangle$ with electron operator $c_{k}$ and coherence factor $g_k$ (See chapter 2-4 in Ref.~\onlinecite{SchriefferBook}).  
In both $\Psi_{\text{sVBS-I}}$ and $|\text{BCS}\rangle$, the fermions alway appear in pairs and the wavefunctions can be expressed by a superposition of such fermion pairs, as  demonstrated by expanding the exponential (See Fig.\ref{expSUSY.fig}).  

\subsection{Type II sVBS states}
\label{eqn:type-II-intro}
The type II sVBS state is an extension of the previous series 
of VBS states (type I) and now contains doped (antisymmetric) bound pairs 
of two species of holes. 
The inclusion of  two species of holes $f$ and $g$ allows us to write down 
a wavefunction more symmetric with respect to the bosonic- 
 and the fermionic degrees of freedom. 
Now, we introduce the type II sVBS states of the form:  
\begin{equation}
|\text{sVBS-II}\rangle=\prod_{\langle ij\rangle} 
(a_i^{\dagger}b_j^{\dagger}-b_i^{\dagger}a_j^{\dagger}-rf_i^{\dagger}g_j^{\dagger}-rg_i^{\dagger}f_j^{\dagger})^M|\text{vac} \rangle,
\label{sVBSstateII}
\end{equation}
which is associated with another matrix:
\begin{equation}
{\cal R}_{\text{II}} = 
\begin{pmatrix}
0 & 1 & 0 & 0\\
-1 & 0 & 0 & 0\\
0 & 0 & 0 &-1 \\
0 & 0 & -1 & 0  
\end{pmatrix}  
\; .
\end{equation}
The new fermion $g_i$ satisfies the standard anti-commutation relations 
$\{g_i,g^{\dagger}_j\}=\delta_{ij}$, $\{f_{i},g_{j}\}=0$, etc. 
Apparently, the type II sVBS state reduces to the type I 
after $g_i\rightarrow f_i$ (and the due rescaling $r\rightarrow \frac{1}{2}r$). 
With inclusion of another species of the (spinless) hole, in the type II VBS states, 
there appear the local sites $f^{\dagger}_i g_i^{\dagger}|0\rangle$ with spin-0, 
which are not realized in the type I sVBS states. 
As we will show in the end of this section, 
the type II sVBS states have the UOSp(2$|$2) symmetry larger than 
UOSp(1$|$2) symmetry of the type I sVBS states. 

We have two species of fermions, and the total particle number at each site $i$ is 
constrained by 
\begin{equation}
zM=a^{\dagger}_ia_i+b^{\dagger}_ib_i+f^{\dagger}_if_i+g^{\dagger}_ig_i, 
\end{equation}
where $z$ is the lattice coordination number. 
Since the eigenvalues of $n_{f}(i){=}f^{\dagger}_{i}f_{i}$ and 
$n_{g}(i){=}g^{\dagger}_{i}g_{i}$ can take either 0 or 1, 
in the type II sVBS chain $(z=2)$, the following four eigenvalues are allowed 
for the local spin quantum number $S_i=\frac{1}{2}(a^{\dagger}_ia_i+b^{\dagger}_ib_i)$: 
\begin{equation}
S_i={M},~ M-\frac{1}{2},~M-\frac{1}{2},~ M-1,
\end{equation}
which respectively correspond to the possible combinations of the fermion numbers:
\begin{equation}
(n_f(i),n_g(i))=(0,0),~(1,0),~(0,1),~(1,1) \; .
\end{equation}
In particular, for the $M=1$ sVBS chain (i.e. $z=2$), the possible values read
\begin{equation}
S_i=1,~\frac{1}{2},~\frac{1}{2},~ 0 \; .
\end{equation}
Therefore, the local Hilbert space is spanned by the following nine basis states 
\begin{align}
&|1\rangle=\frac{1}{\sqrt{2}}{a_i^{\dagger}}^2|\text{vac}\rangle,\quad 
|0\rangle={a_i^{\dagger}}b_i^{\dagger}|\text{vac}\rangle, \quad 
|{-1}\rangle=\frac{1}{\sqrt{2}}{b_i^{\dagger}}^2|\text{vac}\rangle, \nonumber\\
& ~~~~~~~~~|\!\uparrow\rangle=a_i^{\dagger}f_i^{\dagger}|\text{vac}\rangle, \quad 
|\!\downarrow\rangle=b_i^{\dagger}f_i^{\dagger}|\text{vac}\rangle,\nonumber\\
& ~~~~~~~~~|\!\uparrow'\rangle=a_i^{\dagger}g_i^{\dagger}|\text{vac}\rangle, \quad 
|\!\downarrow'\rangle=b_i^{\dagger}g_i^{\dagger}|\text{vac}\rangle,\quad\nonumber\\
& ~~~~~~~~~ ~~~~~~~~~|0'\rangle=g_i^{\dagger}f_i^{\dagger}|\text{vac}\rangle.
\end{align}
The name `type II' is indicative of an $\mathcal{N}{=}2$ SUSY multiplet formed by 
these states.  
Again, $|\text{vac}\rangle$ denotes the vacuum with respect to 
$(a,b,f,g)$.  
The edge states are now given by  
\begin{equation}
\begin{split}
& |\!\uparrow\rrangle=a^{\dagger}|\text{vac}\rangle, \;\;
|\!\downarrow\rrangle=b^{\dagger}|\text{vac}\rangle, \\
& |0\rrangle=f^{\dagger}|\text{vac}\rangle, \;\; |0'\rrangle=g^{\dagger}|\text{vac}\rangle, 
\end{split}
\end{equation}
and, correspondingly, there appear $4\times 4=16$ degenerate ground states 
for the $M=1$ type-II sVBS chain (see section ~\ref{sec:SUSY-VBS-2} for the detail). 

The $M=1$ sVBS chain has the following properties.  
As in the type I sVBS, it reproduces the pure spin VBS state 
for $r\rightarrow 0$: 
\begin{equation}
|\text{sVBS-II}\rangle \rightarrow |\text{VBS}\rangle
=\prod_{i}(a^{\dagger}_i b_{i+1}^{\dagger}-b_i^{\dagger}a^{\dagger}_{i+1})|\text{vac}\rangle 
\; .
\end{equation}
On the other hand, when $r\rightarrow \infty$, it reduces to the  
{\em totally uncorrelated} fermionic (F) state filled with holes:
\begin{equation}
\begin{split}
|\text{sVBS-II}\rangle \rightarrow |\text{F-VBS}\rangle
& \equiv \prod_{i} 
(f^{\dagger}_ig^{\dagger}_{i+1}+
 g^{\dagger}_if^{\dagger}_{i+1})|\text{vac}\rangle \\
& = 
\pm \prod_{i}f^{\dagger}_i g^{\dagger}_i|\text{vac}\rangle 
\end{split}
\end{equation}
(the sign factor depends on both the parity of the system size and the edge states).
Here we have assumed the open boundary condition\footnote{%
If the periodic boundary condition is used, we have zero state for odd-length chains.} 
and the normalization of the edge states given in 
section ~\ref{sec:MPS-type-II}.  

\begin{figure}[!t]
\centering
\includegraphics[width=7cm]{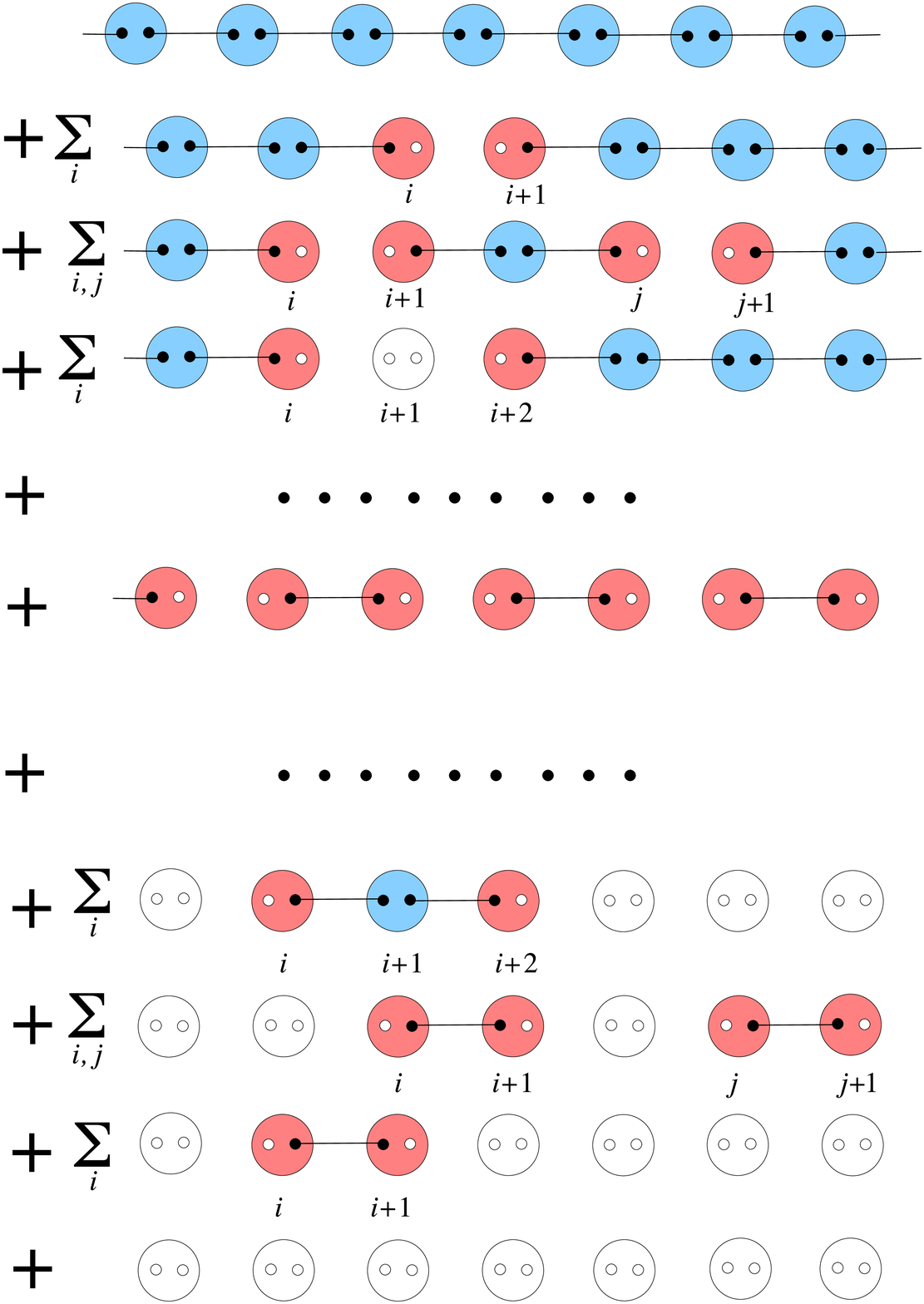}
\caption{(color online) The type II sVBS states are expressed as superposition of the 
hole-pair doped VBS states. Unlike the type I sVBS states, 
the spinless sites, depicted by the large white circles with double holes, appear. 
The MG states are realized in the ``middle'' of the sequence. 
The original VBS state and the hole-VBS state are respectively realized 
in the first and last lines.%
 \label{expNewSUSY.fig} }
\end{figure}

This properties are quite similar to those of the BCS state; 
at $g_k\rightarrow 0$, the BCS state reduces to the electron vacuum (no fermions), 
while for $g_k\rightarrow\infty$, it coincides with the Fermi sphere (filled with electrons).   
In this sense, the type II sVBS states look more similar to the BCS state  
than the type I VBS states. 
As in the previous case, one can pursue the analogy to the BCS wave function by using 
the spin-hole coherent state representation of $|\text{sVBS-II}\rangle$:  
\begin{align}
\Psi_{\text{sVBS-II}}&= \prod_{<ij>}(u_iv_j-v_iu_j-r\theta_i\eta_j-r\eta_i\theta_j)^M\nonumber\\
&=\exp\biggl(-Mr\sum_{<ij>}\frac{\theta_i\eta_j+\eta_i\theta_j}{u_iv_j-v_iu_j}  \biggr)\nonumber\\
&~~\cdot \exp\biggl(-Mr^2\sum_{<ij>}\frac{\theta_i\eta_i \theta_j\eta_j}{(u_iv_j-v_iu_j)^2}   \biggr)          
{\cdot}  \Phi_{\text{VBS}}. 
\end{align}
Expanding the exponentials, one can easily see that with finite $r$, 
the type II sVBS states can be expressed as a superposition of the hole-pair-doped VBS states 
and that the system exhibits the superconducting property.  
However, unlike type I, type II sVBS states have no spin degrees of freedom at $r\rightarrow \infty$. 
The intuitive picture of the $M=1$ type II sVBS chain is depicted in Fig.\ref{expNewSUSY.fig}.

Before concluding this section, we give a remark about the symmetry of the type II sVBS state.  
In Ref.~\onlinecite{Arovas-H-Q-Z-09}, an apparently different form of the  
 sVBS states    
\begin{equation}
|\text{sVBS}^{\prime}\rangle=\prod_{\langle ij \rangle}(a_i^{\dagger}b_j^{\dagger}
-b_i^{\dagger}a_j^{\dagger}-rf_i^{\dagger}f_j^{\dagger}
-rg_i^{\dagger}g_j^{\dagger})^M|\text{vac}\rangle, 
\end{equation}
has been introduced.  
The state $|\text{sVBS}^{\prime}\rangle$ is 
manifestly invariant under the UOSp(2$|$2) transformation, 
since it is constructed by using the UOSp(2$|$2)-invariant matrix
\begin{equation} 
\mathcal{R}'=
\begin{pmatrix}
0 & 1 & 0  & 0 \\
-1 & 0 & 0 & 0 \\
0 & 0 & -1 & 0 \\
0 & 0 & 0 & -1 
\end{pmatrix} \; .
\end{equation}
($r$ can be absorbed in the normalization of $f$ and $g$.)   
In fact, the two sVBS states  $|\text{sVBS-II}\rangle$ and $|\text{sVBS'}\rangle$ 
are physically equivalent.  
By the unitary transformation
\begin{equation}
\begin{pmatrix}
f^{\dagger}\\
g^{\dagger}
\end{pmatrix}
\rightarrow 
\frac{1}{\sqrt{2}}
\begin{pmatrix}
1 & -1 \\
1 & 1 
\end{pmatrix} \begin{pmatrix}
f^{\dagger}\\
g^{\dagger}
\end{pmatrix},
\end{equation}
the fermion-pair part of $|\text{sVBS-II}\rangle$ is transformed to 
\begin{equation}
f_i^{\dagger}g_{j}^{\dagger}+g_i^{\dagger}f_j^{\dagger}\longrightarrow 
f_i^{\dagger}f_{j}^{\dagger}-g_i^{\dagger}g_j^{\dagger}. 
\end{equation}
Then, we flip the sign of either $g_i^{\dagger}$ or $g_j^{\dagger}$ to recover the correct form of 
 the fermion-pair part in $|\text{sVBS}^{\prime}\rangle$: 
\begin{equation}
f_i^{\dagger}f_{j}^{\dagger}-g_i^{\dagger}g_j^{\dagger}\longrightarrow 
f_i^{\dagger}f_{j}^{\dagger}+g_i^{\dagger}g_j^{\dagger}. 
\end{equation}
As the phase of the operators can be chosen arbitrarily, flipping the sign of them 
does not affect physics. 
Therefore, both $|\text{sVBS-II}\rangle$ and $|\text{sVBS}^{\prime}\rangle$ have the same symmetry 
UOSp(2$|$2) in common and are physically equivalent; all physical quantities 
take completely identical values for these two states{\cite{supp}}. 

\section{SUSY-VBS state-I}
\label{sec:SUSY-AKLT-1}
In the following sections, we consider the sVBS states defined 
on one-dimensional (1D) chain.  
A simplest SUSY-extension of the 1D spin-1 AKLT (VBS) 
state\cite{affleck1987rrv,affleck1988vbg} is defined 
as ($M=1$, $z=2$ in eq.(\ref{sVBSstateI})):
\begin{equation}
\begin{split}
|\text{sVBS-I}\rangle 
\equiv & (\cdots)
(a^{\dagger}_{j-1}b^{\dagger}_{j}-b^{\dagger}_{j-1}a^{\dagger}_{j}
-r\, f^{\dagger}_{j-1}f^{\dagger}_{j}) \\
& (a^{\dagger}_{j}b^{\dagger}_{j+1}-b^{\dagger}_{j}a^{\dagger}_{j+1}
-r\, f^{\dagger}_{j}f^{\dagger}_{j+1}) (\cdots)|\text{vac}\rangle \; .
\end{split}
\label{eqn:def-SUSY-AKLT-1}
\end{equation}
The (non-hermitian\footnote{%
As has been argued in Ref.\onlinecite{Arovas-H-Q-Z-09}, 
this non-hermiticity is readily cured by adopting 
$P^{\dagger}_{L}({\cal C}_{j,j+1})P_{L}({\cal C}_{j,j+1})$ 
instead of $P_{L}({\cal C}_{j,j+1})$ itself.}) 
parent Hamiltonian for the SUSY (UOSp($1|2$)) VBS model is given as%
\cite{Arovas-H-Q-Z-09}:
\begin{equation}
\widetilde{{\cal H}}_{l=1\text{ sVBS}}
= \sum_{j}\left\{
V_{3/2}P_{3/2}({\cal C}_{j,j+1})
+ V_{2}P_{2}({\cal C}_{j,j+1})
\right\} \; , 
\label{eqn:parent-Ham}
\end{equation}
where ${\cal C}_{j,j+1}$ and $P_{l}({\cal C}_{j,j+1})$ respectively 
denote 
the UOSp($1|2$) Casimir operator on a two-site cluster $(j,j+1)$ 
(see eqs.(\ref{eqn:def-Casimir},\ref{eqn:Casimir-2site}) 
for the definition of Casimir operators)
and the projection operator onto $l_{\text{tot}}=l$ subspace 
(note that the total superspin $l_{\text{tot}}$ of two $l=1$ superspins 
can take all integer- and half-integer values between 0 and 2; 
see eq.(\ref{eqn:CG-decomp})).  
For the positivity of the Hamiltonian, we require 
$V_{3/2},V_{2} \gneq 0$.    
Specifically, the local Hamiltonian $h_{j,j+1}$ is given by the following 
fourth-order polynomial of the Casimir ${\cal C}_{j,j+1}$:
\begin{multline}
h({\cal C})=
\left(\frac{V_{3/2}}{6}-\frac{V_{2}}{70}\right) {\cal C}
+\left(\frac{3V_{2}}{70}-\frac{43 V_{3/2}}{90}\right) {\cal C}^2 \\
+\left(\frac{14V_{3/2}}{45}-\frac{2V_{2}}{63}\right) {\cal C}^3 
+ \left(\frac{2V_{2}}{315}-\frac{2V_{3/2}}{45}\right) {\cal C}^4 \; .
\end{multline}
\subsection{Matrix-product representation}
\label{sec:MPS-construction}
First let us briefly recapitulate the basic properties of 
a generic (bosonic) matrix-product state of the following form 
(see, for instance, Refs.~\onlinecite{perezgarcia-2007-7,Verstraete-M-C-08} 
for recent reviews of the matrix-product representations):
\begin{subequations}
\begin{equation}
|\text{MPS}\rangle 
= \bigotimes_{j=1}^{L} A_{j}  \; ,
\end{equation}
where the matrix $A_{j}$ consists of state vectors at 
the site-$j$ and its size is determined solely by the size of 
the auxiliary Hilbert space and is independent of the number of sites\footnote{%
Of course, we can construct `polymerized' matrix-product states 
where $m$s alternate with certain periods.}.   
The state $|\text{MPS}\rangle$ in general is not normalized and 
we reserve the notation $|\text{MPS}\rangle$ (and $|\text{sVBS}\rangle$) for the unnormalized 
states. 
Ground states which can be expressed in this form may be 
generically expected to have finite degeneracy.  For example, 
the ground state of the AKLT model, which is expressed by 
the spin-$S$ VBS state, is shown\cite{affleck1987rrv,affleck1988vbg} 
to have $(S{+}1){\times}(S{+}1)$-fold 
degenerate, when the model is defined on a finite {\em open} chain.   
When the system is defined on a periodic chain, we have to take 
the trace over the matrix indices:
\begin{equation}
|\text{MPS}\rangle_{\text{PBC}} 
= \text{Tr}\left\{
\bigotimes_{j=1}^{L} A_{j}  
\right\}
\; ,
\label{eqn:MPS-PBC-1}
\end{equation}
\end{subequations} 
Below, we shall see that the expression eq.(\ref{eqn:MPS-PBC-1}) 
should be modified when $A$ contains {\em both} bosonic degrees 
of freedom and fermionic ones.  

Now let us construct the matrix-product representation%
\cite{Fannes-N-W-89,Fannes-N-W-92} of the type I (UOSp(1$|$2)) VBS state  
(\ref{eqn:def-SUSY-AKLT-1}).  When the Schwinger-boson/fermion 
representation of the state is known, the simplest way\cite{Totsuka-S-94} 
would be to find an operator-valued 
matrix in such a way that everytime when we multiply a new matrix 
(say, $g_{j+1}$) from the right the (SUSY) valence-bond operator 
\[
 (a^{\dagger}_{j}b^{\dagger}_{j+1}-b^{\dagger}_{j}a^{\dagger}_{j+1}
-r\, f^{\dagger}_{j}f^{\dagger}_{j+1})
\] 
is inserted between the previous right edge (site-$j$) 
and the newly added site ($j{+}1$).  
To this end, let us introduce the `spinor':
\begin{equation}
\psi_{j}=(a^{\dagger}_{j}, b^{\dagger}_{j}, \sqrt{r}f_{j}^{\dagger})^{\text{t}}   \; ,
\end{equation}
in terms of which the above UOSp(1$|$2) valence bond can be written 
compactly as:
\begin{equation}
 (a^{\dagger}_{j}b^{\dagger}_{j+1}-b^{\dagger}_{j}a^{\dagger}_{j+1}
-r\, f^{\dagger}_{j}f^{\dagger}_{j+1})
= \psi_{j}^{\text{t}}{\cal R}_{\text{I}}\psi_{j+1} 
\label{eqn:SUSY-VB-spior}
\end{equation}
(`t' denotes the transposition).  The `metric' ${\cal R}_{\text{I}}$ has been defined as
\begin{equation}
{\cal R}_{\text{I}} = 
\begin{pmatrix}
0 & 1 & 0 \\
-1 & 0 & 0 \\
0 & 0 & -1 
\end{pmatrix}  \; .
\label{eqn:metric}
\end{equation}
Then the sVBS state (\ref{eqn:def-SUSY-AKLT-1}) is written as  
a string of 3$\times$3 matrices $(\alpha,\beta=1,2,3)$:
\begin{equation}
\begin{split}
|\text{sVBS}\rangle_{\alpha\beta} &= 
({\cal R}_{\text{I}}\psi_1)^{\alpha} \prod_{i=1}^{L-1}  
(\psi_i^{\text{t}}\mathcal{R}_{\text{I}} \psi_{i+1})  
\psi^{\beta}_L |\text{vac}\rangle \\
& \equiv ({\cal R}_{\text{I}}\psi_1)^{\alpha} \psi_1^{\text{t}}
\cdot \left(\prod_{i=2}^{L-1} \mathcal{R}_{\text{I}} \psi_{i}\psi_i^{\text{t}} \right)
 \cdot ({\cal R}_{\text{I}}\psi_L \psi_L^{\beta}) |\text{vac}\rangle \\
&=(A_1 A_2  \cdots  A_{L})_{\alpha\beta} ,
\end{split} 
\label{eqn:SUSY-MPS}
 \end{equation}
 where 
\begin{equation}
\begin{split}
A_j &= \mathcal{R}_{\text{I}}\psi_j \cdot\psi_j^{\text{t}} |\text{vac}\rangle_{j} \\
&=    
\begin{pmatrix}
a_j^{\dagger}b_j^{\dagger}
& (b_j^{\dagger})^2 &
\sqrt{r}b_j^{\dagger}f_j^{\dagger} \\
-(a_j^{\dagger})^2  & -a_j^{\dagger}b_j^{\dagger} 
& -\sqrt{r} a_j^{\dagger}f_j^{\dagger} \\
-\sqrt{r}f_j^{\dagger}a_j^{\dagger}  & 
-\sqrt{r}f_j^{\dagger}b_j^{\dagger} & 0 
\end{pmatrix} 
|\text{vac}\rangle_{j} \\
&= 
\begin{pmatrix}
|0\rangle_{j}
& \sqrt{2}|{-1}\rangle_{j} &
\sqrt{r}|\!\downarrow\rangle_{j} \\
-\sqrt{2}|1\rangle_{j}  & -|0\rangle_{j}  
& -\sqrt{r} |\!\uparrow\rangle_{j} \\
-\sqrt{r}|\!\uparrow\rangle_{j}  & 
-\sqrt{r}|\!\downarrow\rangle_{j} & 0 
\end{pmatrix} \\
& \equiv \sum_{m=-1,0,1}
\Gamma^{(\text{B})}(m)|m\rangle
+ \sum_{\widetilde{m}=\uparrow,\downarrow}
\Gamma^{(\text{F})}(\widetilde{m})|\widetilde{m}\rangle 
\; .
\end{split}
\label{eqn:SUSY-g-matrix}
\end{equation}
The 3$\times$3 matrices $\Gamma^{(\text{B})}$ and $\Gamma^{(\text{F})}$ 
respectively denote the bosonic- and the fermionic part.  
The edge operators 
${\cal R}_{\text{I}}\psi_{1}=(b_{1}^{\dagger},-a_{1}^{\dagger},-\sqrt{r}f_{1}^{\dagger})^{\text{t}}$ and 
$\psi_{L}$ appearing respectively on the left- and the right edge 
represent the three possible edge states (spin-up/down and hole) on each edge. 

Following the same steps as the above for 
\begin{equation}
\begin{split}
\langle \text{sVBS-I}| =& \langle \text{vac} |
(\cdots)(a_{j}b_{j+1}-b_{j}a_{j+1}-r\, f_{j+1}f_{j})\\
& (a_{j-1}b_{j}-b_{j-1}a_{j}-r\, f_{j}f_{j-1})(\cdots) \; ,
\end{split} 
\end{equation}
we obtain 
\begin{equation}
{}_{\alpha\beta}\langle \text{sVBS-I}| = \left(
A_{L}^{\dagger}A_{L-1}^{\dagger}\cdots A_{2}^{\dagger}A_{1}^{\dagger}
\right)_{\beta\alpha}
\end{equation}
with 
\begin{equation}
\begin{split}
A^{\dagger}_{j}
&= 
{}_{j}\langle \text{vac}|\psi^{\ast}_{j}\psi^{\dagger}_{j}{\cal R}^{\text{t}}
\\
&=
{}_{j}\langle \text{vac}|
\begin{pmatrix}
a_{j}b_{j} & -(a_{j})^{2} & -\sqrt{r}a_{j}f_{j} \\
(b_{j})^{2} & -a_{j}b_{j} & -\sqrt{r}b_{j}f_{j} \\
\sqrt{r}f_{j}b_{j} & -\sqrt{r}f_{j}a_{j} & 0 
\end{pmatrix} \; ,
\end{split}
\end{equation}
where $\psi^{\ast}_{j}\equiv (a_{j},b_{j},\sqrt{r}f_{j})^{\text{t}}$. 

By construction, it is obvious that 
{\em all} the nine matrix elements of 
the following string of $A$-matrices:
\begin{multline}
\bigotimes_{j=1}^{L} A_{j} \\
= 
\begin{pmatrix}
b^{\dagger}_{1} \\ -a^{\dagger}_{1} \\ -\sqrt{r}f^{\dagger}_{1} 
\end{pmatrix}
\left\{
\prod_{j=1}^{L-1}
(a^{\dagger}_{j}b^{\dagger}_{j+1}-b^{\dagger}_{j}a^{\dagger}_{j+1}
-r\, f^{\dagger}_{j}f^{\dagger}_{j+1})
\right\} \\
\times
\begin{pmatrix}
a^{\dagger}_{L} & b^{\dagger}_{L} & \sqrt{r}f^{\dagger}_{L} 
\end{pmatrix}
|\text{vac}\rangle
\label{eqn:MPS-open}
\end{multline}
are the (zero-energy) ground states of the parent Hamiltonian $\sum_{j=1}^{L-1}h_{j,j+1}$.  
That is, the product $\bigotimes_{j=1}^{L} A_{j}$ gives the 
the ground states of the $M=1$ sVBS model  
on an {\em open} chain with length $L$.   
Here it is important to note that we are free to choose 
the polynomials ($\begin{pmatrix}
b^{\dagger}_{1} & -a^{\dagger}_{1} & -\sqrt{r}f^{\dagger}_{1} 
\end{pmatrix}$ from the left edge and 
$\begin{pmatrix}
a^{\dagger}_{L} & b^{\dagger}_{L} & \sqrt{r}f^{\dagger}_{L} 
\end{pmatrix}$ from the right) appearing at the edges.  
As will be discussed in section \ref{sec:edge-states}, 
this leads to a remarkable feature of the VBS-like systems--%
{\em edge states}. 

In constructing the sVBS state on a {\em periodic} chain, one  
has to treat the fermion sign carefully and one sees that 
the trace operation used in the standard MPS representation 
(\ref{eqn:MPS-PBC-1}) should be 
replaced with the {\em supertrace} 
(see Appendix \ref{sec:quick-recipe}):
\begin{subequations}
\begin{equation}
|\text{sVBS}\rangle_{\text{periodic}}
= \text{STr}\left\{
\bigotimes_{j=1}^{L} A_{j}
\right\} \; ,
\label{eqn:MPS-PBC-2}
\end{equation}
where the supertrace here is defined as 
\begin{equation}
\text{STr}({\cal M}) \equiv {\cal M}_{11}+{\cal M}_{22}-{\cal M}_{33} \; .
\label{eqn:STr-1}
\end{equation}
\end{subequations}

From these $A$-matrices, we can calculate the following 9$\times$9 $T$-matrices 
({\em transfer matrix}):
\begin{equation}
\begin{split}
& T(\bar{\alpha},\alpha;\bar{\beta},\beta) \equiv A^{\ast}(\bar{\alpha},\bar{\beta})
A(\alpha,\beta) \\
& = 
\left(
\begin{array}{lllllllll}
 1 & 0 & 0 & 0 & 2 & 0 & 0 & 0 & r \\
 0 & -1 & 0 & 0 & 0 & 0 & 0 & 0 & 0 \\
 0 & 0 & 0 & 0 & 0 & 0 & 0 & -r & 0 \\
 0 & 0 & 0 & -1 & 0 & 0 & 0 & 0 & 0 \\
 2 & 0 & 0 & 0 & 1 & 0 & 0 & 0 & r \\
 0 & 0 & 0 & 0 & 0 & 0 & r & 0 & 0 \\
 0 & 0 & 0 & 0 & 0 & -r & 0 & 0 & 0 \\
 0 & 0 & r & 0 & 0 & 0 & 0 & 0 & 0 \\
 r & 0 & 0 & 0 & r & 0 & 0 & 0 & 0
\end{array}
\right) \\
& \quad  (\bar{\alpha},\alpha,\bar{\beta},\beta=1,2,3) \; ,
\end{split}
\end{equation}  
where $A^{\ast}$ is obtained from $A$ by 
$|\cdot\rangle \mapsto \langle \cdot|$ and complex conjugation. 
The eigenvalues of $T$ are
\begin{multline}
\biggl\{-1(\times 3),-i r(\times 2),i r(\times 2), \\
\frac{1}{2} \left(3-\sqrt{8 r^2+9}\right),\frac{1}{2}
   \left(3+\sqrt{8 r^2+9}\right) \biggr\} \; .
\end{multline}
The largest eigenvalue which is relevant in determining 
the physical quantities in the thermodynamic limit is given, 
for any finite $r$, by 
\begin{equation}
\frac{1}{2}\left(3+\sqrt{8 r^2+9} \right) \; .
\end{equation}
In the limit $r\rightarrow \infty$, another eigenvalue 
$\left(3-\sqrt{8 r^2+9}\right)/2$ becomes degenerate with the above. 

The use of the supertrace in eq.(\ref{eqn:MPS-PBC-2}) modifies 
the expression (\ref{eqn:norm-PBC-1}) of the norm 
for the periodic system to:
\begin{subequations}
\begin{equation}
\VEV{\text{MPS}|\text{MPS}}_{\text{PBC}} 
= \sum_{\alpha,\beta}\text{sgn}(\alpha)\text{sgn}(\beta)
\left\{T^{L}\right\}_{(\alpha,\beta;\alpha,\beta)} \; ,
\label{eqn:norm-PBC-2}
\end{equation}
where 
\begin{equation} 
\text{sgn}(\alpha) =
\begin{cases}
1 & \text{for } \alpha=1,2 \\
-1 & \text{for } \alpha=3 \; .
\end{cases}
\end{equation}
\end{subequations}
\begin{figure}[h]
\begin{center}
\includegraphics[scale=0.7]{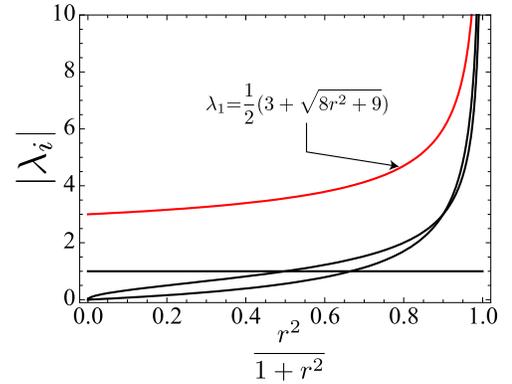}
\caption{(Color online) Plot of absolute values of the five 
different eigenvalues of $T$. The largest eigenvalue is 
always unique and non-degenerate. 
\label{fig:eigenV-SUSY-AKLT1}}
\end{center}
\end{figure}
\subsection{Edge states}
\label{sec:edge-states}
\begin{figure}[h]
\begin{center}
\includegraphics[scale=0.5]{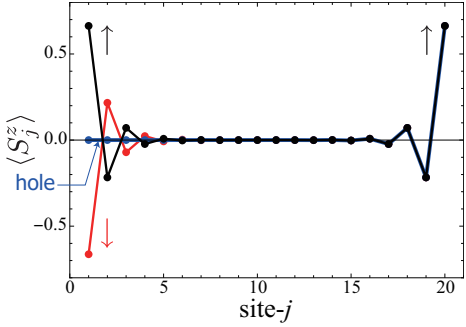}
\caption{(Color online) Plot of $\VEV{S^{z}_{j}}$ ($r=0.3$) for 
various (left) edge states `hole', `$\uparrow$' and 
`$\downarrow$' (with the right edge state fixed 
to $s_{\text{R}}=\uparrow$). The system is non-magnetic in the bulk 
and magnetic moment exists only around the edges of the chain.
\label{fig:edge-spin}}
\end{center}
\end{figure}

Now we would like to mention an important feature of 
the VBS-like states defined on an open chain.   
From the expression (\ref{eqn:MPS-open}), it is clear that 
the nine degenerate ground states correspond to different choices 
of the edge polynomials 
$(b^{\dagger}_{1},\, -a^{\dagger}_{1},\, -\sqrt{r}f^{\dagger}_{1} )$ 
and 
$(a^{\dagger}_{L}, \, b^{\dagger}_{L},\,  \sqrt{r}f^{\dagger}_{L})$.  
In fact, we can explicitly indicate the edge-dependence of 
the ground states as follows:
\begin{equation}
\begin{split}
&|\text{sVBS}\rangle_{\text{open}}
=\bigotimes_{j=1}^{L} A_{j} \\
&= 
\begin{pmatrix}
|s_{\text{L}}{=}\downarrow;s_{\text{R}}{=}\uparrow\rangle & 
|s_{\text{L}}{=}\downarrow;s_{\text{R}}{=}\downarrow\rangle & 
|s_{\text{L}}{=}\downarrow;s_{\text{R}}{=}\circ\rangle \\
|s_{\text{L}}{=}\uparrow;s_{\text{R}}{=}\uparrow\rangle & 
|s_{\text{L}}{=}\uparrow;s_{\text{R}}{=}\downarrow\rangle & 
|s_{\text{L}}{=}\uparrow;s_{\text{R}}{=}\circ\rangle \\
|s_{\text{L}}{=}\circ;s_{\text{R}}{=}\uparrow\rangle & 
|s_{\text{L}}{=}\circ;s_{\text{R}}{=}\downarrow\rangle & 
|s_{\text{L}}{=}\circ;s_{\text{R}}{=}\circ\rangle \\
\end{pmatrix} \; .
\end{split}
\label{eqn:MPS-edge}
\end{equation}
From this, we can readily see that the matrix indices of the MPS 
are directly related to the edge states. 
It is instructive to calculate $\VEV{S^{z}_{j}}$ for various 
edge states
$|\text{sVBS}\rangle_{\text{open}}^{(s_{\text{L}},s_\text{R})}$. 
In Fig.~\ref{fig:edge-spin}, we plot the local magnetization $\VEV{S^{z}_{j}}$ 
for three left edge states $s_{\text{L}}$ (with the right 
edge state $s_{\text{R}}$ fixed).  

A remark is in order here. One may think of the above edge moments 
($s=1/2$ moment or a hole) as independent physical objects and conclude that 
the (SUSY) VBS states are orthogonal with respect to these edge 
states.  However, this is not true; in fact, the above edge moments are 
{\em emergent} objects and sVBS states with different edge states have 
{\em finite} overlaps with each other,  
which are exponentially decreasing as the system size $L$. 
That is, two VBS states with different edge states are orthogonal 
to each other {\em only} in the infinite-size limit.  
In the MPS formulation, this is a direct consequence of the fact 
\begin{equation}
[T^{n}]_{(\alpha_{\text{L}},\beta_{\text{L}};\alpha_{\text{R}},\beta_{\text{R}})}
\xrightarrow{n\nearrow \infty}\delta_{\alpha_{\text{L}},\beta_{\text{L}}}
\delta_{\alpha_{\text{R}},\beta_{\text{R}}} \times 
{\cal F}_{\alpha_{\text{L}},\alpha_{\text{R}}}(r) \; .
\end{equation}  
In fact, this property greatly simplifies the calculations below. 
\subsection{Spin-spin correlation}
Now that we have obtained all the necessary matrices, 
we can follow the steps described in section \ref{sec:cor-func} 
to calculate various correlation functions.  

The ordinary spin-spin correlation function 
$\VEV{S^{a}_{x}S^{a}_{x+n}}$ reads:
\begin{subequations}
\begin{align}
& \frac{2 \left(r^2+3+\sqrt{8 r^2+9}\right)}{\sqrt{8 r^2+9} 
\left(3+\sqrt{8 r^2+9}\right)}
\quad  (\text{for }n=0)  \\
\begin{split}
& \frac{  13 r^2+24 +\left(r^2+8\right) \sqrt{8r^2+9}}
{2\sqrt{8 r^2+9} \left(3+\sqrt{8 r^2+9}\right)}
\left(-\frac{2}{3+\sqrt{8 r^2+9}}\right)^n  \\
& \qquad (\text{for }n>0) \; .
\end{split}
\end{align}
\end{subequations}
The exponentially decaying factor defines the correlation 
length\cite{Arovas-H-Q-Z-09}:
\begin{equation}
\xi_{\text{spin}}(r)^{-1} \equiv 
\log \left\{\frac{3+\sqrt{8 r^2+9}}{2}\right\}  \; ,
\label{eqn:spin-cor-length}
\end{equation}
which is monotonically decreasing in $r$.  
In the pure AKLT-limit $r\rightarrow 0$, it reduces to the well-known 
results\cite{affleck1987rrv,affleck1988vbg}:
\begin{equation}
\VEV{S^{a}_{x}S^{a}_{x+n}}= 
\begin{cases}
\frac{2}{3} \quad & \text{for }n=0 \\
\frac{4}{3}\left(\frac{-1}{3}\right)^{n} 
\quad & \text{for } n>0 \; .
\end{cases}
\end{equation}

\begin{widetext}
For later convenience, we calculate the {\em static} structure factor 
$S^{\alpha\alpha}(k)$.  
The result is given as:
\begin{equation}
S^{zz}(k)=
\frac{4 \left(2 r^4+17 r^2+\left(3 r^2+6\right) 
\sqrt{8 r^2+9}+18\right) (1-\cos (k))}{\sqrt{8
   r^2+9} \left(\sqrt{8 r^2+9}+3\right) 
\left\{4 r^2
+3\sqrt{8 r^2+9}+11
+2 \left(\sqrt{8 r^2+9}+3\right) \cos (k)
\right\}}.
\end{equation}
\end{widetext}
\subsection{superconducting correlation}
\label{sec:SC-correlation-I}
\begin{figure}[h]
\begin{center}
\includegraphics[scale=0.45]{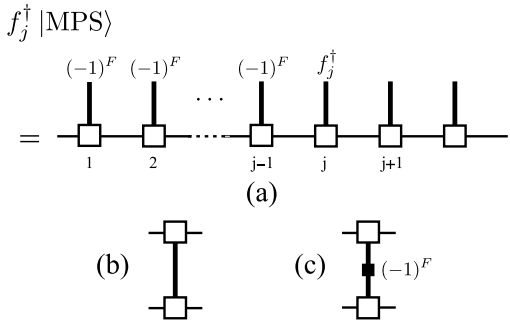}
\caption{Action of fermion operator on the MPS. 
(a): Due to the fermionic anticommutation relation, extra factors $(-1)^{F}$ appear 
in the $A$-matrices on the left of site-$j$. 
Accordingly, a new transfer matrix (c) is necessary as well as the standard one (b) 
when we calculate expectation values containing fermionic operators. 
\label{fig:MPS-diagram}}
\end{center}
\end{figure}

\begin{figure}[h]
\begin{center}
\includegraphics[scale=0.5]{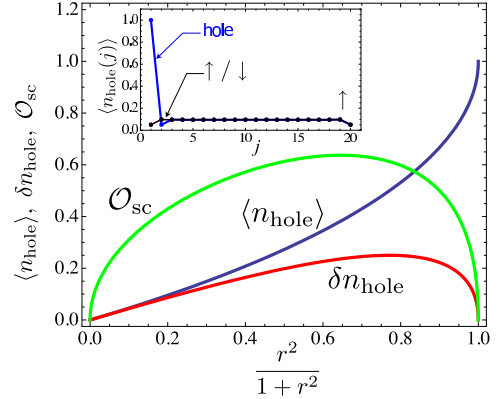}
\caption{(Color online) Plot of 
${\cal O}_{\text{sc}}=\VEV{\Delta_{j}}$, 
the hole density $\VEV{n_{\text{hole}}(j)}=\VEV{f^{\dagger}_{j}f_{j}}$ and 
the hole-number fluctuation 
$\VEV{f^{\dagger}_{j}f_{j}}-\VEV{f^{\dagger}_{j}f_{j}}^{2}$ 
as a function of $r$.  Here the bulk values are plotted.  
Inset: Profile of the hole density ($r=0.5$) 
for a finite system ($L=20$) with different left edge states 
($\uparrow$, $\downarrow$, and `hole').  Only the left edge state 
is changed with the right one fixed to $s_{\text{R}}=\uparrow$. 
The hole density approaches exponentially to the bulk value 
as we move away from the edge. 
\label{fig:SC-orderparam}}
\end{center}
\end{figure}

In order to handle the operators containing fermions, 
we have to generalize the general recipe presented in Appendix  
\ref{sec:quick-recipe}.  Take for example the hole-pair 
creation operator\cite{Arovas-H-Q-Z-09}:
\begin{equation}
\begin{split}
\Delta_{j} & \equiv 
(a_{j}b_{j+1} - b_{j}a_{j+1})f^{\dagger}_{j}f^{\dagger}_{j+1} \\
& = (a_{j}f^{\dagger}_{j})(b_{j+1}f^{\dagger}_{j+1})
- (b_{j}f^{\dagger}_{j})(a_{j+1}f^{\dagger}_{j+1}) \; .
\end{split}
\end{equation}
In order to apply the method presented in sections \ref{sec:norm} and  
\ref{sec:cor-func}, first a string of $A$-matrices 
$A_{1}\otimes\cdots\otimes A_{j}$ has to be moved to the left of 
$f^{\dagger}_{j+1}$ and through this procedure it acquires 
a Jordan-Wigner-like phase $\prod_{k=1}^{j}(-1)^{F_{k}}$ 
($F_{k}$ counts the fermion number 0 or 1 at the site $k$; 
see Fig.\ref{fig:MPS-diagram}):
\begin{equation} 
(-1)^{F_{1}}A_{1}\otimes \cdots \otimes (-1)^{F_{j}}A_{j} \; .
\end{equation} 
Next a string 
$(-1)^{F_{1}}A_{1}\otimes\cdots\otimes(-1)^{F_{j-1}}A_{j-1}$ 
and $f^{\dagger}_{j}$ are interchanged and this multiplies 
the matrices $A_{1}$, \dots, $A_{j-1}$ 
additional $(-1)^{F_{k}}$-factors to remove the fermion sign 
except at the site $j$. 
Therefore, we need four more matrices
\begin{equation}
\begin{split}
& T^{\mathcal{O} f^{\dagger}}(\bar{\alpha},\alpha;\bar{\beta},\beta)\equiv 
A^{\ast}(\bar{\alpha},\bar{\beta})(\mathcal{O} f^{\dagger})A(\alpha,\beta) 
\; , \\
& \widetilde{T}^{\mathcal{O} f^{\dagger}}(\bar{\alpha},\alpha;\bar{\beta},\beta)\equiv 
A^{\ast}(\bar{\alpha},\bar{\beta})\left\{\mathcal{O} f^{\dagger}(-1)^{F}\right\}
A(\alpha,\beta) \\
& \phantom{
\widetilde{T}^{\beta f^{\dagger}}(\bar{m},m;\bar{n},n)\equiv 
A^{\ast}(\bar{m},\bar{n})g(m,n)} 
(\mathcal{O}=a,b) \; .
\end{split}
\end{equation}   
By using these, the numerator of $\VEV{\Delta_{j}}$ 
is calculated as:
\begin{equation}
T^{j-1}\left\{\widetilde{T}^{af^{\dagger}}T^{bf^{\dagger}}
- 
\widetilde{T}^{bf^{\dagger}}T^{af^{\dagger}}\right\}
T^{l-j-1} \; .
\end{equation}
Also interesting are the hole density 
\begin{equation}
\VEV{n_{\text{hole}}}=\VEV{f^{\dagger}_{j}f_{j}}
\end{equation}
and the hole-number fluctuation 
\begin{equation}
\delta n_{\text{hole}}=
\sqrt{\VEV{f^{\dagger}_{j}f_{j}}-\VEV{f^{\dagger}_{j}f_{j}}^{2}} \; .
\end{equation} 
By using the method described above, we can readily calculate 
these quantities.  For instance, the hole density in the bulk system 
is computed as:
\begin{equation}
\VEV{n_{\text{hole}}} = 
\frac{r^2(5 + \sqrt{8 r^2+9})}{8 r^2 +9+\left(r^2+3\right) 
\sqrt{8 r^2+9}}  \; .
\end{equation}
As is clearly seen in the inset of Fig.~\ref{fig:SC-orderparam}, 
near the edges of an open chain, the hole density is different from 
the bulk value and approaches exponentially with the `healing length' 
given by 
\begin{equation}
\xi_{\text{hole}}(r)^{-1} 
= \log\left\{\frac{\sqrt{8 r^2+9}+3}{\sqrt{8 r^2+9}-3}\right\} \; .
\end{equation}
Note that this is different from the spin correlation length 
$\xi_{\text{spin}}(r)$ in eq.(\ref{eqn:spin-cor-length}) and 
the superconducting correlation length 
\begin{equation}
\xi_{\text{sc}}(r) = 1/\log\left\{\frac{\sqrt{8 r^2+9}+3}{2r}\right\}
\end{equation}
defined by the exponential decay of the singlet off-diagonal correlation 
function\cite{Arovas-H-Q-Z-09} 
$\VEV{(a_{j}b_{j+n} - b_{j}a_{j+n})f^{\dagger}_{j}f^{\dagger}_{j+n}}$.  

In Fig.\ref{fig:SC-orderparam}, we plot the expectation value 
of the hole-pair creation operator:
\begin{equation}
{\cal O}_{\text{sc}} \equiv 
\VEV{\Delta_{j}}
\end{equation}
together with the hole density 
$\VEV{n_{\text{hole}}}=\VEV{f^{\dagger}_{j}f_{j}}$ 
and the hole-number fluctuation 
$\delta n_{\text{hole}}=
\sqrt{\VEV{f^{\dagger}_{j}f_{j}}-\VEV{f^{\dagger}_{j}f_{j}}^{2}}$. 
From $r=0$ ($S=1$ VBS limit) to $r\rightarrow \infty$ 
($S=1/2$ Majumdar-Ghosh limit), the hole density is monotonically 
increasing.  When $r=0$ and $r\rightarrow \infty$, 
the hole number fluctuation is suppressed ($n_{\text{hole}}$ 
takes definite values 0 and 1, respectively) 
and consequently the `superconducting correlation' 
becomes zero. This is consistent with what we expect from the analogy to 
the BCS wave function pointed out in Ref.\onlinecite{Arovas-H-Q-Z-09}.    

\section{Hidden order}
\label{sec:hidden-order}
\subsection{Generalized Hidden Order in sVBS states}
The hidden order is a generalized concept of the N\'{e}el order.
For $S=1$ antiferromagnetic spin chain, the N\'{e}el order 
is depicted as  
\begin{equation}
\cdots ~ + ~ - ~ + ~ - ~ + ~ - ~ + ~ - ~ + ~  \cdots 
\end{equation}
Here, $+$ stands for $S_z=+1$, and $-$ for $S_z=-1$. In the sequence, 
$+$ and $-$ are alternating, representing the classical antiferromagnets. 
A typical  $S_z$ sequence of VBS chain is given by   
\begin{equation}
\cdots  ~ + ~ - ~ + ~ 0 ~ - ~ + ~ - ~  0 ~ 0 ~ + ~ - ~ 0 ~ + ~\cdots
\end{equation}
When we remove zeros in the sequence, we arrive at the usual 
N\'{e}el order.  This is the hidden (string) order observed 
in gapped antiferromagnetic spin liquids\cite{denNijs-R-89,Tasaki-91}. 
The hidden order is a non-local order, since the removing zeros is a global procedure. 
Since in the sVBS states one-hole states carry one-half spins at each site, $S_z=1/2$ 
and $-1/2$ generally appear in the sequence. 
The locations of such one-half-spins are, however, not completely random; 
The following procedure reveal the existence of a generalized hidden order in the sVBS states. 
A typical $S_z$ sequence of sVBS states is given by  
\begin{equation}
\cdots  ~ 0 ~ \underbrace{\uparrow  ~ \uparrow} ~ 0 ~ 0 ~ 
\underbrace{\downarrow ~ \downarrow} ~ 
+ ~ - ~  0 ~ 0 ~ \underbrace{\uparrow ~ \downarrow} ~ 
+ ~ \underbrace{\downarrow ~  \uparrow} ~ 
\underbrace{\downarrow ~ \downarrow} ~ 0 ~\cdots  
\end{equation}
First, we search the spin-half sites from the left and whenever we encounter a pair of 
spin-half sites we sum the two $S_{z}$-values to replace the pair with a single site 
having the effective $S_{z}$ (e.g. $\downarrow \; \downarrow \, \mapsto - $):
\begin{equation}
\cdots  ~ 0 ~ + ~ 0 ~ 0 ~ - ~ 
+ ~ - ~  0 ~ 0 ~ 0 ~ 
+ ~ 0 ~ - ~ 0 ~\cdots  
\end{equation}
Then, we remove the zeros in the sequence to obtain the standard N\'{e}el pattern:    
\begin{equation}
\cdots ~ + ~ - ~ + ~ -  ~ + ~ -  ~\cdots 
\end{equation}
This argument leads us to conclude the existence of (generalized) hidden order 
in the sVBS states. By the SU(2)-invariance of the sVBS state, the same is true for 
the $S_{x}$-sequence as well.  
The hidden order is ``measured'' by the non-local string order parameter\cite{denNijs-R-89}. 
In sections \ref{subsec:stringcorrelationI} and \ref{subsec:stringcorrelationII}, 
we explicitly calculate the string order for the type I and the type II sVBS states, 
respectively. 
\subsection{Matrix-product representation and hidden order}
Before proceeding to the actual calculation of the string correlation, 
we delve the hidden order inherent in the sVBS state from the MPS point of view.  
Since the condition for the string correlators to have finite values is known in 
a general and mathematical manner\cite{Garcia-W-S-V-C-08}, 
we give here a more physical argument. 

To clarify this hidden structure in the spin configuration, 
let us pick up an arbitrary site $j$ and consider the partial sum 
of $S^{z}_{k}$s contained in the block between the left edge and the site-$j$:
\begin{equation}
{\cal S}^{z}_{\text{tot}}(j) \equiv 
\sum_{k=1}^{j}S_{k}^{z}  \; .
\end{equation}
In considering the possible values of ${\cal S}^{z}_{\text{tot}}(j)$, 
it is convenient to consider the MPS for the block:
\begin{equation}
\left\{A_{1}\otimes\cdots \otimes A_{j}\right\} .  
\end{equation}
Since the sVBS state on any finite subsystem (\ref{eqn:SUSY-MPS}) 
is made up of a product of (SUSY) valence bonds (\ref{eqn:SUSY-VB-spior}) 
carrying $S^{z}=0$, 
the above ${\cal S}^{z}_{\text{tot}}(j)$ is determined {\em only} 
by the edge states of the subsystem
\begin{multline}
\left\{A_{1}\otimes\cdots \otimes A_{j}\right\} \\
= 
\begin{pmatrix}
|{\cal S}^{z}_{\text{tot}}(j){=}0\rangle 
& |{\cal S}^{z}_{\text{tot}}(j){=}-1\rangle &
|{\cal S}^{z}_{\text{tot}}(j){=}-1/2\rangle \\
|{\cal S}^{z}_{\text{tot}}(j){=} 1\rangle 
& |{\cal S}^{z}_{\text{tot}}(j){=}0\rangle &
|{\cal S}^{z}_{\text{tot}}(j){=} 1/2\rangle \\
|{\cal S}^{z}_{\text{tot}}(j){=} 1/2\rangle 
& |{\cal S}^{z}_{\text{tot}}(j){=}-1/2\rangle &
|{\cal S}^{z}_{\text{tot}}(j){=}0\rangle 
\end{pmatrix}
\; .
\label{eqn:left-subsys}
\end{multline}

\begin{figure}[H]
\begin{center}
\includegraphics[scale=0.45]{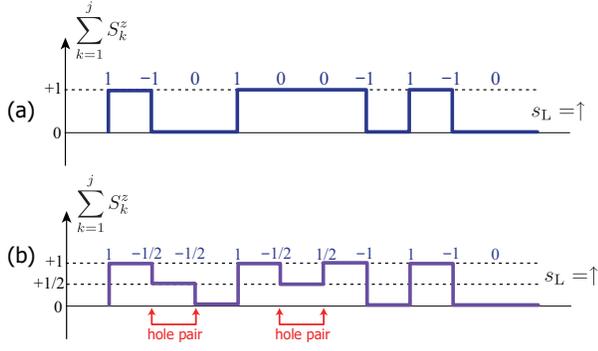}
\caption{(Color online) Height plot of typical spin configurations 
in spin-1 VBS state (a) and $M=1$ sVBS state (b).  
Note that heights are confined within a region of width 1.  
Although simple `diluted' N\'{e}el picture does not hold 
because of the presence of hole pairs, still we can find 
{\em string order} when hole pairs are grouped together in (b). 
\label{fig:step-VBS-1}}
\end{center}
\end{figure}

To see what (\ref{eqn:left-subsys}) implies, 
it is suggesting to plot ${\cal S}^{z}_{\text{tot}}(j)$ as 
a sequence of steps.  Namely, we assign a local height variable 
$h_{j}={\cal S}^{z}_{\text{tot}}(j)$ to a bond to the right of the site $j$.  
Then, the local spin value $S_{j}^{z}$ is expressed 
as a step $h_{j}-h_{j-1}$ between the adjacent heights.  
It is obvious that this 
height plot is in one-to-one correspondence to the original 
$\{S^{z}\}$ configuration.  
Eq.(\ref{eqn:left-subsys}) shows a set of possible 
heights (i.e. ${\cal S}^{z}_{\text{tot}}(j)$) at a given site $j$.  
For instance, if the left edge state is $\uparrow$, 
the corresponding states are contained in the first row of 
(\ref{eqn:left-subsys}) and one readily sees that 
only 0, 1 and 1/2 are allowed for the sVBS state.     
Fig.\ref{fig:step-VBS-1} shows a typical 
height configuration corresponding to the usual VBS state\footnote{%
the usual spin-1 VBS state is obtained by picking up the top-left 
2$\times$2 block.} (a) and 
its SUSY counterpart (b). Strikingly, the height configuration 
is always meandering between the height-0 and the height-1 
(although the absolute height of the meandering line depends 
on the left edge states, the height 
configuration is always confined within a region of width 1). 
The same reasoning applies to the general spin-$S$ VBS cases and 
we can show\cite{Totsuka-S-mpg-95} that the height configurations are 
confined within a region of width $S$.  
This is highly non-trivial since in the ferromagnetic state 
we have an ever going-up steps.  
This `almost flat' feature of the VBS state has been first 
realized by den Nijs and Rommelse\cite{denNijs-R-89} 
for the $S=1$ case.  

In the case of $S=1$, one can strengthen this statement; 
in any spin (or height) configurations satisfying the above 
property, $S^{z}=1$ and $-1$ occur in an alternating manner when 
the intervening 0s are neglected (see Fig.\ref{fig:step-VBS-1}(a)).  
This may be viewed as a {\em diluted} N\'{e}el order.  
In the standard N\'{e}el state, we can insert an alternating phase 
$(-1)^{j-i}$ to make the correlation between the two spins 
$S^{z}_{i}$ and $S^{z}_{j}$ ferromagnetic.  In the diluted case, 
on the other hand, we can easily see that the string operator 
$\prod_{k=i}^{j-1}\exp(i\pi S^{z}_{k})$ will do the job and that 
one can use the following order parameter 
({\em string order parameter}) detects the Haldane 
state\cite{denNijs-R-89,Tasaki-91}:
\begin{equation}
{\cal O}_{\text{string}}^{\infty}
\equiv 
\lim_{n\nearrow \infty}
\Biggl\langle S_{j}^{z}\, \prod_{k=j}^{j+n-1} 
\exp\left\{
i\pi S^{z}_{k}
\right\} S^{z}_{j+n}\Biggr\rangle \; .
\end{equation}
For the spin-1 VBS state, it is evaluated\cite{Kennedy-T-92a} 
exactly as 
$(2/3)^{2}$ (`$2/3$' comes from the probability of having 
non-zero $S^{z}$).   

In the SUSY case, the situation is slightly more complicated 
since we have height-1/2s corresponding to sites with one hole.  
However, if we note that the holes appear always in pairs, 
we can easily see that the  insertion of hole-pairs 
(which carry $S^{z}=1/2$) does not affect the string 
part
\begin{equation*}
\prod_{k=j}^{j+n-1} 
\exp\left[
i\pi S^{z}_{k}
\right] = \exp\left\{
i\pi \sum_{k=j}^{j+n-1}S^{z}_{k}
\right\}
\end{equation*} 
and we may expect that string order persists in the SUSY case 
($r\neq 0$) as well (see Fig.\ref{fig:step-VBS-1}(b)).    
\subsection{String correlation}\label{subsec:stringcorrelationI}
\begin{figure}[H]
\begin{center}
\includegraphics[scale=0.5]{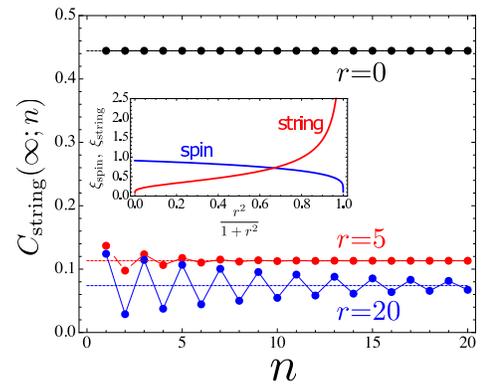}
\caption{(Color online) String correlation function in the bulk 
$C_{\text{string}}(\infty;n)$ 
for various values of $r$: 
(i) $r=0$ (top; pure spin AKLT), (ii) $r=5.0$ (middle) and 
(iii) $r=20.0$ (bottom).  Note that for the pure spin AKLT model ($r=0$), 
the string correlation function is constant $4/9$.  For $r\neq 0$, 
the string correlation functions exponentially approach to the limiting 
values shown by dashed lines. 
\label{fig:string-AKLT-2a}}
\end{center}
\end{figure}

The finite-distance string correlation function\cite{denNijs-R-89} 
\begin{equation}
C_{\text{string}}(j;n) \equiv 
\Biggl\langle S_{j}^{z}\, \exp\left[
i\pi \sum_{k=j}^{j+n-1}S^{z}_{k}
\right]S^{z}_{j+n}\Biggr\rangle
\end{equation}
can be evaluated in a similar 
manner.  In the case of open chains, it suffers from the boundary 
effects.  However, if we consider the case where both end points 
$j$ and $j+n$ are infinitely far from the chain edges, the expression 
simplifies a lot. 
In general, it contains exponentially decaying parts 
\begin{equation}
(-1)^{n}
\left\{\frac{\sqrt{8 r^2+9}-3}{\sqrt{8 r^2+9}+3}\right\}^n
\end{equation}
as well as the constant (i.e. long-range-ordered) one (see Fig.~\ref{fig:string-AKLT-2a}):
\begin{equation}
{\cal O}_{\text{string}}^{\infty}(r)=
\frac{4 \left\{ r^4+14 r^2+18 +2 \left(r^2+3\right) \sqrt{8 r^2+9}
\right\}}{\left(8 r^2+9\right)
   \left(\sqrt{8 r^2+9}+3\right)^2}  \; .
\end{equation}
Only in the limit $r\rightarrow 0$, the exponentially decaying 
parts disappear and the string correlation function becomes 
constant $4/9$ (perfect string correlation).  
Note that the correlation length 
$\xi_{\text{string}}$ is different from that ($\xi_{\text{spin}}$) 
for the spin-spin correlation.  
With increase of the hole-doping parameter $r$, 
the effective spin magnitude gets reduced by the increase of the spin-half sites 
and accordingly the string order parameter monotonically decreases 
(see Fig.~\ref{fig:string-AKLT-2a}).   

At $r\rightarrow \infty$, the type I sVBS chain $(M=1)$ realizes 
the Majumdar-Ghosh dimer states with one-half spin degrees of freedom at each site and 
the string order parameter ${\cal O}_{\text{string}}^{\infty}$ 
reaches its finite minimum $1/16$, which implies that the string 
order survives even in the $r\nearrow \infty$ limit.   
This agrees with the observation that the spin-1 Haldane state is 
adiabatically connected to the spin-1/2 dimer state\cite{Hida-92a}. 
Meanwhile, 
the type II sVBS chain $(M=1)$ is reduced to the hole-VBS chain with  
no spin degree of freedom at $r\rightarrow \infty$, and hence the string order 
vanishes completely in this limit. 

\begin{figure}[H]
\begin{center}
\includegraphics[scale=0.6]{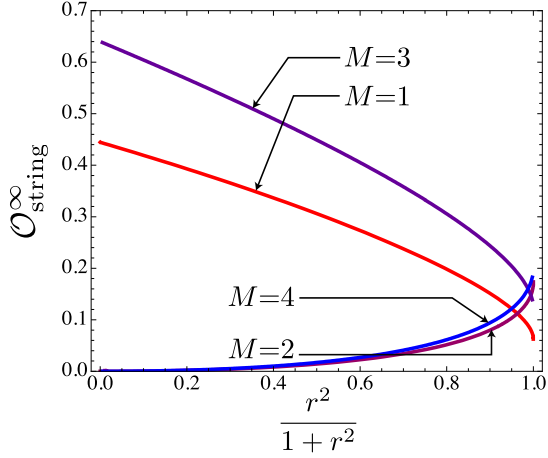}
\caption{(Color online) The infinite-distance limit of 
the string correlation function 
${\cal O}_{\text{string}}^{\infty}=
\lim_{n\rightarrow \infty}C_{\text{string}}(\infty;n)$
for several values of $M$ plotted as a function of $r$.  
Note that ${\cal O}^{\infty}_{\text{string}}(r{=}0)=0$ 
for even-$M$ corresponding to the vanishing of string order parameter 
for even-$S$. 
\label{fig:Ostring-model1}}
\end{center}
\end{figure}

In Ref.~\onlinecite{Arovas-H-Q-Z-09}, a SUSY-analogue of 
the higher-$S$ VBS states is discussed as well.  
The ordinary spin-$S$ VBS states obtained in the zero hole-density ($r\rightarrow 0$) 
limit are known to exhibit different topological properties according to 
the parity of spin-$S$; 
the string order parameter vanishes for 
the even-spin VBS states while it is finite for odd-$S$\cite{Oshikawa-92,Totsuka-S-mpg-95}.  
In this sense, it would be interesting 
to calculate the string order parameter 
${\cal O}_{\text{string}}^{\infty}$ for the generalized sVBS states.  
As is seen in eq.(\ref{sVBSstateI}), the role of spin $S$ is played by 
an integer $M$ (superspin) in the SUSY case.  For all $M$, we can construct 
the matrix-product representation of the $M$-sVBS state by using 
$(2M+1){\times}(2M+1)$ matrices (see Appendix \ref{sec:general-M-MPS})
and after straightforward evaluation 
we obtain the results shown in Fig.~\ref{fig:Ostring-model1}.    
As is expected from the previous studies, 
the $r=0$ value of ${\cal O}_{\text{string}}^{\infty}$ 
vanishes for even-$M$.   
When the hole pairs are doped, on the other hand, 
the string order revives.  In section \ref{sec:symmetry-protected}, 
we will interpret this from the point of view of 
symmetry-protected topological order.
\section{Single-mode approximation}\label{sec:singlemodeappro}
In this section, we consider the dynamical quantities, i.e. low-lying excitation 
spectra by using {\em single-mode approximation}.  
As is easily verified, the so-called Lieb-Schultz-Mattis 
twist\cite{Lieb-S-M-61}, 
which provides a basic picture of gapless low-lying excitations in 
half-odd-integer spin chains, does not work 
in the usual VBS state\cite{Totsuka-S-mpg-95}. Instead, an excited triplet bond 
({\em crackion}-- a `crack' in a solid of valence bonds) 
in the valence-bond solid gives, to good approximation, 
a physical low-lying excitation.  
As has been shown by Fath and S\'{o}lyom\cite{Fath-S-93b}, 
the crackions are equivalent to 
the triplon excitations created by spin operators $\bolS(\bolk)$. 
\subsection{Spin excitations}
\begin{figure}[H]
\begin{center}
\includegraphics[scale=0.5]{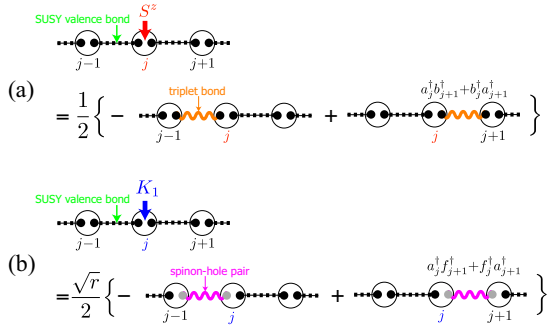}
\caption{(Color online) Action of local spin operator $S^{z}$ (a) and 
fermionic generator $K_{1}$ (b) onto 
the sVBS state. The local operators $S^{a}_{j}$ ($a=x,y,z$) 
and $K_{1,2}$ respectively create 
a triplet bond and a spinon-hole pair ({\em crackion}) 
on either of the two adjacent bonds $(j-1,j)$ and  
$(j,j+1)$.  %
\label{fig:crackion}}
\end{center}
\end{figure}

Let us start by investigating the action of local spin operators 
\begin{equation}
S^{+}_{j}=a^{\dagger}_{j}b_{j} \, , \; 
S^{-}_{j}=b^{\dagger}_{j}a_{j} \, , \; 
S^{z}_{j}=\frac{1}{2}(a^{\dagger}_{j}a_{j}-b^{\dagger}_{j}b_{j}) 
\end{equation}
on the sVBS state.  A little algebra shows that these spin operators 
create triplet bonds around the site $j$ (see Fig.\ref{fig:crackion}):
\begin{subequations}
\begin{align}
& S^{+}_{j}|\text{sVBS-I}\rangle 
= |\psi^{(1)}_{j-1}\rangle -|\psi^{(1)}_{j}\rangle   \\
& S^{z}_{j}|\text{sVBS-I}\rangle 
= \frac{1}{2}\left\{
- |\psi^{(0)}_{j-1}\rangle + |\psi^{(0)}_{j}\rangle 
\right\} \; ,
\label{eqn:z-crackion}
\end{align}
\end{subequations}
where $|\psi^{(1)}_{j}\rangle$ and $|\psi^{(0)}_{j}\rangle$ are 
obtained by replacing the SUSY valence bond 
$(a^{\dagger}_{j}a^{\dagger}_{j+1}-b^{\dagger}_{j}b^{\dagger}_{j+1}
-r\, f^{\dagger}_{j}f^{\dagger}_{j+1})$ by triplet bonds 
$a^{\dagger}_{j}a^{\dagger}_{j+1}$ and 
$(a^{\dagger}_{j}b^{\dagger}_{j+1}+b^{\dagger}_{j}a^{\dagger}_{j+1})$, 
respectively.  This implies that the triplon-crackion equivalence 
holds in the sVBS case as well.   

The single-mode approximation to the magnetic excitations is given 
by
\begin{equation}
\begin{split}
\omega_{\text{SMA}}^{\text{s},\alpha}(k) &= -\frac{1}{2}
\frac{\langle\text{sVBS-I}|\, [[{\cal H}, S^{\alpha}(k)],
S^{\alpha}(-k)]\, |\text{sVBS-I}\rangle}{\langle\text{sVBS-I}|S^{\alpha}(k)
S^{\alpha}(-k)|\text{sVBS-I}\rangle} \\
& = \frac{\langle\text{sVBS-I}| S^{\alpha}(k){\cal H}\,
S^{\alpha}(-k)|\text{sVBS-I}\rangle}{\langle\text{sVBS-I}|S^{\alpha}(k)
S^{\alpha}(-k)|\text{sVBS-I}\rangle} \\
& \geq \omega_{\text{true}}^{\text{s},\alpha}(k) \; .
\end{split}
\label{eqn:SMA-formula-1}
\end{equation}
By the SU(2) symmetry, it suffices to evaluate $\omega_{\text{SMA}}$ 
only for $\alpha=z$ and the spin index $\alpha$ will be suppressed hereafter.  
Using eq.(\ref{eqn:z-crackion}), the denominator (static structure factor)
$\langle\text{sVBS-I}|S^{\alpha}(k)S^{\alpha}(-k)|\text{sVBS-I}\rangle$ 
is evaluated as:
\begin{multline}
\langle\text{sVBS-I}|S^{z}(k)
S^{z}(-k)|\text{sVBS-I}\rangle \;\; (\equiv S^{zz}(k))\\
= \frac{1}{2}(1-\cos k)\VEV{\psi^{(0)}(k)|\psi^{(0)}(k)} \; ,
\label{eqn:SofK}
\end{multline}
where $|\psi^{(0)}(k)\rangle$ denotes the Fourier transform 
\[
 |\psi^{(0)}(k)\rangle = \frac{1}{\sqrt{L}}\sum_{r}
\be^{-ikr}|\psi^{(0)}_{r}\rangle \; .
\]

Similarly, the local property of the sVBS states 
\begin{equation}
\langle \text{sVBS-I}|h_{j,j+1}= h_{j,j+1} |\text{sVBS-I}\rangle=0 
\;\; (\forall j)
\label{eqn:HtosVBS}
\end{equation}
implies that only the diagonal part survives:
\begin{equation}
\VEV{\psi^{(0)}_{i}|{\cal H}|\psi^{(0)}_{j}} 
= \delta_{i,j}\VEV{\psi^{(0)}_{j}|h_{j,j+1}|\psi^{(0)}_{j}} \; .
\end{equation}
From this, one deduces:
\begin{equation}
\begin{split}
& \langle\text{sVBS-I}| S^{\alpha}(k){\cal H}\,
S^{\alpha}(-k)|\text{sVBS-I}\rangle \\
& \quad 
= \frac{1}{2}(1-\cos k)\VEV{\psi^{(0)}_{j}|h_{j,j+1}|\psi^{(0)}_{j}} \\
&\quad 
= \frac{1}{2}(1-\cos k)\VEV{\psi^{(0)}(k)|{\cal H}|\psi^{(0)}(k)}  \; .
\end{split}
\label{eqn:SHS}
\end{equation}
Eqs.(\ref{eqn:SofK}) and (\ref{eqn:SHS}) are combined to give 
\begin{equation}
\begin{split}
\omega^{\text{s}}_{\text{SMA}}(k) &= 
\frac{\VEV{\psi^{(0)}(k)|{\cal H}|\psi^{(0)}(k)}}{%
\VEV{\psi^{(0)}(k)|\psi^{(0)}(k)}
} \;\;(\equiv \omega_{\text{crackion}}(k))
\\
&= \frac{\VEV{\psi^{(0)}_{x}|h_{x,x+1}|\psi^{(0)}_{x}}}{%
\VEV{\psi^{(0)}(k)|\psi^{(0)}(k)}
} \\
&= \frac{1}{2}\frac{(1-\cos k)}{S^{zz}(k)}
\VEV{\psi^{(0)}_{x}|h_{x,x+1}|\psi^{(0)}_{x}} \; .
\label{eqn:SMA-3}
\end{split}
\end{equation}
At this point, one may note a peculiar feature of the VBS-like 
states. Normally, a local excitation created by physical operators (e.g. $S^{\alpha}_{j}$) 
propagates on a lattice by using the off-diagonal matrix elements:
\begin{equation}
\VEV{\psi^{(0)}_{i}|{\cal H}|\psi^{(0)}_{j}} \;\; 
(i\neq j)\; .
\end{equation} 
In the VBS-like models, on the other hand, 
$\VEV{\psi^{(0)}_{i}|{\cal H}|\psi^{(0)}_{j}}$ is diagonal 
by construction (all the diagonal elements are given by  
$\VEV{\psi^{(0)}_{j}|h_{j,j+1}|\psi^{(0)}_{j}}$) 
and excitations cannot use this channel.  
Rather the non-trivial $k$-dependence of 
$\omega_{\text{SMA}}(k)$ comes only from the non-trivial overlap 
between the crackion states:
\begin{subequations}
\begin{align}
& \VEV{\psi^{(0)}_{i}|\psi^{(0)}_{j}} =
\frac{3+ \sqrt{8 r^2+9}}{2\sqrt{8r^2+9}}
\left(-\frac{2}{3+\sqrt{8 r^2+9}}\right)^{|i-j |}   
\label{eqn:crackion-overlap-1} \\
& \left\{
\VEV{\psi^{(0)}(k)|\psi^{(0)}(k)}
\right\}^{-1} \propto \frac{(1-\cos k)}{S^{zz}(k)} \; .
\label{eqn:crackion-overlap-2}
\end{align} 
\end{subequations}
An important conclusion can be drawn from eq.(\ref{eqn:SMA-3}); 
the physical triplon excitation energy 
$\omega^{\text{s}}(k)(\leq \omega^{\text{s}}_{\text{SMA}}(k))$ 
becomes zero (i.e. gapless) 
as $k\rightarrow 0$ {\em unless} the static structure 
factor $S^{zz}(k)$ behaves like $k^{2}$ ($k\sim 0$). 
For any spin-$S$ VBS states and the sVBS states, we have checked 
that $S^{zz}(k)$ contains a factor $(1-\cos k)\sim k^{2}$, 
which opens a gap at $k=0$.   
\subsection{Hole excitations}
A similar analysis can be done for the charged (hole) excitations 
which are always accompanied by spinon-like (i.e. $S=1/2$) objects. 
These excitations are created by applying the two fermionic generators 
of UOSp(1$|$2)
\begin{equation}
\begin{split}
& K_{1}(j)= \frac{1}{2}(x^{-1}f_{j}a_{j}^{\dagger}+x f_{j}^{\dagger}b_{j}) \\
& K_{2}(j)= \frac{1}{2}(x^{-1}f_{j}b_{j}^{\dagger}-x
 f_{j}^{\dagger}a_{j}) \;\;
(x\equiv \sqrt{r})
\end{split}
\end{equation}
to the VBS ground state. 
By using the explicit form of the ground-state wavefunction, 
it is easy to show 
\begin{equation}
K_{1}(j)|\text{sVBS-I}\rangle 
= \frac{\sqrt{r}}{2}\left\{
|\psi_{j-1}^{(1/2)}\rangle - |\psi_{j}^{(1/2)}\rangle 
\right\} \; ,
\end{equation}
where the crackion state 
$|\psi_{j}^{(1/2)}\rangle$ is obtained by replacing 
the SUSY valence bond 
$(a_{j}^{\dagger}b_{j+1}^{\dagger}-b_{j}^{\dagger}a_{j+1}^{\dagger}
-r f_{j}^{\dagger}f_{j+1}^{\dagger})$ on the bond $(j,j+1)$ with 
a spinon-hole pair 
$(a_{j}^{\dagger}f_{j}^{\dagger}+f_{j}^{\dagger}a_{j+1}^{\dagger})$ 
(see Fig.~\ref{fig:crackion}(b)). 
The excited state $K_{2}|\text{sVBS}\rangle$ is defined similarly 
with $a^{\dagger}$ in the above expression replaced with 
$b^{\dagger}$.  Then, the SMA excitation energy is given by an
expression 
similar to eq.(\ref{eqn:SMA-3}):
\begin{equation}
\begin{split} 
\omega^{\text{h}}_{\text{SMA}}(k) &= 
\frac{\langle \text{sVBS-I}|K_{1}(-k)^{\dagger}{\cal H}
K_{1}(-k)|\text{sVBS-I}\rangle}%
{\langle \text{sVBS-I}|K_{1}(-k)^{\dagger}K_{1}(-k)|
\text{sVBS-I}\rangle} \\
&= \frac{\langle \psi^{(1/2)}(j)|{\cal H}|\psi^{(1/2)}(j)\rangle}%
{\langle \psi^{(1/2)}(k)|\psi^{(1/2)}(k)\rangle} \; .
\end{split}
\label{eqn:SMA-4}
\end{equation}
\subsection{Fixing parent Hamiltonian}
Before calculating the SMA spectra 
(\ref{eqn:SMA-3}) and (\ref{eqn:SMA-4}), we have to fix the form of 
the parent Hamiltonian.  
As has been mentioned in section \ref{sec:SUSY-AKLT-1}, 
the non-hermitian parent Hamiltonian for 
the SUSY (UOSp($1|2$)) VBS model is 
given\cite{Arovas-H-Q-Z-09} by eq.(\ref{eqn:parent-Ham}):
\begin{equation*}
\widetilde{\cal H}_{L=1\text{ sVBS}}
= \sum_{j}\left\{
V_{3/2}P_{3/2}({\cal C}_{j,j+1})
+ V_{2}P_{2}({\cal C}_{j,j+1})
\right\}
\end{equation*}
with the coupling constants $V_{3/2},V_{2} \gneq 0$ positive.    

\begin{figure}[H]
\begin{center}
\includegraphics[scale=0.6]{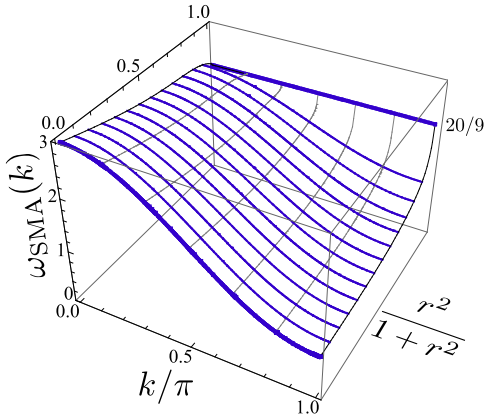}
\caption{(Color online) The spin excitation (triplon) 
spectrum $\omega^{\text{s}}_{\text{SMA}}(k)$ obtained by single-mode approximation (SMA). 
At $r=0$, it reduces to the well-known 
dispersion $\omega_{\text{SMA}}(k)= 10(5+3 \cos k)/27$ of the spin-1 VBS model. 
When $r\nearrow \infty$ (Majumdar-Ghosh limit), 
dispersion becomes flat. 
\label{fig:SMA-spectrum}}
\end{center}
\end{figure}

The above form is not very convenient 
since it breaks hermiticity necessary for eq.(\ref{eqn:HtosVBS}) and 
one still has one free parameter even after the overall 
energy scale is fixed\footnote{%
In the usual spin-1 VBS (AKLT) model, the overall energy scale 
fixes the parent Hamiltonian uniquely. For $S \geq 2$, the energy 
scale alone is not enough to determine the unique parent Hamiltonian.}. 
Instead of using $\widetilde{\cal H}_{L=1\text{ sVBS}}$, 
one may adopt 
\begin{equation}
{\cal H}_{L=1\text{ sVBS}}
=\widetilde{\cal H}_{L=1\text{ sVBS}}^{\dagger}
\widetilde{\cal H}_{L=1\text{ sVBS}}
\end{equation}
as the hermitian Hamiltonian\footnote{%
Using $P^{\dagger}_{3/2}P_{2}=P^{\dagger}_{2}P_{3/2}=0$, one can see that 
this definition is essentially equivalent to replacing the projection operators $P_{l}$ 
with $P^{\dagger}_{l}P_{l}$. }.  
One way to fix the remaining coupling is to require that the SUSY 
parent Hamiltonian should reduce in the $r\rightarrow \infty$ 
to the standard (SU(2)) VBS Hamiltonian\cite{affleck1988vbg}
\begin{equation}
{\cal H}_{S=1\text{ VBS}}
= \sum_{j}\left\{
\bolS_{j}{\cdot}\bolS_{j+1} 
+ \frac{1}{3}(\bolS_{j}{\cdot}\bolS_{j+1})^{2} + \frac{2}{3}
\right\} \; .
\label{eqn:S1-VBS-Ham}
\end{equation}
However, this still has a problem; since some of the matrix elements 
in the fermionic sector have a factor $1/r$, the limit 
$r\rightarrow \infty$ is divergent.  Fortunately, this is not 
so serious.  If we note that the ground states contain no fermion 
in the $r\rightarrow\infty$ limit, the most natural way is 
to require that the SUSY parent Hamiltonian {\em projected onto 
the bosonic sector} should coincide with the spin-1 VBS 
Hamiltonian (\ref{eqn:S1-VBS-Ham}).  This fixes the two coupling 
constants as\footnote{%
In fact, we can freely add any function $V(r)$ satisfying 
$V(r)\gneq 0$ ($r>0$) and $V(r)\rightarrow 0$ ($r\rightarrow 0$). 
The simplest choice, which is regular even 
in the $r\rightarrow \infty$ limit, would be $V(r)=\tanh r$.    
}:
\begin{equation}
V_{3/2}=\tanh r \; , \;\; V_{2}=\sqrt{2} \; .
\label{eqn:VBS-coupling}
\end{equation}
The spin-excitation (`crackion') spectrum obtained by using 
(\ref{eqn:SMA-3}) and (\ref{eqn:VBS-coupling}) is shown 
in Fig.\ref{fig:SMA-spectrum}.  
At $r=0$ (AKLT-limit), the dispersion reduces to the well-known 
one\cite{Arovas-A-H-88}: 
\begin{equation}
\omega^{\text{s}}_{\text{SMA}}(k) = 
\frac{10}{27} (5 + 3 \cos k)  \; .
\end{equation}
For $r\nearrow \infty$, on the other hand, the spin excitation loses 
the dispersion.  This is easily understood since the ground-state 
in this limit reduces 
to the translationally invariant combination of two Majumdar-Ghosh 
states (see Fig.~\ref{expSUSY.fig}) and the overlap between crackion 
states, which gives the dispersion of the spin excitations, 
trivializes (see (\ref{eqn:SMA-3}) and (\ref{eqn:crackion-overlap-1})):
\begin{equation}
\VEV{\psi^{(0)}_{i}|\psi^{(0)}_{j}} \propto \delta_{i,j} \; , \;\; 
\VEV{\psi^{(0)}(k)|\psi^{(0)}(k)} =\text{const.} \; .
\end{equation} 

\begin{figure}[H]
\begin{center}
\includegraphics[scale=0.6]{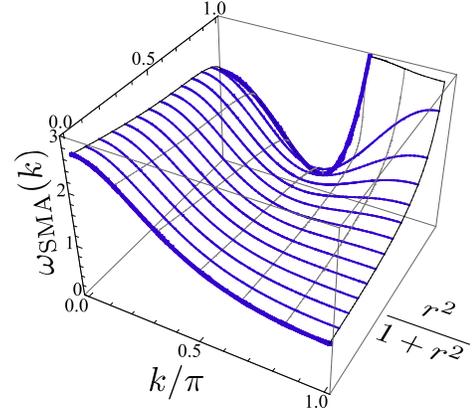}
\caption{(Color online) The excitation spectrum $\omega^{\text{h}}_{\text{SMA}}(k)$ of 
a spinon-hole pair obtained by single-mode approximation 
(eq.(\ref{eqn:SMA-4})). 
This spinon-hole pair state is created by fermionic generator 
$K_{1}$ except at $r=0$, where the transition matrix elements of $K_{1}$ 
from the ground state vanish. 
\label{fig:SMA-spectrum-hole}}
\end{center}
\end{figure}

The charge excitation spectrum is calculated similarly by using 
eq.(\ref{eqn:SMA-4}).  The result is shown in
Fig.~\ref{fig:SMA-spectrum-hole}. 
For $r=0$, the spectrum is given by%
\begin{equation}
\omega_{\text{SMA}}^{\text{h}}(k)
= \frac{8}{3(2-\cos k)} \; .
\end{equation}
A remark is in order here about the existence of the two different spectra 
$\omega^{\text{s}}(k)$ and $\omega^{\text{h}}(k)$.  One may naively 
expect $\omega^{\text{s}}(k)=\omega^{\text{h}}(k)$ as the supersymmetry 
relates the bosonic generators $\mathbf{S}$ and the fermionic ones $K_{\alpha}$.  
However, this relies 
on the existence of a `unitary' transformation which linearly transforms 
the set of the SUSY generators onto themselves (adjoint representation). 
Since no such transformation exists here, we generally expect different spectra 
for the spin- and the charge sector as has been shown above. 
\section{SUSY-VBS state II}
\label{sec:SUSY-VBS-2}
Now let us add one more fermion species and 
consider yet another SUSY-VBS wavefunction which now 
includes two holes $f$ and $g$.  
As has been mentioned in section \ref{eqn:type-II-intro}, 
the state contains two (spin) bosons $(a,b)$ and two fermions $(f,g)$, and 
we may expect it to exhibit clearer spin-charge symmetry with respect to $r=1$. 

The second generalized sVBS wavefunction (the case $M=1$ of eq.(\ref{sVBSstateII})) 
is defined by:
\begin{equation}
\begin{split}
& |\text{sVBS-II}\rangle \\
& \equiv  (\cdots)
\left\{a^{\dagger}_{j-1}b^{\dagger}_{j}-b^{\dagger}_{j-1}a^{\dagger}_{j}
-r(f^{\dagger}_{j-1}g^{\dagger}_{j}
+g^{\dagger}_{j-1}f^{\dagger}_{j})\right\} \\
& \quad 
\left\{a^{\dagger}_{j}b^{\dagger}_{j+1}-b^{\dagger}_{j}a^{\dagger}_{j+1}
-r(f^{\dagger}_{j}g^{\dagger}_{j+1}
+g^{\dagger}_{j}f^{\dagger}_{j+1})\right\}
(\cdots)|0\rangle \; .
\end{split}
\label{eqn:def-SUSY-AKLT-2}
\end{equation}
As we have seen in section \ref{eqn:type-II-intro}, this state is based 
on the algebra UOSp(2$|$2) and one can construct 
the parent Hamiltonian in a similar manner to the type I case (based on 
 UOSp(1$|$2)) 
(we do not give the explicit form here. The interested readers may refer 
the online supplementary material\cite{supp}.).  
\subsection{Matrix-product representation}
\label{sec:MPS-type-II}
We follow the same steps as in section \ref{sec:SUSY-AKLT-1} 
with a different metric matrix
\begin{equation}
{\cal R}_{\text{II}} = 
\begin{pmatrix}
0 & 1 & 0 & 0\\
-1 & 0 & 0 & 0\\
0 & 0 & 0 &-1 \\
0 & 0 & -1 & 0  
\end{pmatrix}  
\label{eqn:metric-II}
\end{equation}
and the spinor 
$(a^{\dagger}_{j},b^{\dagger}_{j}, \sqrt{r}f^{\dagger}_{j},\sqrt{r}g^{\dagger}_{j})^{\text{t}}$
to obtain the MPS representation for the second sVBS state:
\begin{subequations}
\begin{equation}
\begin{split}
A_{j} &= 
\begin{pmatrix}
b^{\dagger}_{j} \\ -a^{\dagger}_{j} \\ 
-\sqrt{r}g^{\dagger}_{j} \\ -\sqrt{r}f^{\dagger}_{j}
\end{pmatrix}
\begin{pmatrix}
a^{\dagger}_{j} & b^{\dagger}_{j} & 
\sqrt{r}f^{\dagger}_{j} & \sqrt{r}g^{\dagger}_{j} 
\end{pmatrix}
|\text{vac}\rangle_{j} \\
&= 
\begin{pmatrix}
a^{\dagger}_{j}b^{\dagger}_{j} & (b^{\dagger}_{j})^{2} 
& \sqrt{r}b^{\dagger}_{j}f^{\dagger}_{j}
& \sqrt{r}b^{\dagger}_{j}g^{\dagger}_{j} \\
-(a^{\dagger}_{j})^{2} & -a^{\dagger}_{j}b^{\dagger}_{j} 
& -\sqrt{r}a^{\dagger}_{j}f^{\dagger}_{j}
& -\sqrt{r}a^{\dagger}_{j}g^{\dagger}_{j} \\
-\sqrt{r}g^{\dagger}_{j}a^{\dagger}_{j}
& -\sqrt{r}g^{\dagger}_{j}b^{\dagger}_{j} 
& -r g^{\dagger}_{j}f^{\dagger}_{j} & 0 \\
-\sqrt{r}f^{\dagger}_{j}a^{\dagger}_{j}
& -\sqrt{r}f^{\dagger}_{j}b^{\dagger}_{j} 
& 0 & -r f^{\dagger}_{j}g^{\dagger}_{j}
\end{pmatrix}
| \text{vac} \rangle_{j}
\end{split}
\end{equation}
\begin{equation}
\begin{split}
A^{\dagger}_{j} &= {}_{j}\langle \text{vac}|
\begin{pmatrix}
a_{j} \\ b_{j} \\ \sqrt{r}f_{j} \\ \sqrt{r}g_{j} 
\end{pmatrix}
\begin{pmatrix}
b_{j} & -a_{j} & -\sqrt{r}g_{j} & -\sqrt{r}f_{j}
\end{pmatrix} \\
&= {}_{j}\langle \text{vac}|
\begin{pmatrix}
a_{j}b_{j} & -(a_{j})^{2} & -\sqrt{r}a_{j}g_{j}& -\sqrt{r}a_{j}f_{j} \\
(b_{j})^{2} & -a_{j}b_{j} & -\sqrt{r}b_{j}g_{j}& -\sqrt{r}b_{j}f_{j} \\
\sqrt{r}f_{j}b_{j}& -\sqrt{r}f_{j}a_{j} & -r f_{j}g_{j} & 0 \\
\sqrt{r}g_{j}b_{j}& -\sqrt{r}g_{j}a_{j} & 0 & -r g_{j}f_{j}
\end{pmatrix}
\end{split}
\end{equation}
\end{subequations}

\begin{figure}[h]
\begin{center}
\includegraphics[scale=0.8]{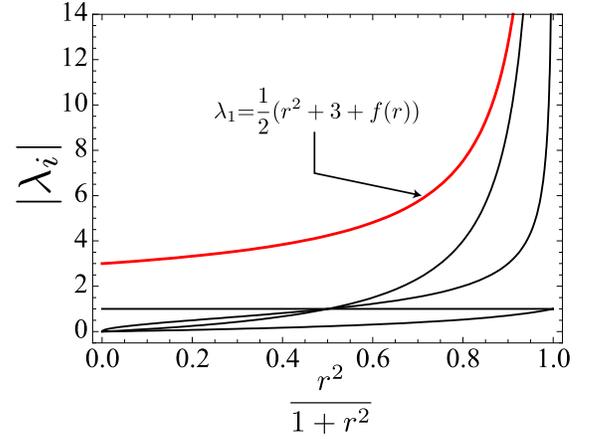}
\caption{(color online) Plot of absolute values $|\lambda_{i}|$ 
of the seven different eigenvalues of $G$. Since $|\lambda_{i}|$s 
are symmetric with respect to $r\mapsto -r$, only the $r>0$ part 
is shown.  
\label{fig:eigenV-SUSY-AKLT2}}
\end{center}
\end{figure}

As in the first sVBS state, the supertrace is necessary for the periodic 
system:
\begin{equation}
|\text{sVBS-II}\rangle = \text{STr}\left\{
\bigotimes_{j=1}^{L} A_{j}
\right\} \; ,
\label{eqn:STr-2}
\end{equation}
where $\text{STr}(\mathcal{M}) \equiv \mathcal{M}_{11}+\mathcal{M}_{22}
-\mathcal{M}_{33}-\mathcal{M}_{44}$.   
The $T$-matrix is a 16$\times$16 matrix 
and has seven different eigenvalues $\lambda_{i}$ 
(see Fig.~\ref{fig:eigenV-SUSY-AKLT2}): 
\begin{equation}
\begin{split}
\{ \lambda_{i} \}
=&
\bigl\{-1(\times 3),-i r(\times 4),+i r(\times 4),
-r^2(\times 2),r^2, \\
& \frac{1}{2} \left(r^2+3-f(r)\right),
\frac{1}{2} \left(r^2+3+f(r)\right)\bigr\} \; ,
\end{split}
\end{equation}
where $f(r)\equiv \sqrt{r^4+10r^2+9}$.  
Regardless of the value of $r$, the eigenvalue with largest modulus is:
\begin{equation}
\lambda_{1} = 
\frac{1}{2} \left(r^2+3+f(r)\right)  \; .
\end{equation}
Since the set of eigenvalues is invariant under $r\leftrightarrow -r$, 
we can restrict ourselves to $r\geq 0$.  
\subsection{spin-spin correlation}
Let us begin with the spin-spin correlation function. 
By using the method described in Appendix \ref{sec:cor-func}, 
it is straightforward to calculate 
the correlation function $\VEV{S^{a}_{j}S^{a}_{j+n}}$: 
\begin{equation}
\begin{split}
& \VEV{S^{a}_{j}S^{a}_{j+n}} \\
&=
\begin{cases}
\frac{2}{f(r)} & \; \text{for } n=0 \\
\frac{r^2+5+f(r)}{2 f(r)}
\left(-\frac{2}{r^2+3+f(r)}\right)^n 
& \; \text{for } n>0  \; . 
\end{cases}
\end{split}
\end{equation}

In obtaining these expressions, it has been assumed that 
both end points ($x$ and $x+n$) are infinitely far from the edges 
(otherwise there will be another decaying factor coming from 
the edge effects). 
From these, we can read off the spin-spin correlation length:
\begin{equation}
\xi_{\text{spin}}(r) =
1/\log \left\{
\left(r^2+3+f(r)\right)/2 \right\}  \; ,
\end{equation}
which monotonically decreases from $1/\ln(3)$ ($r=0$) 
to 0 ($r\nearrow\infty$). 

The existence of the edge states may be best illustrated by 
plotting the local magnetization $\VEV{S^{z}_{j}}$. 
\begin{figure}[H]
\begin{center}
\includegraphics[scale=0.6]{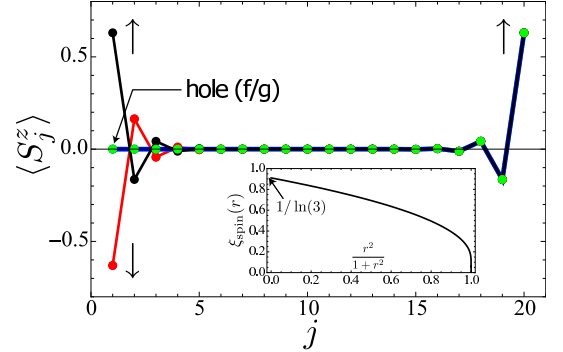}
\caption{(color online) Plot of local magnetization profile 
$\VEV{S^{z}_{j}}$ for different (left) edge states (with right 
edge fixed). (inset) Spin correlation length $\xi_{\text{spin}}(r)$ 
as a function of $r$. It monotonically decreases as $r$ is increased 
and approaches to zero like 
$\xi_{\text{spin}}\sim 1/\log \left(1+r^2\right)$.   
\label{fig:SzSz-AKLT-2a}}
\end{center}
\end{figure}
\subsection{Superconducting correlation}
\begin{figure}[H]
\begin{center}
\includegraphics[scale=0.7]{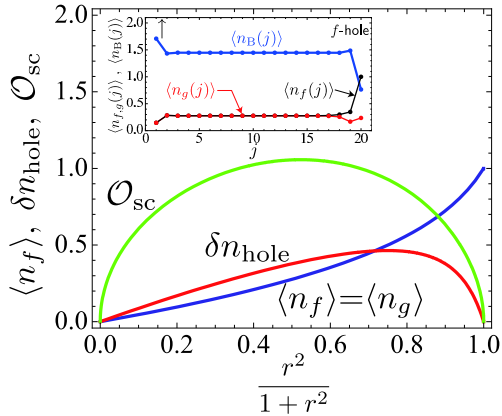}
\caption{(Color online) Plot of 
${\cal O}_{\text{sc}}=\VEV{\Delta_{j}}$, 
the hole density $\VEV{n_{\text{hole}}(j)}=\VEV{f^{\dagger}_{j}f_{j}}$ and 
the hole-number fluctuation 
$\VEV{f^{\dagger}_{j}f_{j}}-\VEV{f^{\dagger}_{j}f_{j}}^{2}$ 
as a function of $r$.  Here the bulk values are plotted.  
(Inset): Profile of the hole density ($r=0.5$ or $r^{2}/(1+r^{2})=0.2$) 
for a finite system ($L=20$).  Only the left edge state 
is changed with the right one fixed to $s_{\text{R}}=\uparrow$. 
The hole density approaches exponentially to the bulk value 
as we move away from the edge. 
\label{fig:SC-orderparam-2}}
\end{center}
\end{figure}

Since the type-II sVBS state (\ref{sVBSstateII}) contains 
hole pairs on adjacent sites, we may expect that the pair 
amplitudes take finite expectation values. 
As in section \ref{sec:SC-correlation-I}, we may define 
the following order parameters on general grounds:
\begin{subequations}
\begin{align}
\Delta^{ff}_{j} & \equiv 
(a_{j}b_{j+1} - b_{j}a_{j+1})f^{\dagger}_{j}f^{\dagger}_{j+1} \\
\Delta^{gg}_{j} & \equiv 
(a_{j}b_{j+1} - b_{j}a_{j+1})g^{\dagger}_{j}g^{\dagger}_{j+1} \\
\Delta^{fg}_{j} & \equiv 
(a_{j}b_{j+1} - b_{j}a_{j+1})
(f^{\dagger}_{j}g^{\dagger}_{j+1}
+ g^{\dagger}_{j}f^{\dagger}_{j+1}) \; .
\end{align}
\end{subequations}
However, the first two are identically zero by construction 
of $|\text{sVBS-II}\rangle$.  
The only non-vanishing superconducting order parameter 
\begin{equation}
{\cal O}_{\text{sc}}=\VEV{\Delta^{fg}_{j}}
\end{equation}
is plotted in Fig.~\ref{fig:SC-orderparam-2} 
for various values of $r$.  Also plotted are 
the hole ($f$ and $g$) number $\VEV{n_{f,g}}$ 
and the hole-number fluctuation $\delta n_{\text{hole}}$:
\begin{equation}
\begin{split}
& \VEV{n_{f}}=\VEV{f_{j}^{\dagger}f_{j}}=
\VEV{g^{\dagger}_{j}g_{j}}=\VEV{n_{g}} \, , \\
&\delta n_{\text{hole}}=\VEV{n^{2}_{\text{hole}}} 
- \VEV{n_{\text{hole}}}^{2} \; 
(n_{\text{hole}}\equiv n_{f}+n_{g}) \; .
\end{split}
\end{equation}
The superconducting order parameter ${\cal O}_{\text{SC}}$ 
is maximal at $r\approx 1.05$ (or, $r^{2}/(1+r^{2})\approx 0.52$).   

The superconducting correlation (hole-hole correlation) 
\begin{equation}
C^{fg}_{\text{sc}}(n) \equiv 
(a_{j}b_{j+n} - b_{j}a_{j+n})
(f^{\dagger}_{j}g^{\dagger}_{j+n}
+ g^{\dagger}_{j}f^{\dagger}_{j+n})
\end{equation}
decays exponentially with the correlation length
\begin{equation}
\xi_{\text{sc}}(r)
=\log^{-1} \left\{\frac{r^2+3+f(r)}{2 r}\right\} \; .
\end{equation}
\begin{figure}[H]
\begin{center}
\includegraphics[scale=0.8]{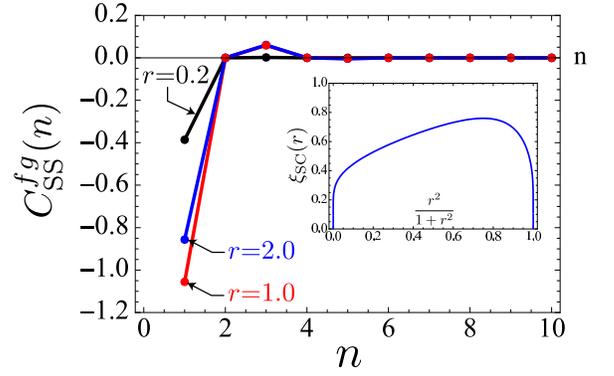}
\caption{(Color online) Plot of hole correlation 
$C^{fg}_{\text{SC}}(n) =
(a_{j}b_{j+n} - b_{j}a_{j+n})
(f^{\dagger}_{j}g^{\dagger}_{j+n}
+ g^{\dagger}_{j}f^{\dagger}_{j+n})$ for various $r$.  
Due to the form of the wave function, hole correlation identically 
vanishes when the distance $n$ is even. 
Inset: correlation length $\xi_{\text{SC}}(r)$ of the hole correlation. 
\label{fig:SC-AKLT-2}}
\end{center}
\end{figure}
\subsection{String correlation}\label{subsec:stringcorrelationII}
Then, we proceed to the string correlation function. 
As in the previous case (type I sVBS), the string correlation 
explicitly depends on the distance between the two end points through 
the exponentially decaying factor:
\begin{equation}
(-1)^{n}\left\{\frac{f(r)-(r^2+3)}{
f(r)+(r^2+3)}\right\}^{n}  \; .
\end{equation}
These expressions imply that the correlation lengths 
($\xi_{\text{string}}$) for the string correlation are different from 
$\xi_{\text{spin}}$ for the spin-spin correlation function.  

\begin{figure}[H]
\begin{center}
\includegraphics[scale=0.7]{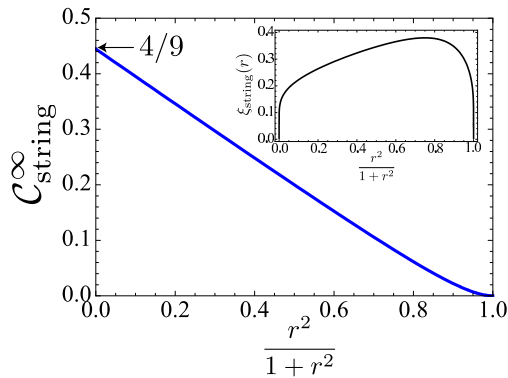}
\caption{The infinite-distance limit of 
the string correlation function 
$\VEV{S_{x}^{z}\exp[i\pi\sum_{j=x}^{x+n-1}S_{j}^{z}]S^{z}_{x+n}}$ 
($n\nearrow \infty$) as a function of $r$.  
The value of string correlation smoothly decreases from the AKLT value 
$4/9$ to 0 (no spins left). 
\label{fig:string-AKLT-2c}}
\end{center}
\end{figure}

The infinite-distance limit of the string correlation is given as:
\begin{equation}
{\cal O}^{\infty}_{\text{string}}
=\frac{4}{\left(r^2+1\right) \left(r^2+9\right)} \; .
\label{eqn:string-AKLT-2a}
\end{equation}
It is easy to check that when $r=0$ eq.(\ref{eqn:string-AKLT-2a}) reproduces 
the value $4/9$ of the spin-1 AKLT model\cite{Kennedy-T-92a}.  
The results are plotted in Fig.~\ref{fig:string-AKLT-2c} 
together with the correlation length $\xi_{\text{string}}(r)$.  
In contrast to the first case $|\text{sVBS-I}\rangle$ 
(see Fig.~\ref{fig:Ostring-model1}),  
the $r\nearrow \infty$ limit of ${\cal O}_{\text{string}}^{\infty}$ 
is zero since spins disappear from the state $|\text{sVBS-II}\rangle$ 
in this limit.  
\section{Symmetry-protected topological order}
\label{sec:symmetry-protected}
Though the string-order parameter captures the diluted N\'{e}el order of the Haldane phase, 
the string-order itself is fragile under small perturbations \cite{Anfuso-R-07-2,Gu-W-09}.  
Recently, Li and Haldane proposed \cite{Li-H-08} to use 
the structure of the low-lying part of the entanglement spectrum 
(the logarithm of the eigenvalues of the reduced density matrix 
for either of the two partitioned systems)  
as the signature of topological order inherent in the state.    
Pollmann et al. \cite{Pollmann-T-B-O-10,pollmann-2009} have 
investigated the relation between the level structure (e.g. degeneracy) 
of the entanglement spectrum  and discrete symmetries of the system;   
they showed that, for  odd-$S$ spin chains, the existence of (at least one of) the three 
discrete symmetries (time-reversal symmetry, link-inversion, 
and $\mathbb{Z}_2\times \mathbb{Z}_2$ symmetry) guarantees (at least two-fold) degeneracy 
in each entanglement level, while  
for even-$S$ spin chains, the existence of the above discrete symmetries tells nothing 
about degeneracy. 
By this observation, they have argued that the Haldane phase 
in odd-$S$ spin chains is a stable topological phase protected by discrete symmetries.  

Such arguments can also be applicable to the stability discussion of the Haldane-like phase 
of the present SUSY spin models. For instance, the type I sVBS states contains      
the UOSp(1$|$2) superspin-$M$ multiplet that consists of two SU(2) spin multiplets 
whose spins differ by $1/2$.  
By partitioning a superspin-$M$ sVBS infinite chain to two semi-infinite segments,  
there appear two SU(2) spins $M/2$ and $(M-1)/2$ 
on the ``edge'' of each of two sVBS chain segments 
(hence $(2M+1)$ edge states instead of $(S+1)$ ones in the usual spin-$S$ VBS states). 
It is noted that, regardless of the parity of the bulk superspin $M$, 
the sVBS state accommodates a half-integer SU(2) spin on the edge. 
Therefore, for any integer-superspin sVBS states,  
the entanglement spectrum always contains a sector consisting of at least doubly degenerate levels  
which come from the half-integer SU(2) spin sector of the entanglement Hilbert space.   
For example, the entanglement spectrum of the $M=2$ sVBS state consists of a doubly degenerate level 
corresponding to the doubly degenerate fermionic sector and a bosonic level with three-fold degeneracy. 
In fact, we can show that if one of  the discrete symmetries (link inversion and time-reversal) 
is present in the SUSY spin chains, there is always a sector in the entanglement spectrum 
each of whose levels is at least doubly degenerate.   
This implies that the `Haldane phase' is stabilized {\em regardless} of the parity of the bulk (integer) 
superspins. We will report the details elsewhere.  

\section{Summary}
\label{sec:summary}
In the present paper, we have constructed a supersymmetric extension of the matrix-product 
states (sMPS) for two different types (I and II)  of supersymmetric VBS (sVBS) states 
and exactly evaluated various physical quantities. 
The sMPS constructed here contains the fermionic elements as well as the usual bosonic 
(i.e. commuting) ones and this slightly complicates the treatment (for instance, 
instead of the trace, the supertrace is used for the periodic systems).  
We investigated the hole-doping behaviors of various correlation functions 
(spin-spin and superconducting) and the spin- and the hole excitation spectrum.   

In the charge sector, the type I sVBS chains exhibit insulating behavior 
at zero and infinite concentrations of the doped holes and the superconducting 
order parameter is finite only for finite doping. 
In the spin sector, the type I sVBS chains interpolate between the usual VBS state 
and the inhomogeneous VBS state (in the simplest case, it reduces to the MG dimer state) 
at the two extremal limits of hole-doping $r=0$ and $r=\infty$, respectively. 
The single-mode approximation has been applied to obtain the spin- and the charge excitation 
spectrum. 
There are two types of low-lying excitations, $i.e.$ the triplon and the spinon-hole pair, 
created respectively by the bosonic and fermionic generators of the super Lie algebra.  
The spinon-hole pair is peculiar to the sVBS states;  
it simultaneously possesses the property of the spin-$1/2$ spinon and 
the unpaired hole in the superconducting background.  
We have found that the spinon-hole pair can be the lowest excitation 
in some parameter region of the hole-doping.  

As another class of sVBS states based on a larger ($\mathcal{N}{=}2$) SUSY, 
we have introduced the type II sVBS states.   
In the high-doping limit ($r\rightarrow \infty$), 
the superspin-1 ($M=1$) type II sVBS state reduces to the totally uncorrelated hole-VBS state, 
while it reproduces the spin VBS state in the zero-doping limit. 
The type II sVBS state displays qualitatively similar behaviors in the spin- and the charge 
properties except that now physical quantities are more symmetric with respect to 
the point $r=1$ reflecting that the model contains the equal numbers of bosons and fermions.  
 
We have demonstrated the existence of a hidden order in the sVBS states 
(both type I and II) by calculating 
the non-local string correlations.  What is remarkable is that the string correlation revives 
upon hole doping although it vanishes in the pure-spin limit $r\rightarrow 0$ 
when the spin $S=M$ is even integer.  
This may be understood as an example of symmetry-protected topological order in SUSY spin chains. 

Though the present work is restricted to 1D chains, the sVBS states themselves 
can be formulated on any lattice in arbitrary dimensions, 
and may generally exhibit resonating-valence-bond (RVB) features at finite hole doping.  
For instance, an $M=2$ sVBS state with three species of holes simulates the Rokhsar-Kivelson 
RVB\cite{Rokhsar-K-88} in the high-doping limit. 
Such higher dimensional analyses are interesting both theoretically and experimentally, 
and may be carried by a supersymmetric extension of the tensor network method.

\section*{Acknowledgement}
We wish to thank the organizers 
of the workshop 
{\it Topological Aspects of Solid State Physics} 
held at The Institute for Solid State Physics
where this work was initiated.  
We are also grateful to Frank Pollmann for helpful discussions. 
K.H. would like to thank warm hospitality of the condensed matter group in YITP during his stay, and acknowledges supports from the GCOE  visitor program of Kyoto University.  
He is also deeply grateful to D.P. Arovas, X.L. Qi and S.C. Zhang for the precedent collaboration on which the present work is  based.  
K.T. was supported in part by 
Grant-in-Aids for Scientific Research 
(C) 20540375, and Priority Areas ``Novel States of Matter 
Induced by Frustration'' (No.19052003) from MEXT, Japan and 
by the global COE (GCOE) program `The next generation of 
physics, spun from universality and emergence' of Kyoto University.  
\appendix
\section{A crash course on supersymmetry}
\label{crashsusy}
\subsection{UOSp(1$|$2) and UOSp(2$|$2)}
\label{subsec:uosp12}
The superalgebra UOSP(1$|$2) consists of the following five 
generators 
\begin{subequations}
\begin{align}
\begin{split}
& S^{x}=(a^{\dagger}b+b^{\dagger}a)/2 \;, \; 
S^{y}=(a^{\dagger}b-b^{\dagger}a)/(2i), \\ 
& S^{z}=(a^{\dagger}a-b^{\dagger}b)/2 \;\; (\text{bosonic})
\end{split}
\\
\begin{split}
& K_{1}= \frac{1}{2}(x^{-1}fa^{\dagger}+x f^{\dagger}b) \\
& K_{2}= \frac{1}{2}(x^{-1}fb^{\dagger}-x f^{\dagger}a) \;\;
(\text{fermionic})\label{osp12generators}
\end{split}
\end{align}
satisfying the (anti)commutation relations:
\begin{equation}
\begin{split}
& [S^{a}\, , \,S^{b}]=i\epsilon^{abc}S^{c} \quad
(a,b,c=x,y,z) \\
& [S^{a}\, , \, K_{\mu}] = \frac{1}{2}K_{\nu}(\sigma^{a})_{\nu\mu}
 \quad  (\mu,\nu=1,2) \\
& \{ K_{\mu}\, , \, K_{\nu}\} 
= \frac{1}{2}(i\sigma^{y}\sigma^{a})_{\mu\nu}S^{a} \; .
\end{split}\label{uosp12algebra}
\end{equation}
\end{subequations}
At this stage, the parameter $x$, which defines a one-parameter 
deformation of UOSp(1$|$2), is arbitrary.   
The second equation implies that the fermionic generators 
$K_{1}$ and $K_{2}$ span a two-dimensional spinor representation 
of SU(2).  

Any irreducible representation of UOSp(1$|$2) is specified by 
{\em superspin} $l(=0,1/2,1,\ldots)$.  A convenient way of 
constructing a superspin-$l$ representation is to use 
Schwinger operators (bosons $a$, $b$ and fermion $f$) satisfying 
\begin{equation}
n_{a}+n_{b}+n_{f}\equiv a^{\dagger}a + b^{\dagger}b + f^{\dagger}f 
=2l \; (\in \mathbb{Z}) \; .
\end{equation}
Then, the Casimir operator ${\cal C}$ is calculated as: 
\begin{equation}
{\cal C} \equiv \bolS^{2} + (K_{1}K_{2} - K_{2}K_{1}) 
= l(l+1/2) \; .
\label{eqn:def-Casimir}
\end{equation}

The SU(2) subalgebra depends only on $a$ and $b$:
\begin{equation}
\begin{split}
& \bolS^{2} = \left\{(n_{a}+n_{b})^{2}
+2(n_{a}+n_{b})\right\}/4 = S(S+1) \\
& S=(n_{a}+n_{b})/2  \; .
\end{split}
\end{equation}
Since $n_{f}=0,1$, a $(4l{+}1)$-dimensional 
superspin-$l$ representation splits into 
two SU(2) irreducible representations:
\begin{equation}
\begin{split}
& \text{(i)} \; S=l \;\; (n_{f}=0) \cdots (2l+1)\text{-dim} \\
& \text{(ii)} \; S=l-1/2 \;\; (n_{f}=1) \cdots 2l\text{-dim} \; ,
\end{split}
\end{equation}
which are connected to each other by the fermionic generators 
$K_{1,2}$.   
For instance, the five states in the $l=1$ representation are:
\begin{equation}
\begin{split}
& \text{(i)} \;
|+\rangle=\frac{1}{2}{a_i^{\dagger}}^2|\text{vac}\rangle,
\quad |0\rangle={a_i^{\dagger}b_i^{\dagger}}|\text{vac}\rangle, 
\quad |-\rangle=\frac{1}{2}{b_i^{\dagger}}^2|\text{vac}\rangle, \\
& \text{(ii)} \; |\!\uparrow\rangle=
 a_i^{\dagger}f_i^{\dagger}|\text{vac}\rangle, 
\quad |\!\downarrow \rangle =b_i^{\dagger}f_i^{\dagger}|\text{vac}\rangle  \; .
\end{split}
\end{equation}
In constructing the sVBS states, we identify (ii) as a one-hole 
state.  The $l=1/2$ case is relevant in realizing 
the so-called {\em superqubit}\cite{borsten-2010-81}.   

A two-site system can be treated in the same manner as in SU(2); 
we just define $\bolS^{\text{tot}}=\bolS^{(1)}+\bolS^{(2)}$,  
$K_{1,2}^{\text{tot}}=K_{1,2}^{(1)}+K_{1,2}^{(2)}$ and 
the corresponding Casimir operator by 
\begin{equation}
\begin{split}
{\cal C}_{1,2} &\equiv \bolS^{\text{tot}}{\cdot}\bolS^{\text{tot}}
+\epsilon_{\mu\nu}K^{\text{tot}}_{\mu}K^{\text{tot}}_{\nu} \\
&= {\cal C}^{(1)}+{\cal C}^{(2)} 
+ 2\left\{
\bolS^{(1)}{\cdot}\bolS^{(2)} 
+\epsilon_{\mu\nu} K_{\mu}^{(1)}K_{\nu}^{(2)}
\right\} \\
&\equiv {\cal C}^{(1)}+{\cal C}^{(2)} 
+ 2{\cal S}^{(1)}{\cdot}{\cal S}^{(2)} \; .
\end{split}
\label{eqn:Casimir-2site}
\end{equation}
The Clebsch-Gordan decomposition is simply given as:
\begin{equation}
l\otimes l \simeq 0\oplus \frac{1}{2} \oplus 1 \oplus 
\cdots \oplus (2l-1/2) \oplus 2l  \; .
\label{eqn:CG-decomp}
\end{equation}
So far, the deformation parameter $x$ is arbitrary.  
However, in order for 
$(a_{1}^{\dagger}b^{\dagger}_{2}-b_{1}^{\dagger}a^{\dagger}_{2} 
-r f_{1}^{\dagger}f^{\dagger}_{2})$ to behave as a 
UOSp(1$|$2)-singlet, $x^{2}=r$ is required.   
 
By flipping the relative signs of the first and second terms in $K_{\mu}$ (\ref{osp12generators}), one may define ``new'' fermionic operators:  
\begin{align}
& D_{1}= \frac{1}{2}(-x^{-1}fa^{\dagger}+x f^{\dagger}b) \nonumber\\
& D_{2}= -\frac{1}{2}(x^{-1}fb^{\dagger}+x f^{\dagger}a).
\end{align}
The type I sVBS states are not invariant under the transformation generated by $D_{\mu}$. 
(Thus, the largest symmetry of the type I sVBS states is UOSp(1$|$2).) 
With inclusion of $D_{\mu}$, the UOSp(1$|$2) generators satisfy the  UOSp(2$|$2) algebra 
\begin{equation}
\begin{split}
&[S_a,S_b]=i\epsilon_{abc}S_c\, , \;\;
\{K_{\mu},K_{\nu}\}=\frac{1}{2}(\epsilon\sigma_a)_{\mu\nu}S_a\, , \\
&\{D_{\mu},D_{\nu}\}=-\frac{1}{2}{(\epsilon\sigma_a)}_{\mu\nu}S_a\, , \\
&[S_a,K_{\mu}]=\frac{1}{2}(\sigma_a)_{\nu\mu}K_{\nu} \, , \;\;
[S_a,D_{\mu}]=\frac{1}{2}(\sigma_a)_{\nu\mu}D_{\nu} \, , \\
&\{K_{\mu},D_{\nu}\}=-{\frac{1}{4}}\epsilon_{\mu\nu}\Gamma \, , \\
&[S_a,\Gamma]=0 \, , \;\; 
[K_{\mu},\Gamma]=-D_{\mu} \, , \;\;
[D_{\mu},\Gamma]=-K_{\mu} \, , 
\label{osp22algebra}
\end{split}
\end{equation}
where $\Gamma$ is defined by 
\begin{equation}
\Gamma=a^{\dagger}a+b^{\dagger}b+2f^{\dagger}f \; .
\end{equation}
\section{A quick recipe for matrix-formalism}
\label{sec:quick-recipe}
In this section, we extend the standard 
formalism for bosonic matrix-product states so that 
we can handle fermionic states as well.  
\subsection{Norm}
\label{sec:norm}
We begin with the computation of the norm of $|\text{MPS}\rangle$. 
Since we consider cases where $A_{j}$ is made up with {\em both} 
bosonic- and fermionic states, a special care has to be taken and 
we proceed step by step.   If we write the matrix indicies explicitly, 
$|\text{MPS}\rangle$ reads:
\begin{equation}
\begin{split}
& |\text{MPS}\rangle_{(\alpha,\gamma)} \\
&= \sum_{\{\beta_{j}\}}
A_{1}(\alpha,\beta_{1})A_{2}(\beta_{1},\beta_{2})\dots \\
& \phantom{\sum_{\{\beta_{j}\}}}
\underrightarrow{%
\dots 
A_{j}(\beta_{j-1},\beta_{j})A_{j+1}(\beta_{j},\beta_{j+1})\dots 
}
A_{L}(\beta_{L-1},\gamma)  \; ,
\end{split}
\end{equation}
where the arrow indicates how the order of matrix multiplication 
and the site indices ($1,2,\ldots,l$) are related. 
If the parent Hamiltonian ${\cal H}=\sum_{j}h_{j,j+1}$ 
is defined in such a way that 
\begin{equation}
h_{j,j+1}(A_{j}\otimes A_{j+1})=0 \;\; 
\text{(for all matrix elements)} \; ,
\end{equation}
the matrix indices are physically related to some zero-energy 
degrees of freedom localized at the boundaries ({\em edge states}).  
It is important to keep the order ($\rightarrow$) 
of the string of matrices. 
If we adopt the following convention for the hermitian conjugation 
of fermionic operators:
\begin{equation}
\left(f_{1}f_{2}\dots f_{i-1}f_{i}\right)^{\dagger} 
\equiv f^{\dagger}_{i}f^{\dagger}_{i-1}\dots f^{\dagger}_{2}
f^{\dagger}_{1} \; ,
\end{equation}
then the dual of $|\text{MPS}\rangle$ reads 
\begin{equation}
\begin{split}
& \langle \text{MPS}|_{(\alpha,\gamma)} \\
&= \sum_{\{\bar{\beta}_{j}\}} 
A^{\ast}_{L}(\bar{\beta}_{L-1},\gamma)\dots \\
& \qquad 
A^{\ast}_{j+1}(\bar{\beta}_{j},\bar{\beta}_{j+1})
A^{\ast}_{j}(\bar{\beta}_{j-1},\bar{\beta}_{j}) 
\dots 
A^{\ast}_{2}(\bar{\beta}_{1},\bar{\beta}_{2})
A^{\ast}_{1}(\alpha,\bar{\beta}_{1}) \\
&= \sum_{\{\bar{\beta}_{j}\}} 
A^{\dagger}_{L}(\gamma,\bar{\beta}_{L-1})\dots \\
& \qquad \underleftarrow{%
A^{\dagger}_{j+1}(\bar{\beta}_{j+1},\bar{\beta}_{j})
A^{\dagger}_{j}(\bar{\beta}_{j},\bar{\beta}_{j-1}) 
\dots 
}
A^{\dagger}_{2}(\bar{\beta}_{2},\bar{\beta}_{1})
A^{\dagger}_{1}(\bar{\beta}_{1},\alpha) \; ,
\end{split}
\end{equation}
where $A^{\ast}_{j}$ and $A^{\dagger}_{j}$ denotes a matrix obtained by 
replacing $|\cdot\rangle \mapsto \langle \cdot|$ in $A_{j}$ and 
its transposition.  

For a periodic chain, the fermion sign has to be treated carefully. 
Using the identity 
$\psi_L^t\mathcal{R}\psi_1=\text{STr}(\mathcal{R}\psi_1\psi_L^t)$ 
(the supertrace STr is defined in such a way that extra minus signs are multiplied 
for the fermionic sectors. See (\ref{eqn:STr-1}) and (\ref{eqn:STr-2}), for instance),  
we can express the supersymmetric MPS (sMPS) as  
\begin{equation}
\begin{split}
|\text{sMPS}\rangle_{\text{PBC}}
&=    
\underbrace{%
\prod_{i=1}^{L-1}  
(\psi_i^t\mathcal{R} \psi_{i+1})
}_{\text{Grassmann even}}
\psi^{\alpha}_L ({\cal R}\psi_1)^{\alpha} |0\rangle  \\
&= \text{STr} \left\{
({\cal R}\psi_1) \prod_{i=1}^{L-1}  
(\psi_i^t\mathcal{R} \psi_{i+1})
\psi_L^{t}
\right\}  \\
&=\text{STr}(A_1 A_2 \cdots A_{L}).  
\end{split}
\end{equation}

Since the overlap 
$A_{j}^{\ast}(\bar{\beta}_{j-1},\bar{\beta}_{j})A_{j}(\beta_{j-1},\beta_{j})$ 
is a commuting $c$-number ({\em transfer matrix}), it is straightforward to show, 
by proceeding term by term from the inner most overlap to the outer, 
the following equation: 
\begin{subequations}
\begin{equation}
\VEV{\text{MPS}|\text{MPS}}_{(\alpha,\gamma)} 
= \{
T_{1}\cdots T_{j} \cdots T_{L}
\}_{(\alpha,\alpha;\gamma,\gamma)} \; ,
\end{equation}
where 
\begin{equation}
\begin{split}
& T_{1}(\alpha,\alpha;\bar{\beta}_{1},\beta_{1}) \equiv 
A_{1}^{\ast}(\alpha,\bar{\beta}_{1})A_{1}(\alpha,\beta_{1}) \\
& T_{L}(\bar{\beta}_{L-1},\beta_{L-1};\bar{\gamma},\gamma) \equiv 
A_{L}^{\ast}(\bar{\beta}_{L-1},\bar{\gamma})A_{L}(\beta_{L-1},\gamma) \\
& T_{j}(\bar{\beta}_{j-1},\beta_{j-1};\bar{\beta}_{j},\beta_{j}) \equiv 
A_{j}^{\ast}(\bar{\beta}_{j-1},\bar{\beta}_{j})A_{j}(\beta_{j-1},\beta_{j}) \; .
\end{split}
\end{equation}
\end{subequations}
For the purpose of calculating various correlation functions, 
it is convenient to consider generalized overlaps of the following 
form:
\begin{equation}
\begin{split}
{}_{(\alpha,\beta)}\VEV{\text{MPS}|\text{MPS}}_{(\gamma,\delta)} 
&= \{
T_{1} \cdots T_{j} \cdots T_{L}
\}_{(\alpha,\gamma;\beta,\delta)}\\
&= \{T^{L}\}_{(\alpha,\gamma;\beta,\delta)} \; ,
\end{split}
\label{eqn:overlap-by-MPS}
\end{equation} 
which are not necessarily proportional to $\delta_{\alpha,\gamma}\delta_{\beta,\delta}$ for finite-$L$.  
If the periodic boundary condition is imposed, 
the norm corresponding to the bosonic MPS eq.(\ref{eqn:MPS-PBC-1}) reads
\begin{equation}
\VEV{\text{MPS}|\text{MPS}}_{\text{PBC}} 
= \sum_{\alpha,\beta}\left\{T^{L}\right\}_{(\alpha,\beta;\alpha,\beta)} 
= \text{Tr}\, T^{L} \; . 
\label{eqn:norm-PBC-1}
\end{equation}
In the case of sMPS, the above expression should be replaced with
eq.(\ref{eqn:norm-PBC-2}).  
\subsection{Correlation functions}
\label{sec:cor-func}
Having established the way of evaluating overlaps, it is 
straightforward to extend it to correlation functions. 
For simplicity, we only consider bosonic operators here 
(we will generalize the calculation to fermionic operators as well). 
 
Let us consider first the ordinary two-point correlation function:
\begin{equation}
\VEV{{\cal O}^{A}_{x} {\cal O}^{B}_{y}}_{(\alpha,\gamma)}
=\frac{
\VEV{\text{MPS}|{\cal O}^{A}_{x} {\cal O}^{B}_{y}|\text{MPS}}_{(\alpha,\gamma)}}
{\VEV{\text{MPS}|\text{MPS}}_{(\alpha,\gamma)}}
\; .
\label{eqn:def-core-fn}
\end{equation}
Since the two physical operators ${\cal O}^{A}_{x}$ and 
${\cal O}^{B}_{y}$ are bosonic, the calculation goes in almost 
the same manner as in the case of norms except that here 
we have two new matrices:
\begin{equation}
\begin{split}
& T^{{\cal O}^{A}}_{x}
(\bar{\beta}_{x-1},\beta_{x-1};\bar{\beta}_{x},\beta_{x}) \equiv 
A_{x}^{\ast}(\bar{\beta}_{x-1},\bar{\beta}_{x}){\cal O}^{A}_{x}
A_{x}(\beta_{x-1},\beta_{x}) \\
& T^{{\cal O}^{B}}_{y}
(\bar{\beta}_{y-1},\beta_{y-1};\bar{\beta}_{y},\beta_{y})\equiv 
A_{y}^{\ast}(\bar{\beta}_{y-1},\bar{\beta}_{y}){\cal O}^{B}_{y}
A_{y}(\beta_{y-1},\beta_{y})  
\end{split}
\end{equation}
instead of $T_{x}$ and $T_{y}$.   
Then, by using (\ref{eqn:overlap-by-MPS}), 
the numerator of eq.(\ref{eqn:def-core-fn}) may be expressed as:
\begin{subequations}
\begin{equation}
\left\{T^{x-1} T^{{\cal O}^{A}} T^{y-x-1}
T^{{\cal O}^{B}}T^{L-y} \right\}_{(\alpha,\alpha;\gamma,\gamma)}  \; .
\end{equation}
Therefore, the matrix-product expression of the correlation function 
is given by:
\begin{equation}
\VEV{{\cal O}^{A}_{x} {\cal O}^{B}_{y}}_{(\alpha,\gamma)}
= \frac{\left\{T^{x-1} T^{{\cal O}^{A}} T^{y-x-1}
T^{{\cal O}^{B}}T^{L-y} \right\}_{(\alpha,\alpha;\gamma,\gamma)}}
{\left\{\widetilde{T}T^{L-1}\right\}_{(\alpha,\alpha;\gamma,\gamma)}}  \; .
\label{eqn:2ptcor-by-mpg}
\end{equation}
\end{subequations}

In physical applications, we will encounter the following 
string-like correlation functions:
\begin{subequations}
\begin{multline}
\Biggl\langle
{\cal O}^{A}_{x}\left(
\prod_{j=x+1}^{y-1}{\cal O}^{C_{j}}_{j} 
\right) {\cal O}^{B}_{y}
\Biggr\rangle_{(\alpha,\gamma)}  \\
=\frac{
\VEV{\text{MPS}|{\cal O}^{A}_{x}\left(
\prod_{j=x+1}^{y-1}{\cal O}^{C_{j}}_{j} 
\right)
{\cal O}^{B}_{y}|\text{MPS}}_{(\alpha,\gamma)}}
{\VEV{\text{MPS}|\text{MPS}}_{(\alpha,\gamma)}}
\; .
\label{eqn:def-stringcore-fn}
\end{multline}
It is straightforward to obtain:
\begin{equation}
\begin{split}
& \Biggl\langle
{\cal O}^{A}_{x}\left(
\prod_{j=x+1}^{y-1}{\cal O}^{C_{j}}_{j} 
\right) {\cal O}^{B}_{y}
\Biggr\rangle_{(\alpha,\gamma)}  \\
&= \frac{\left\{T^{x-1} T^{{\cal O}^{A}} 
(T^{{\cal O}^{C}})^{y-x-1}
T^{{\cal O}^{B}}T^{L-y} \right\}_{(\alpha,\alpha;\gamma,\gamma)}}
{\left\{T^{L}\right\}_{(\alpha,\alpha;\gamma,\gamma)}}  \; .
\end{split}
\label{eqn:stringcor-by-mpg}
\end{equation}
\end{subequations}
In order to calculate the so-called string correlation function 
(see section \ref{sec:hidden-order}), 
we should take:
\begin{equation}
{\cal O}^{A}=S^{x,z}\be^{i\pi S^{x,z}} 
\; , \;\; {\cal O}^{B}=S^{x,z} \; , \;\; 
{\cal O}^{C} = \be^{i\pi S^{x,z}} \; .
\end{equation}

When we consider the expectation values involving fermionic operators, 
the calculation is slightly more complicated as we have seen in section \ref{sec:SC-correlation-I}. 

\section{MPS for Type I sVBS with general $M$}
\label{sec:general-M-MPS}
The construction of MPS for the $M=1$ type-I VBS state in section \ref{sec:MPS-construction} 
can be readily generalized to general $M$.  
To this end, it is helpful to note that 
\begin{multline}
(a^{\dagger}_{j}b^{\dagger}_{j+1}-b^{\dagger}_{j}a^{\dagger}_{j+1} 
-r f^{\dagger}_{j}f^{\dagger}_{j+1})^{M} \\
= (a^{\dagger}_{j}b^{\dagger}_{j+1}-b^{\dagger}_{j}a^{\dagger}_{j+1} )^{M}
-Mr (a^{\dagger}_{j}b^{\dagger}_{j+1}-b^{\dagger}_{j}a^{\dagger}_{j+1} )^{M-1}
f^{\dagger}_{j}f^{\dagger}_{j+1} \; .
\end{multline}
Since each term on RHS can be written in terms of matrices with dimensions $M+1$ or $M$, 
the valence-bond operator on LHS may be expressed by a block-diagonal 
$(2M+1)$-dimensional matrix (a generalization of ${\cal R}$ in (\ref{eqn:metric})).  
Following the same steps as in eqs.(\ref{eqn:SUSY-MPS}) and (\ref{eqn:SUSY-g-matrix}), 
we obtain the followings:
\begin{equation}
A_{\alpha\beta}(j)={\cal F}_{\alpha}^{\text{L}}(a^{\dagger}_{j},b^{\dagger}_{j},f^{\dagger}_{j})
{\cal F}_{\beta}^{\text{R}}(a^{\dagger}_{j},b^{\dagger}_{j},f^{\dagger}_{j})
|\text{V}\rangle_{j}  \; ,
\end{equation}
where the polynomials ${\cal F}^{\text{R/L}}_{\alpha}$ $(\alpha=1,\ldots,2M{+}1)$ are given by
\begin{subequations}
\begin{equation}
\begin{split}
& {\cal F}_{\alpha}^{\text{L}}(x,y,z)\\ 
& \equiv 
\begin{cases}
(-1)^{\alpha-1}\sqrt{ {}_M \text{C}_{\alpha-1}} 
x^{\alpha-1}y^{M-(\alpha-1)}   \qquad \\
 \phantom{\sqrt{Mr}\sqrt{{}_{M-1}\text{C}_{a-(M+2)}}}   
 \text{for } \alpha=1,\dots, M+1  & \\
(-1)^{M-(\alpha-1)}\sqrt{Mr}\sqrt{{}_{M-1}\text{C}_{\alpha-(M+2)}} 
x^{\alpha-(M+2)}y^{(2M+1)-\alpha} z  & \\
 \phantom{\sqrt{Mr}\sqrt{{}_{M-1}\text{C}_{\alpha-(M+2)}}} 
 \text{for } \alpha=M+2 ,\dots, 2M+1 & 
\end{cases}
\end{split}
\end{equation}
and 
\begin{equation}
\begin{split}
& {\cal F}_{\beta}^{\text{R}}(x,y,z)\\ 
& \equiv 
\begin{cases}
\sqrt{ {}_M \text{C}_{\beta-1}} 
x^{M-\beta+1}y^{\beta-1}   \qquad 
 \text{for } \beta=1,\dots, M+1  & \\
\sqrt{Mr}\sqrt{{}_{M-1}\text{C}_{\beta-(M+2)}} 
x^{(2M+1)-\beta}y^{\beta-(M+2)} z  & \\
 \phantom{\sqrt{Mr}\sqrt{{}_{M-1}\text{C}_{\alpha-(M+2)}}}  
 \text{for } \beta=M+2 ,\dots, 2M+1 & 
\end{cases}
\end{split}
\end{equation}
with the standard binomial coefficient ${}_{m}\text{C}_{n}\equiv\begin{pmatrix}m \\ n\end{pmatrix}$.
\end{subequations}

\bibliographystyle{apsrev4-1}

\begin{thebibliography}{69}%
\makeatletter
\providecommand \@ifxundefined [1]{%
 \@ifx{#1\undefined}
}%
\providecommand \@ifnum [1]{%
 \ifnum #1\expandafter \@firstoftwo
 \else \expandafter \@secondoftwo
 \fi
}%
\providecommand \@ifx [1]{%
 \ifx #1\expandafter \@firstoftwo
 \else \expandafter \@secondoftwo
 \fi
}%
\providecommand \natexlab [1]{#1}%
\providecommand \enquote  [1]{``#1''}%
\providecommand \bibnamefont  [1]{#1}%
\providecommand \bibfnamefont [1]{#1}%
\providecommand \citenamefont [1]{#1}%
\providecommand \href@noop [0]{\@secondoftwo}%
\providecommand \href [0]{\begingroup \@sanitize@url \@href}%
\providecommand \@href[1]{\@@startlink{#1}\@@href}%
\providecommand \@@href[1]{\endgroup#1\@@endlink}%
\providecommand \@sanitize@url [0]{\catcode `\\12\catcode `\$12\catcode
  `\&12\catcode `\#12\catcode `\^12\catcode `\_12\catcode `\%12\relax}%
\providecommand \@@startlink[1]{}%
\providecommand \@@endlink[0]{}%
\providecommand \url  [0]{\begingroup\@sanitize@url \@url }%
\providecommand \@url [1]{\endgroup\@href {#1}{\urlprefix }}%
\providecommand \urlprefix  [0]{URL }%
\providecommand \Eprint [0]{\href }%
\providecommand \doibase [0]{http://dx.doi.org/}%
\providecommand \selectlanguage [0]{\@gobble}%
\providecommand \bibinfo  [0]{\@secondoftwo}%
\providecommand \bibfield  [0]{\@secondoftwo}%
\providecommand \translation [1]{[#1]}%
\providecommand \BibitemOpen [0]{}%
\providecommand \bibitemStop [0]{}%
\providecommand \bibitemNoStop [0]{.\EOS\space}%
\providecommand \EOS [0]{\spacefactor3000\relax}%
\providecommand \BibitemShut  [1]{\csname bibitem#1\endcsname}%
\let\auto@bib@innerbib\@empty
\bibitem [{\citenamefont {Affleck}\ \emph {et~al.}(1987)\citenamefont
  {Affleck}, \citenamefont {Kennedy}, \citenamefont {Lieb},\ and\ \citenamefont
  {Tasaki}}]{affleck1987rrv}%
  \BibitemOpen
  \bibfield  {author} {\bibinfo {author} {\bibfnamefont {I.}~\bibnamefont
  {Affleck}}, \bibinfo {author} {\bibfnamefont {T.}~\bibnamefont {Kennedy}},
  \bibinfo {author} {\bibfnamefont {E.}~\bibnamefont {Lieb}}, \ and\ \bibinfo
  {author} {\bibfnamefont {H.}~\bibnamefont {Tasaki}},\ }\href@noop {}
  {\bibfield  {journal} {\bibinfo  {journal} {Phys. Rev. Lett.}\ }\textbf
  {\bibinfo {volume} {59}},\ \bibinfo {pages} {799} (\bibinfo {year}
  {1987})}\BibitemShut {NoStop}%
\bibitem [{\citenamefont {Affleck}\ \emph {et~al.}(1988)\citenamefont
  {Affleck}, \citenamefont {Kennedy}, \citenamefont {Lieb},\ and\ \citenamefont
  {Tasaki}}]{affleck1988vbg}%
  \BibitemOpen
  \bibfield  {author} {\bibinfo {author} {\bibfnamefont {I.}~\bibnamefont
  {Affleck}}, \bibinfo {author} {\bibfnamefont {T.}~\bibnamefont {Kennedy}},
  \bibinfo {author} {\bibfnamefont {E.}~\bibnamefont {Lieb}}, \ and\ \bibinfo
  {author} {\bibfnamefont {H.}~\bibnamefont {Tasaki}},\ }\href@noop {}
  {\bibfield  {journal} {\bibinfo  {journal} {Commun. Math. Phys.}\ }\textbf
  {\bibinfo {volume} {115}},\ \bibinfo {pages} {477} (\bibinfo {year}
  {1988})}\BibitemShut {NoStop}%
\bibitem [{\citenamefont {Haldane}(1983{\natexlab{a}})}]{Haldane-83a}%
  \BibitemOpen
  \bibfield  {author} {\bibinfo {author} {\bibfnamefont {F.}~\bibnamefont
  {Haldane}},\ }\href@noop {} {\bibfield  {journal} {\bibinfo  {journal}
  {Phys.Lett.}\ }\textbf {\bibinfo {volume} {{\bf 93A}}},\ \bibinfo {pages}
  {464} (\bibinfo {year} {1983}{\natexlab{a}})}\BibitemShut {NoStop}%
\bibitem [{\citenamefont {Haldane}(1983{\natexlab{b}})}]{Haldane-83b}%
  \BibitemOpen
  \bibfield  {author} {\bibinfo {author} {\bibfnamefont {F.}~\bibnamefont
  {Haldane}},\ }\href@noop {} {\bibfield  {journal} {\bibinfo  {journal}
  {Phys.Rev.Lett.}\ }\textbf {\bibinfo {volume} {{\bf 50}}},\ \bibinfo {pages}
  {1153} (\bibinfo {year} {1983}{\natexlab{b}})}\BibitemShut {NoStop}%
\bibitem [{\citenamefont {Kennedy}\ and\ \citenamefont
  {Tasaki}(1992{\natexlab{a}})}]{Kennedy-T-92a}%
  \BibitemOpen
  \bibfield  {author} {\bibinfo {author} {\bibfnamefont {T.}~\bibnamefont
  {Kennedy}}\ and\ \bibinfo {author} {\bibfnamefont {H.}~\bibnamefont
  {Tasaki}},\ }\href@noop {} {\bibfield  {journal} {\bibinfo  {journal}
  {Phys.Rev.}\ }\textbf {\bibinfo {volume} {{\bf B45}}},\ \bibinfo {pages}
  {304} (\bibinfo {year} {1992}{\natexlab{a}})}\BibitemShut {NoStop}%
\bibitem [{\citenamefont {Kennedy}\ and\ \citenamefont
  {Tasaki}(1992{\natexlab{b}})}]{Kennedy-T-92b}%
  \BibitemOpen
  \bibfield  {author} {\bibinfo {author} {\bibfnamefont {T.}~\bibnamefont
  {Kennedy}}\ and\ \bibinfo {author} {\bibfnamefont {H.}~\bibnamefont
  {Tasaki}},\ }\href@noop {} {\bibfield  {journal} {\bibinfo  {journal}
  {Commun.Math.Phys.}\ }\textbf {\bibinfo {volume} {{\bf 147}}},\ \bibinfo
  {pages} {431} (\bibinfo {year} {1992}{\natexlab{b}})}\BibitemShut {NoStop}%
\bibitem [{\citenamefont {Arovas}\ \emph {et~al.}(1988)\citenamefont {Arovas},
  \citenamefont {Auerbach},\ and\ \citenamefont {Haldane}}]{Arovas-A-H-88}%
  \BibitemOpen
  \bibfield  {author} {\bibinfo {author} {\bibfnamefont {D.~P.}\ \bibnamefont
  {Arovas}}, \bibinfo {author} {\bibfnamefont {A.}~\bibnamefont {Auerbach}}, \
  and\ \bibinfo {author} {\bibfnamefont {F.}~\bibnamefont {Haldane}},\
  }\href@noop {} {\bibfield  {journal} {\bibinfo  {journal} {Phys. Rev. Lett.}\
  }\textbf {\bibinfo {volume} {60}},\ \bibinfo {pages} {531} (\bibinfo {year}
  {1988})}\BibitemShut {NoStop}%
\bibitem [{\citenamefont {Girvin}\ and\ \citenamefont
  {Arovas}(1989)}]{Girvin-A-89}%
  \BibitemOpen
  \bibfield  {author} {\bibinfo {author} {\bibfnamefont {S.}~\bibnamefont
  {Girvin}}\ and\ \bibinfo {author} {\bibfnamefont {D.}~\bibnamefont
  {Arovas}},\ }\href@noop {} {\bibfield  {journal} {\bibinfo  {journal}
  {Phys.Scr.}\ }\textbf {\bibinfo {volume} {{\bf T 27}}},\ \bibinfo {pages}
  {156} (\bibinfo {year} {1989})}\BibitemShut {NoStop}%
\bibitem [{\citenamefont {Hatsugai}(1992)}]{hatsugaiJPSJ1992}%
  \BibitemOpen
  \bibfield  {author} {\bibinfo {author} {\bibfnamefont {Y.}~\bibnamefont
  {Hatsugai}},\ }\href@noop {} {\bibfield  {journal} {\bibinfo  {journal}
  {Journal of the Physical Society of Japan}\ }\textbf {\bibinfo {volume}
  {61}},\ \bibinfo {pages} {3856} (\bibinfo {year} {1992})}\BibitemShut
  {NoStop}%
\bibitem [{\citenamefont {Oshikawa}(1992)}]{Oshikawa-92}%
  \BibitemOpen
  \bibfield  {author} {\bibinfo {author} {\bibfnamefont {M.}~\bibnamefont
  {Oshikawa}},\ }\href@noop {} {\bibfield  {journal} {\bibinfo  {journal}
  {J.Phys. Condens.Matter}\ }\textbf {\bibinfo {volume} {{\bf 4}}},\ \bibinfo
  {pages} {7469} (\bibinfo {year} {1992})}\BibitemShut {NoStop}%
\bibitem [{\citenamefont {Totsuka}\ and\ \citenamefont
  {Suzuki}(1994)}]{Totsuka-S-94}%
  \BibitemOpen
  \bibfield  {author} {\bibinfo {author} {\bibfnamefont {K.}~\bibnamefont
  {Totsuka}}\ and\ \bibinfo {author} {\bibfnamefont {M.}~\bibnamefont
  {Suzuki}},\ }\href {\doibase 10.1088/0305-4470/27/19/017} {\bibfield
  {journal} {\bibinfo  {journal} {J. Phys. A: Math. Gen.}\ }\textbf {\bibinfo
  {volume} {27}},\ \bibinfo {pages} {6443} (\bibinfo {year}
  {1994})}\BibitemShut {NoStop}%
\bibitem [{\citenamefont {Totsuka}\ and\ \citenamefont
  {Suzuki}(1995)}]{Totsuka-S-mpg-95}%
  \BibitemOpen
  \bibfield  {author} {\bibinfo {author} {\bibfnamefont {K.}~\bibnamefont
  {Totsuka}}\ and\ \bibinfo {author} {\bibfnamefont {M.}~\bibnamefont
  {Suzuki}},\ }\href@noop {} {\bibfield  {journal} {\bibinfo  {journal}
  {J.Phys.:condens.matter}\ }\textbf {\bibinfo {volume} {{\bf 7}}},\ \bibinfo
  {pages} {1639} (\bibinfo {year} {1995})}\BibitemShut {NoStop}%
\bibitem [{\citenamefont {Qi}\ and\ \citenamefont {Zhang}(2010)}]{qi-2010-63}%
  \BibitemOpen
  \bibfield  {author} {\bibinfo {author} {\bibfnamefont {X.-L.}\ \bibnamefont
  {Qi}}\ and\ \bibinfo {author} {\bibfnamefont {S.-C.}\ \bibnamefont {Zhang}},\
  }\href@noop {} {\bibfield  {journal} {\bibinfo  {journal} {Physics Today}\
  }\textbf {\bibinfo {volume} {63}},\ \bibinfo {pages} {33} (\bibinfo {year}
  {2010})}\BibitemShut {NoStop}%
\bibitem [{\citenamefont {Wolf}\ \emph {et~al.}(2006)\citenamefont {Wolf},
  \citenamefont {Ortiz}, \citenamefont {Verstraete},\ and\ \citenamefont
  {Cirac}}]{wolf-2006-97}%
  \BibitemOpen
  \bibfield  {author} {\bibinfo {author} {\bibfnamefont {M.~M.}\ \bibnamefont
  {Wolf}}, \bibinfo {author} {\bibfnamefont {G.}~\bibnamefont {Ortiz}},
  \bibinfo {author} {\bibfnamefont {F.}~\bibnamefont {Verstraete}}, \ and\
  \bibinfo {author} {\bibfnamefont {J.~I.}\ \bibnamefont {Cirac}},\ }\href@noop
  {} {\bibfield  {journal} {\bibinfo  {journal} {Physical Review Letters}\
  }\textbf {\bibinfo {volume} {97}},\ \bibinfo {pages} {110403} (\bibinfo
  {year} {2006})}\BibitemShut {NoStop}%
\bibitem [{\citenamefont {Perez-Garcia}\ \emph {et~al.}(2007)\citenamefont
  {Perez-Garcia}, \citenamefont {Verstraete}, \citenamefont {Wolf},\ and\
  \citenamefont {Cirac}}]{perezgarcia-2007-7}%
  \BibitemOpen
  \bibfield  {author} {\bibinfo {author} {\bibfnamefont {D.}~\bibnamefont
  {Perez-Garcia}}, \bibinfo {author} {\bibfnamefont {F.}~\bibnamefont
  {Verstraete}}, \bibinfo {author} {\bibfnamefont {M.~M.}\ \bibnamefont
  {Wolf}}, \ and\ \bibinfo {author} {\bibfnamefont {J.~I.}\ \bibnamefont
  {Cirac}},\ }\href@noop {} {\bibfield  {journal} {\bibinfo  {journal} {Quantum
  Inf..Comput.}\ }\textbf {\bibinfo {volume} {7}},\ \bibinfo {pages} {401}
  (\bibinfo {year} {2007})}\BibitemShut {NoStop}%
\bibitem [{\citenamefont {Pollmann}\ \emph {et~al.}()\citenamefont {Pollmann},
  \citenamefont {Berg}, \citenamefont {Turner},\ and\ \citenamefont
  {Oshikawa}}]{pollmann-2009}%
  \BibitemOpen
  \bibfield  {author} {\bibinfo {author} {\bibfnamefont {F.}~\bibnamefont
  {Pollmann}}, \bibinfo {author} {\bibfnamefont {E.}~\bibnamefont {Berg}},
  \bibinfo {author} {\bibfnamefont {A.~M.}\ \bibnamefont {Turner}}, \ and\
  \bibinfo {author} {\bibfnamefont {M.}~\bibnamefont {Oshikawa}},\ }\href@noop
  {} {\bibinfo  {journal} {arXiv:0909.4059}\ }\BibitemShut {NoStop}%
\bibitem [{\citenamefont {Pollmann}\ \emph {et~al.}(2010)\citenamefont
  {Pollmann}, \citenamefont {Turner}, \citenamefont {Berg},\ and\ \citenamefont
  {Oshikawa}}]{Pollmann-T-B-O-10}%
  \BibitemOpen
\bibfield  {journal} {  }\bibfield  {author} {\bibinfo {author} {\bibfnamefont
  {F.}~\bibnamefont {Pollmann}}, \bibinfo {author} {\bibfnamefont {A.~M.}\
  \bibnamefont {Turner}}, \bibinfo {author} {\bibfnamefont {E.}~\bibnamefont
  {Berg}}, \ and\ \bibinfo {author} {\bibfnamefont {M.}~\bibnamefont
  {Oshikawa}},\ }\href {http://link.aps.org/doi/10.1103/PhysRevB.81.064439}
  {\bibfield  {journal} {\bibinfo  {journal} {Phys. Rev. B}\ }\textbf {\bibinfo
  {volume} {81}},\ \bibinfo {pages} {064439} (\bibinfo {year}
  {2010})}\BibitemShut {NoStop}%
\bibitem [{\citenamefont {Katsura}\ \emph {et~al.}(2007)\citenamefont
  {Katsura}, \citenamefont {Hirano},\ and\ \citenamefont
  {Hatsugai}}]{katsura-2007-76}%
  \BibitemOpen
  \bibfield  {author} {\bibinfo {author} {\bibfnamefont {H.}~\bibnamefont
  {Katsura}}, \bibinfo {author} {\bibfnamefont {T.}~\bibnamefont {Hirano}}, \
  and\ \bibinfo {author} {\bibfnamefont {Y.}~\bibnamefont {Hatsugai}},\
  }\href@noop {} {\bibfield  {journal} {\bibinfo  {journal} {Physical Review
  B}\ }\textbf {\bibinfo {volume} {76}},\ \bibinfo {pages} {012401} (\bibinfo
  {year} {2007})}\BibitemShut {NoStop}%
\bibitem [{\citenamefont {Katsura}\ \emph {et~al.}(2008)\citenamefont
  {Katsura}, \citenamefont {Hirano},\ and\ \citenamefont
  {Korepin}}]{katsura-2008-41}%
  \BibitemOpen
  \bibfield  {author} {\bibinfo {author} {\bibfnamefont {H.}~\bibnamefont
  {Katsura}}, \bibinfo {author} {\bibfnamefont {T.}~\bibnamefont {Hirano}}, \
  and\ \bibinfo {author} {\bibfnamefont {V.~E.}\ \bibnamefont {Korepin}},\
  }\href@noop {} {\bibfield  {journal} {\bibinfo  {journal} {J. Phys. A:Math
  and Theor.}\ }\textbf {\bibinfo {volume} {41}},\ \bibinfo {pages} {135304}
  (\bibinfo {year} {2008})}\BibitemShut {NoStop}%
\bibitem [{\citenamefont {Xu}\ \emph {et~al.}(2008)\citenamefont {Xu},
  \citenamefont {Katsura}, \citenamefont {Hirano},\ and\ \citenamefont
  {Korepin}}]{xu-2008-133}%
  \BibitemOpen
  \bibfield  {author} {\bibinfo {author} {\bibfnamefont {Y.}~\bibnamefont
  {Xu}}, \bibinfo {author} {\bibfnamefont {H.}~\bibnamefont {Katsura}},
  \bibinfo {author} {\bibfnamefont {T.}~\bibnamefont {Hirano}}, \ and\ \bibinfo
  {author} {\bibfnamefont {V.~E.}\ \bibnamefont {Korepin}},\ }\href@noop {}
  {\bibfield  {journal} {\bibinfo  {journal} {J. Stat. Phys.}\ }\textbf
  {\bibinfo {volume} {133}},\ \bibinfo {pages} {347} (\bibinfo {year}
  {2008})}\BibitemShut {NoStop}%
\bibitem [{\citenamefont {Affleck}\ \emph {et~al.}(1991)\citenamefont
  {Affleck}, \citenamefont {Arovas}, \citenamefont {Marston},\ and\
  \citenamefont {Rabson}}]{affleck1991qae}%
  \BibitemOpen
  \bibfield  {author} {\bibinfo {author} {\bibfnamefont {I.}~\bibnamefont
  {Affleck}}, \bibinfo {author} {\bibfnamefont {D.}~\bibnamefont {Arovas}},
  \bibinfo {author} {\bibfnamefont {J.}~\bibnamefont {Marston}}, \ and\
  \bibinfo {author} {\bibfnamefont {D.}~\bibnamefont {Rabson}},\ }\href@noop {}
  {\bibfield  {journal} {\bibinfo  {journal} {Nucl. Phys. B}\ }\textbf
  {\bibinfo {volume} {366}},\ \bibinfo {pages} {467} (\bibinfo {year}
  {1991})}\BibitemShut {NoStop}%
\bibitem [{\citenamefont {Arovas}(2008)}]{arovas2008sss}%
  \BibitemOpen
  \bibfield  {author} {\bibinfo {author} {\bibfnamefont {D.}~\bibnamefont
  {Arovas}},\ }\href@noop {} {\bibfield  {journal} {\bibinfo  {journal} {Phys.
  Rev. B}\ }\textbf {\bibinfo {volume} {77}},\ \bibinfo {pages} {104404}
  (\bibinfo {year} {2008})}\BibitemShut {NoStop}%
\bibitem [{\citenamefont {Greiter}\ \emph {et~al.}(2007)\citenamefont
  {Greiter}, \citenamefont {Rachel},\ and\ \citenamefont
  {Schuricht}}]{greiter2007ers}%
  \BibitemOpen
  \bibfield  {author} {\bibinfo {author} {\bibfnamefont {M.}~\bibnamefont
  {Greiter}}, \bibinfo {author} {\bibfnamefont {S.}~\bibnamefont {Rachel}}, \
  and\ \bibinfo {author} {\bibfnamefont {D.}~\bibnamefont {Schuricht}},\
  }\href@noop {} {\bibfield  {journal} {\bibinfo  {journal} {Phys. Rev. B}\
  }\textbf {\bibinfo {volume} {75}},\ \bibinfo {pages} {60401} (\bibinfo {year}
  {2007})}\BibitemShut {NoStop}%
\bibitem [{\citenamefont {Greiter}\ and\ \citenamefont
  {Rachel}(2007)}]{greiter2007vbs}%
  \BibitemOpen
  \bibfield  {author} {\bibinfo {author} {\bibfnamefont {M.}~\bibnamefont
  {Greiter}}\ and\ \bibinfo {author} {\bibfnamefont {S.}~\bibnamefont
  {Rachel}},\ }\href@noop {} {\bibfield  {journal} {\bibinfo  {journal} {Phys.
  Rev. B}\ }\textbf {\bibinfo {volume} {75}},\ \bibinfo {pages} {184441}
  (\bibinfo {year} {2007})}\BibitemShut {NoStop}%
\bibitem [{\citenamefont {Schuricht}\ and\ \citenamefont
  {Rachel}(2008)}]{schuricht2008vbs}%
  \BibitemOpen
  \bibfield  {author} {\bibinfo {author} {\bibfnamefont {D.}~\bibnamefont
  {Schuricht}}\ and\ \bibinfo {author} {\bibfnamefont {S.}~\bibnamefont
  {Rachel}},\ }\href@noop {} {\bibfield  {journal} {\bibinfo  {journal} {Phys.
  Rev. B}\ }\textbf {\bibinfo {volume} {78}},\ \bibinfo {pages} {014430}
  (\bibinfo {year} {2008})}\BibitemShut {NoStop}%
\bibitem [{\citenamefont {Tu}\ \emph {et~al.}(2008{\natexlab{a}})\citenamefont
  {Tu}, \citenamefont {Zhang},\ and\ \citenamefont {Xiang}}]{Tu08041685}%
  \BibitemOpen
  \bibfield  {author} {\bibinfo {author} {\bibfnamefont {H.}~\bibnamefont
  {Tu}}, \bibinfo {author} {\bibfnamefont {G.}~\bibnamefont {Zhang}}, \ and\
  \bibinfo {author} {\bibfnamefont {T.}~\bibnamefont {Xiang}},\ }\href@noop {}
  {\bibfield  {journal} {\bibinfo  {journal} {J. Phys. A}\ }\textbf {\bibinfo
  {volume} {41}},\ \bibinfo {pages} {415201} (\bibinfo {year}
  {2008}{\natexlab{a}})}\BibitemShut {NoStop}%
\bibitem [{\citenamefont {Tu}\ \emph {et~al.}(2008{\natexlab{b}})\citenamefont
  {Tu}, \citenamefont {Zhang},\ and\ \citenamefont {Xiang}}]{Tu08061839}%
  \BibitemOpen
  \bibfield  {author} {\bibinfo {author} {\bibfnamefont {H.}~\bibnamefont
  {Tu}}, \bibinfo {author} {\bibfnamefont {G.}~\bibnamefont {Zhang}}, \ and\
  \bibinfo {author} {\bibfnamefont {T.}~\bibnamefont {Xiang}},\ }\href@noop {}
  {\bibfield  {journal} {\bibinfo  {journal} {Phys. Rev. B}\ }\textbf {\bibinfo
  {volume} {78}},\ \bibinfo {pages} {094404} (\bibinfo {year}
  {2008}{\natexlab{b}})}\BibitemShut {NoStop}%
\bibitem [{\citenamefont {Rachel}(2009)}]{rachel-2009-86}%
  \BibitemOpen
  \bibfield  {author} {\bibinfo {author} {\bibfnamefont {S.}~\bibnamefont
  {Rachel}},\ }\href@noop {} {\bibfield  {journal} {\bibinfo  {journal}
  {Europhys. Lett.}\ }\textbf {\bibinfo {volume} {86}},\ \bibinfo {pages}
  {37005} (\bibinfo {year} {2009})}\BibitemShut {NoStop}%
\bibitem [{\citenamefont {Zang}\ \emph {et~al.}(2010)\citenamefont {Zang},
  \citenamefont {Jiang}, \citenamefont {Weng},\ and\ \citenamefont
  {Zhang}}]{zang-2010-81}%
  \BibitemOpen
  \bibfield  {author} {\bibinfo {author} {\bibfnamefont {J.}~\bibnamefont
  {Zang}}, \bibinfo {author} {\bibfnamefont {H.-C.}\ \bibnamefont {Jiang}},
  \bibinfo {author} {\bibfnamefont {Z.-Y.}\ \bibnamefont {Weng}}, \ and\
  \bibinfo {author} {\bibfnamefont {S.-C.}\ \bibnamefont {Zhang}},\ }\href@noop
  {} {\bibfield  {journal} {\bibinfo  {journal} {Physical Review B}\ }\textbf
  {\bibinfo {volume} {81}},\ \bibinfo {pages} {224430} (\bibinfo {year}
  {2010})}\BibitemShut {NoStop}%
\bibitem [{\citenamefont {Zheng}\ \emph {et~al.}()\citenamefont {Zheng},
  \citenamefont {Zhang}, \citenamefont {Xiang},\ and\ \citenamefont
  {Lee}}]{zheng-2010}%
  \BibitemOpen
  \bibfield  {author} {\bibinfo {author} {\bibfnamefont {D.}~\bibnamefont
  {Zheng}}, \bibinfo {author} {\bibfnamefont {G.-M.}\ \bibnamefont {Zhang}},
  \bibinfo {author} {\bibfnamefont {T.}~\bibnamefont {Xiang}}, \ and\ \bibinfo
  {author} {\bibfnamefont {D.-H.}\ \bibnamefont {Lee}},\ }\href@noop {}
  {\bibinfo  {journal} {arXiv:1002.0171}\ }\BibitemShut {NoStop}%
\bibitem [{\citenamefont {Arovas}\ \emph {et~al.}(2009)\citenamefont {Arovas},
  \citenamefont {Hasebe}, \citenamefont {Qi},\ and\ \citenamefont
  {Zhang}}]{Arovas-H-Q-Z-09}%
  \BibitemOpen
\bibfield  {journal} {  }\bibfield  {author} {\bibinfo {author} {\bibfnamefont
  {D.~P.}\ \bibnamefont {Arovas}}, \bibinfo {author} {\bibfnamefont
  {K.}~\bibnamefont {Hasebe}}, \bibinfo {author} {\bibfnamefont {X.-L.}\
  \bibnamefont {Qi}}, \ and\ \bibinfo {author} {\bibfnamefont {S.-C.}\
  \bibnamefont {Zhang}},\ }\href
  {http://link.aps.org/doi/10.1103/PhysRevB.79.224404} {\bibfield  {journal}
  {\bibinfo  {journal} {Phys. Rev. B}\ }\textbf {\bibinfo {volume} {79}},\
  \bibinfo {pages} {224404} (\bibinfo {year} {2009})}\BibitemShut {NoStop}%
\bibitem [{Note1()}]{Note1}%
  \BibitemOpen
  \bibinfo {note} {SVBS states are the spin-chain counterpart of the
  supersymmetric quantum Hall effect\cite {hasebe2005PRL}.}\BibitemShut {Stop}%
\bibitem [{\citenamefont {Anderson}(1987)}]{anderson1987rvb}%
  \BibitemOpen
  \bibfield  {author} {\bibinfo {author} {\bibfnamefont {P.}~\bibnamefont
  {Anderson}},\ }\href@noop {} {\bibfield  {journal} {\bibinfo  {journal}
  {Science}\ }\textbf {\bibinfo {volume} {235}},\ \bibinfo {pages} {1196}
  (\bibinfo {year} {1987})}\BibitemShut {NoStop}%
\bibitem [{\citenamefont {Zhang}(1997)}]{Zhang1997Science}%
  \BibitemOpen
  \bibfield  {author} {\bibinfo {author} {\bibfnamefont {S.}~\bibnamefont
  {Zhang}},\ }\href@noop {} {\bibfield  {journal} {\bibinfo  {journal}
  {Science}\ }\textbf {\bibinfo {volume} {275}},\ \bibinfo {pages} {1089}
  (\bibinfo {year} {1997})}\BibitemShut {NoStop}%
\bibitem [{\citenamefont {Zhang}\ and\ \citenamefont
  {Arovas}(1989)}]{Zhang-A-89}%
  \BibitemOpen
  \bibfield  {author} {\bibinfo {author} {\bibfnamefont {S.}~\bibnamefont
  {Zhang}}\ and\ \bibinfo {author} {\bibfnamefont {D.}~\bibnamefont {Arovas}},\
  }\href@noop {} {\bibfield  {journal} {\bibinfo  {journal} {Phys. Rev. B}\
  }\textbf {\bibinfo {volume} {40}},\ \bibinfo {pages} {2708} (\bibinfo {year}
  {1989})}\BibitemShut {NoStop}%
\bibitem [{\citenamefont {Penc}\ and\ \citenamefont {Shiba}(1995)}]{Penc-S-95}%
  \BibitemOpen
  \bibfield  {author} {\bibinfo {author} {\bibfnamefont {K.}~\bibnamefont
  {Penc}}\ and\ \bibinfo {author} {\bibfnamefont {H.}~\bibnamefont {Shiba}},\
  }\href {http://link.aps.org/doi/10.1103/PhysRevB.52.R715} {\bibfield
  {journal} {\bibinfo  {journal} {Phys. Rev. B}\ }\textbf {\bibinfo {volume}
  {52}},\ \bibinfo {pages} {R715} (\bibinfo {year} {1995})}\BibitemShut
  {NoStop}%
\bibitem [{\citenamefont {Xu}\ \emph {et~al.}(2000)\citenamefont {Xu},
  \citenamefont {Aeppli}, \citenamefont {Bisher}, \citenamefont {Broholm},
  \citenamefont {DiTusa}, \citenamefont {Frost}, \citenamefont {Ito},
  \citenamefont {Oka}, \citenamefont {Paul}, \citenamefont {Takagi} \emph
  {et~al.}}]{xu2000hqs}%
  \BibitemOpen
  \bibfield  {author} {\bibinfo {author} {\bibfnamefont {G.}~\bibnamefont
  {Xu}}, \bibinfo {author} {\bibfnamefont {G.}~\bibnamefont {Aeppli}}, \bibinfo
  {author} {\bibfnamefont {M.}~\bibnamefont {Bisher}}, \bibinfo {author}
  {\bibfnamefont {C.}~\bibnamefont {Broholm}}, \bibinfo {author} {\bibfnamefont
  {J.}~\bibnamefont {DiTusa}}, \bibinfo {author} {\bibfnamefont
  {C.}~\bibnamefont {Frost}}, \bibinfo {author} {\bibfnamefont
  {T.}~\bibnamefont {Ito}}, \bibinfo {author} {\bibfnamefont {K.}~\bibnamefont
  {Oka}}, \bibinfo {author} {\bibfnamefont {R.}~\bibnamefont {Paul}}, \bibinfo
  {author} {\bibfnamefont {H.}~\bibnamefont {Takagi}},  \emph {et~al.},\
  }\href@noop {} {\bibfield  {journal} {\bibinfo  {journal} {Science}\ }\textbf
  {\bibinfo {volume} {289}},\ \bibinfo {pages} {419} (\bibinfo {year}
  {2000})}\BibitemShut {NoStop}%
\bibitem [{\citenamefont {Fannes}\ \emph {et~al.}(1992)\citenamefont {Fannes},
  \citenamefont {Nachtergaele},\ and\ \citenamefont {Werner}}]{Fannes-N-W-92}%
  \BibitemOpen
  \bibfield  {author} {\bibinfo {author} {\bibfnamefont {M.}~\bibnamefont
  {Fannes}}, \bibinfo {author} {\bibfnamefont {B.}~\bibnamefont
  {Nachtergaele}}, \ and\ \bibinfo {author} {\bibfnamefont {R.}~\bibnamefont
  {Werner}},\ }\href@noop {} {\bibfield  {journal} {\bibinfo  {journal}
  {Commun.Math.Phys.}\ }\textbf {\bibinfo {volume} {{\bf 144}}},\ \bibinfo
  {pages} {443} (\bibinfo {year} {1992})}\BibitemShut {NoStop}%
\bibitem [{\citenamefont {Kl$\ddot{\mbox{u}}$mper}\ \emph
  {et~al.}(1992)\citenamefont {Kl$\ddot{\mbox{u}}$mper}, \citenamefont
  {Schadschneider},\ and\ \citenamefont {Zittartz}}]{Klumper-S-Z-92}%
  \BibitemOpen
  \bibfield  {author} {\bibinfo {author} {\bibfnamefont {A.}~\bibnamefont
  {Kl$\ddot{\mbox{u}}$mper}}, \bibinfo {author} {\bibfnamefont
  {A.}~\bibnamefont {Schadschneider}}, \ and\ \bibinfo {author} {\bibfnamefont
  {J.}~\bibnamefont {Zittartz}},\ }\href@noop {} {\bibfield  {journal}
  {\bibinfo  {journal} {Z.Phys.}\ }\textbf {\bibinfo {volume} {{\bf B87}}},\
  \bibinfo {pages} {281} (\bibinfo {year} {1992})}\BibitemShut {NoStop}%
\bibitem [{\citenamefont {Verstraete}\ and\ \citenamefont
  {Cirac}(2006)}]{Verstraete-C-06}%
  \BibitemOpen
  \bibfield  {author} {\bibinfo {author} {\bibfnamefont {F.}~\bibnamefont
  {Verstraete}}\ and\ \bibinfo {author} {\bibfnamefont {J.~I.}\ \bibnamefont
  {Cirac}},\ }\href {http://link.aps.org/doi/10.1103/PhysRevB.73.094423}
  {\bibfield  {journal} {\bibinfo  {journal} {Phys. Rev. B}\ }\textbf {\bibinfo
  {volume} {73}},\ \bibinfo {pages} {094423} (\bibinfo {year}
  {2006})}\BibitemShut {NoStop}%
\bibitem [{\citenamefont {Hastings}(2007)}]{Hastings-area-law-07}%
  \BibitemOpen
  \bibfield  {author} {\bibinfo {author} {\bibfnamefont {M.~B.}\ \bibnamefont
  {Hastings}},\ }\href {http://stacks.iop.org/1742-5468/2007/P08024} {\bibfield
   {journal} {\bibinfo  {journal} {Journal of Statistical Mechanics: Theory and
  Experiment}\ }\textbf {\bibinfo {volume} {2007}},\ \bibinfo {pages} {P08024}
  (\bibinfo {year} {2007})}\BibitemShut {NoStop}%
\bibitem [{Note2()}]{Note2}%
  \BibitemOpen
  \bibinfo {note} {In Ref.\protect \rev@citealpnum {Arovas-H-Q-Z-09}, the
  symmetry is referred to OSp(1$|$2), but OSp(1$|$2) can also denote a
  non-compact supergroup whose bosonic subgroup is Sp(2,$\protect \mathbb
  {R}$)$\simeq $SU(1,1) or Sp(2,$\protect \mathbb {C}$)$\simeq $SO(3,1). To
  avoid possible confusions, we utilize the more precise terminology,
  UOSp(1$|$2), in the present paper.}\BibitemShut {Stop}%
\bibitem [{\citenamefont {Majumdar}\ and\ \citenamefont
  {Ghosh}(1969)}]{Majumdar-G-69}%
  \BibitemOpen
  \bibfield  {author} {\bibinfo {author} {\bibfnamefont {C.}~\bibnamefont
  {Majumdar}}\ and\ \bibinfo {author} {\bibfnamefont {D.}~\bibnamefont
  {Ghosh}},\ }\href@noop {} {\bibfield  {journal} {\bibinfo  {journal}
  {J.Math.Phys.}\ }\textbf {\bibinfo {volume} {{\bf 10}}},\ \bibinfo {pages}
  {1388,1399} (\bibinfo {year} {1969})}\BibitemShut {NoStop}%
\bibitem [{\citenamefont {Majumdar}(1970)}]{Majumdar-70}%
  \BibitemOpen
  \bibfield  {author} {\bibinfo {author} {\bibfnamefont {C.}~\bibnamefont
  {Majumdar}},\ }\href@noop {} {\bibfield  {journal} {\bibinfo  {journal}
  {J.Phys.}\ }\textbf {\bibinfo {volume} {{\bf C3}}},\ \bibinfo {pages} {911}
  (\bibinfo {year} {1970})}\BibitemShut {NoStop}%
\bibitem [{Note3()}]{Note3}%
  \BibitemOpen
  \bibinfo {note} {The open boundary condition has been implicitly assumed
  here; if the periodic boundary condition had been used, the two states would
  have been summed up with a minus sign due to the anti-commutating property of
  the holes.}\BibitemShut {Stop}%
\bibitem [{\citenamefont {Auerbach}(1994)}]{auerbach1994iea}%
  \BibitemOpen
  \bibfield  {author} {\bibinfo {author} {\bibfnamefont {A.}~\bibnamefont
  {Auerbach}},\ }\href@noop {} {\emph {\bibinfo {title} {{Interacting Electrons
  and Quantum Magnetism}}}}\ (\bibinfo  {publisher} {Springer-Verlag, Berlin},\
  \bibinfo {year} {1994})\BibitemShut {NoStop}%
\bibitem [{\citenamefont {Schrieffer}(1999)}]{SchriefferBook}%
  \BibitemOpen
  \bibfield  {author} {\bibinfo {author} {\bibfnamefont {J.~R.}\ \bibnamefont
  {Schrieffer}},\ }\href@noop {} {\emph {\bibinfo {title} {Theory Of
  Superconductivity}}}\ (\bibinfo  {publisher} {Westview Press},\ \bibinfo
  {year} {1999})\BibitemShut {NoStop}%
\bibitem [{Note4()}]{Note4}%
  \BibitemOpen
  \bibinfo {note} {If the periodic boundary condition is used, we have zero
  state for odd-length chains.}\BibitemShut {Stop}%
\bibitem [{sup()}]{supp}%
  \BibitemOpen
  \href@noop {} {}\bibinfo {note} {See supplementary material at XXXX for more
  details about SUSY.}\BibitemShut {Stop}%
\bibitem [{Note5()}]{Note5}%
  \BibitemOpen
  \bibinfo {note} {As has been argued in Ref.\protect \rev@citealpnum
  {Arovas-H-Q-Z-09}, this non-hermiticity is readily cured by adopting
  $P^{\dagger }_{L}({\protect \cal C}_{j,j+1})P_{L}({\protect \cal C}_{j,j+1})$
  instead of $P_{L}({\protect \cal C}_{j,j+1})$ itself.}\BibitemShut {Stop}%
\bibitem [{\citenamefont {Verstraete}\ \emph {et~al.}(2008)\citenamefont
  {Verstraete}, \citenamefont {Murg},\ and\ \citenamefont
  {Cirac}}]{Verstraete-M-C-08}%
  \BibitemOpen
  \bibfield  {author} {\bibinfo {author} {\bibfnamefont {F.}~\bibnamefont
  {Verstraete}}, \bibinfo {author} {\bibfnamefont {V.}~\bibnamefont {Murg}}, \
  and\ \bibinfo {author} {\bibfnamefont {J.~I.}\ \bibnamefont {Cirac}},\ }\href
  {http://search.ebscohost.com/login.aspx?direct=true&db=a9h&AN=32990557&site=%
ehost-live} {\bibfield  {journal} {\bibinfo  {journal} {Advances in Physics}\
  }\textbf {\bibinfo {volume} {57}},\ \bibinfo {pages} {143 } (\bibinfo {year}
  {2008})}\BibitemShut {NoStop}%
\bibitem [{Note6()}]{Note6}%
  \BibitemOpen
  \bibinfo {note} {Of course, we can construct `polymerized' matrix-product
  states where $m$s alternate with certain periods.}\BibitemShut {Stop}%
\bibitem [{\citenamefont {Fannes}\ \emph {et~al.}(1989)\citenamefont {Fannes},
  \citenamefont {Nachtergaele},\ and\ \citenamefont {Werner}}]{Fannes-N-W-89}%
  \BibitemOpen
  \bibfield  {author} {\bibinfo {author} {\bibfnamefont {M.}~\bibnamefont
  {Fannes}}, \bibinfo {author} {\bibfnamefont {B.}~\bibnamefont
  {Nachtergaele}}, \ and\ \bibinfo {author} {\bibfnamefont {R.}~\bibnamefont
  {Werner}},\ }\href@noop {} {\bibfield  {journal} {\bibinfo  {journal}
  {Europhys. Lett.}\ }\textbf {\bibinfo {volume} {{\bf 10}}},\ \bibinfo {pages}
  {633} (\bibinfo {year} {1989})}\BibitemShut {NoStop}%
\bibitem [{\citenamefont {den Nijs}\ and\ \citenamefont
  {Rommelse}(1989)}]{denNijs-R-89}%
  \BibitemOpen
  \bibfield  {author} {\bibinfo {author} {\bibfnamefont {M.}~\bibnamefont {den
  Nijs}}\ and\ \bibinfo {author} {\bibfnamefont {K.}~\bibnamefont {Rommelse}},\
  }\href {http://link.aps.org/doi/10.1103/PhysRevB.40.4709} {\bibfield
  {journal} {\bibinfo  {journal} {Phys. Rev. B}\ }\textbf {\bibinfo {volume}
  {40}},\ \bibinfo {pages} {4709} (\bibinfo {year} {1989})}\BibitemShut
  {NoStop}%
\bibitem [{\citenamefont {Tasaki}(1991)}]{Tasaki-91}%
  \BibitemOpen
  \bibfield  {author} {\bibinfo {author} {\bibfnamefont {H.}~\bibnamefont
  {Tasaki}},\ }\href@noop {} {\bibfield  {journal} {\bibinfo  {journal}
  {Phys.Rev.Lett.}\ }\textbf {\bibinfo {volume} {{\bf 66}}},\ \bibinfo {pages}
  {798} (\bibinfo {year} {1991})}\BibitemShut {NoStop}%
\bibitem [{\citenamefont {P{\'{e}}rez-Garc{\'{i}}a}\ \emph
  {et~al.}(2008)\citenamefont {P{\'{e}}rez-Garc{\'{i}}a}, \citenamefont {Wolf},
  \citenamefont {Sanz}, \citenamefont {Verstraete},\ and\ \citenamefont
  {Cirac}}]{Garcia-W-S-V-C-08}%
  \BibitemOpen
  \bibfield  {author} {\bibinfo {author} {\bibfnamefont {D.}~\bibnamefont
  {P{\'{e}}rez-Garc{\'{i}}a}}, \bibinfo {author} {\bibfnamefont {M.~M.}\
  \bibnamefont {Wolf}}, \bibinfo {author} {\bibfnamefont {M.}~\bibnamefont
  {Sanz}}, \bibinfo {author} {\bibfnamefont {F.}~\bibnamefont {Verstraete}}, \
  and\ \bibinfo {author} {\bibfnamefont {J.~I.}\ \bibnamefont {Cirac}},\ }\href
  {http://link.aps.org/doi/10.1103/PhysRevLett.100.167202} {\bibfield
  {journal} {\bibinfo  {journal} {Phys. Rev. Lett.}\ }\textbf {\bibinfo
  {volume} {100}},\ \bibinfo {pages} {167202} (\bibinfo {year}
  {2008})}\BibitemShut {NoStop}%
\bibitem [{Note7()}]{Note7}%
  \BibitemOpen
  \bibinfo {note} {The usual spin-1 VBS state is obtained by picking up the
  top-left 2$\times $2 block.}\BibitemShut {Stop}%
\bibitem [{\citenamefont {Hida}(1992)}]{Hida-92a}%
  \BibitemOpen
  \bibfield  {author} {\bibinfo {author} {\bibfnamefont {K.}~\bibnamefont
  {Hida}},\ }\href {http://link.aps.org/doi/10.1103/PhysRevB.45.2207}
  {\bibfield  {journal} {\bibinfo  {journal} {Phys. Rev. B}\ }\textbf {\bibinfo
  {volume} {45}},\ \bibinfo {pages} {2207} (\bibinfo {year}
  {1992})}\BibitemShut {NoStop}%
\bibitem [{\citenamefont {Lieb}\ \emph {et~al.}(1961)\citenamefont {Lieb},
  \citenamefont {Schultz},\ and\ \citenamefont {Mattis}}]{Lieb-S-M-61}%
  \BibitemOpen
  \bibfield  {author} {\bibinfo {author} {\bibfnamefont {E.}~\bibnamefont
  {Lieb}}, \bibinfo {author} {\bibfnamefont {T.}~\bibnamefont {Schultz}}, \
  and\ \bibinfo {author} {\bibfnamefont {D.}~\bibnamefont {Mattis}},\
  }\href@noop {} {\bibfield  {journal} {\bibinfo  {journal} {Ann.Phys.(N.Y.)}\
  }\textbf {\bibinfo {volume} {{\bf 16}}},\ \bibinfo {pages} {407} (\bibinfo
  {year} {1961})}\BibitemShut {NoStop}%
\bibitem [{\citenamefont {F\'{a}th}\ and\ \citenamefont
  {J.S\'{o}lyom}(1993)}]{Fath-S-93b}%
  \BibitemOpen
  \bibfield  {author} {\bibinfo {author} {\bibfnamefont {G.}~\bibnamefont
  {F\'{a}th}}\ and\ \bibinfo {author} {\bibnamefont {J.S\'{o}lyom}},\
  }\href@noop {} {\bibfield  {journal} {\bibinfo  {journal} {J.Phys.:condensed
  matter}\ }\textbf {\bibinfo {volume} {{\bf 5}}},\ \bibinfo {pages} {8983}
  (\bibinfo {year} {1993})}\BibitemShut {NoStop}%
\bibitem [{Note8()}]{Note8}%
  \BibitemOpen
  \bibinfo {note} {In the usual spin-1 VBS (AKLT) model, the overall energy
  scale fixes the parent Hamiltonian uniquely. For $S \geq 2$, the energy scale
  alone is not enough to determine the unique parent Hamiltonian.}\BibitemShut
  {Stop}%
\bibitem [{Note9()}]{Note9}%
  \BibitemOpen
  \bibinfo {note} {Using $P^{\dagger }_{3/2}P_{2}=P^{\dagger }_{2}P_{3/2}=0$,
  one can see that this definition is essentially equivalent to replacing the
  projection operators $P_{l}$ with $P^{\dagger }_{l}P_{l}$.}\BibitemShut
  {Stop}%
\bibitem [{Note10()}]{Note10}%
  \BibitemOpen
  \bibinfo {note} {In fact, we can freely add any function $V(r)$ satisfying
  $V(r)\gneq 0$ ($r>0$) and $V(r)\rightarrow 0$ ($r\rightarrow 0$). The
  simplest choice, which is regular even in the $r\rightarrow \infty $ limit,
  would be $V(r)=\protect \qopname \relax o{tanh}r$.}\BibitemShut {Stop}%
\bibitem [{\citenamefont {Anfuso}\ and\ \citenamefont
  {Rosch}(2007)}]{Anfuso-R-07-2}%
  \BibitemOpen
  \bibfield  {author} {\bibinfo {author} {\bibfnamefont {F.}~\bibnamefont
  {Anfuso}}\ and\ \bibinfo {author} {\bibfnamefont {A.}~\bibnamefont {Rosch}},\
  }\href {http://link.aps.org/doi/10.1103/PhysRevB.76.085124} {\bibfield
  {journal} {\bibinfo  {journal} {Phys. Rev. B}\ }\textbf {\bibinfo {volume}
  {76}},\ \bibinfo {pages} {085124} (\bibinfo {year} {2007})}\BibitemShut
  {NoStop}%
\bibitem [{\citenamefont {Gu}\ and\ \citenamefont {Wen}(2009)}]{Gu-W-09}%
  \BibitemOpen
  \bibfield  {author} {\bibinfo {author} {\bibfnamefont {Z.-C.}\ \bibnamefont
  {Gu}}\ and\ \bibinfo {author} {\bibfnamefont {X.-G.}\ \bibnamefont {Wen}},\
  }\href {http://link.aps.org/abstract/PRB/v80/e155131} {\bibfield  {journal}
  {\bibinfo  {journal} {Phys. Rev. B}\ }\textbf {\bibinfo {volume} {80}},\
  \bibinfo {pages} {155131} (\bibinfo {year} {2009})}\BibitemShut {NoStop}%
\bibitem [{\citenamefont {Li}\ and\ \citenamefont {Haldane}(2008)}]{Li-H-08}%
  \BibitemOpen
  \bibfield  {author} {\bibinfo {author} {\bibfnamefont {H.}~\bibnamefont
  {Li}}\ and\ \bibinfo {author} {\bibfnamefont {F.~D.~M.}\ \bibnamefont
  {Haldane}},\ }\href {http://link.aps.org/doi/10.1103/PhysRevLett.101.010504}
  {\bibfield  {journal} {\bibinfo  {journal} {Phys. Rev. Lett.}\ }\textbf
  {\bibinfo {volume} {101}},\ \bibinfo {pages} {010504} (\bibinfo {year}
  {2008})}\BibitemShut {NoStop}%
\bibitem [{\citenamefont {Rokhsar}\ and\ \citenamefont
  {Kivelson}(1988)}]{Rokhsar-K-88}%
  \BibitemOpen
  \bibfield  {author} {\bibinfo {author} {\bibfnamefont {D.~S.}\ \bibnamefont
  {Rokhsar}}\ and\ \bibinfo {author} {\bibfnamefont {S.~A.}\ \bibnamefont
  {Kivelson}},\ }\href {http://link.aps.org/abstract/PRL/v61/p2376} {\bibfield
  {journal} {\bibinfo  {journal} {Phys. Rev. Lett.}\ }\textbf {\bibinfo
  {volume} {61}},\ \bibinfo {pages} {2376} (\bibinfo {year}
  {1988})}\BibitemShut {NoStop}%
\bibitem [{\citenamefont {Borsten}\ \emph {et~al.}(2010)\citenamefont
  {Borsten}, \citenamefont {Dahanayake}, \citenamefont {Duff},\ and\
  \citenamefont {Rubens}}]{borsten-2010-81}%
  \BibitemOpen
  \bibfield  {author} {\bibinfo {author} {\bibfnamefont {L.}~\bibnamefont
  {Borsten}}, \bibinfo {author} {\bibfnamefont {D.}~\bibnamefont {Dahanayake}},
  \bibinfo {author} {\bibfnamefont {M.~J.}\ \bibnamefont {Duff}}, \ and\
  \bibinfo {author} {\bibfnamefont {W.}~\bibnamefont {Rubens}},\ }\href@noop {}
  {\bibfield  {journal} {\bibinfo  {journal} {Physical Review D}\ }\textbf
  {\bibinfo {volume} {81}},\ \bibinfo {pages} {105023} (\bibinfo {year}
  {2010})}\BibitemShut {NoStop}%
\bibitem [{\citenamefont {Hasebe}(2005)}]{hasebe2005PRL}%
  \BibitemOpen
  \bibfield  {author} {\bibinfo {author} {\bibfnamefont {K.}~\bibnamefont
  {Hasebe}},\ }\href@noop {} {\bibfield  {journal} {\bibinfo  {journal} {Phys.
  Rev. Lett.}\ }\textbf {\bibinfo {volume} {94}},\ \bibinfo {pages} {206802}
  (\bibinfo {year} {2005})}\BibitemShut {NoStop}%
\end{thebibliography}%
\end{document}